\documentclass[a4paper, 12pt]{report}

\usepackage[centertags]{amsmath}
\usepackage{amscd}
\usepackage{amsfonts}
\usepackage{amssymb}
\usepackage{amsthm}
\usepackage{graphicx}
\usepackage[spanish]{babel}

\usepackage{cite}

\newcommand{\espacio}{\qquad\quad}
\newcommand{\bra}{\left\langle}
\newcommand{\ket}{\right\rangle}
\newcommand{\bracket}{\left\langle\,\right\rangle}
\newcommand{\npart}{|\NS\rangle}
\newcommand{\vacio}{|\ceroS\rangle}
\newcommand{\f}{_\text{F}}
\newcommand{\unmedio}{{\scriptstyle\frac{1}{2}}}
\newcommand{\eff}{_{\text{ef}}}
\newcommand{\Infinity}{\infty}
\newcommand{\flip}{_{\text{flip}}}
\newcommand{\bos}{_{0\text{B}}}
\newcommand{\bosonic}{_{\text{bos}}}
\newcommand{\ferm}{_{0\text{F}}}
\newcommand{\trial}{_{\text{0}}}
\newcommand{\true}{_{\text{}}}
\newcommand{\Tint}{_{\text{int}}}

\newcommand{\sign}{\operatorname{sign}}

\newcommand{\tr}{\operatorname{tr}}
\newcommand{\diag}{\operatorname{diag}}

\newcommand{\psid}{\psi^{\dagger}}
\newcommand{\cd}{c^{\dagger}}
\newcommand{\Fd}{F^{\dagger}}

\newcommand{\bd}{b^{\dagger}}
\newcommand{\varphid}{\varphi^{\dagger}}

\renewcommand{\O}{\mathcal{O}}

\renewcommand{\L}{\mathcal{L}}
\newcommand{\D}{\mathcal{D}}
\newcommand{\Z}{\mathcal{Z}}
\newcommand{\N}{\mathcal{N}}
\renewcommand{\H}{\mathcal{H}}
\newcommand{\R}{\mathcal{R}}
\newcommand{\V}{\mathcal{V}}
\newcommand{\F}{\mathcal{F}}

\newcommand{\NS}{\mathsf{N}}
\newcommand{\ceroS}{\mathsf{0}}

\newcommand{\ch}{\hat{c}}
\newcommand{\phih}{\hat{\phi}}
\newcommand{\etah}{\hat{\eta}}

\newcommand{\lambdah}{\hat{\lambda}}

\newcommand{\Nh}{\hat{N}}
\newcommand{\sigmah}{\hat{\sigma}}
\newcommand{\rhoh}{\hat{\rho}}

\newcommand{\cdh}{\hat{c}^{\dagger}}

\newcommand{\zb}{\bar{z}}

\newcommand{\psib}{\bar{\psi}}
\newcommand{\Psib}{\bar{\Psi}}
\newcommand{\chib}{\bar{\chi}}

\newcommand{\slp}{\raise.15ex\hbox{$/$}\kern-.57em\hbox{$\partial$}}
\newcommand{\slA}{\raise.15ex\hbox{$/$}\kern-.63em\hbox{$A$}}

\newcommand{\difp}{\frac{d^2p}{(2\pi)^2}\,}

\newtheorem{propiedad}{Prop.}
\newtheorem{teorema}{Teorema}

\begin{document}

\thispagestyle{empty}

\begin{center}

\vskip 3cm

{\bf\LARGE Bosonization of Luttinger liquids: spin\\[.2cm]
flipping interactions and spin-orbit coupling}\\[.2cm]
(in spanish)\\[5 cm]

{\bf\large A thesis submitted to\\[.1cm]
Universidad Nacional de La Plata\\[.1cm]
for the\\[.1cm]
Ph.D. degree in Physics}\\[4cm]

{\bf\large by\\[.1cm]
An\'ibal Iucci\\[.1cm]
Departamento de F\'{\i}sica\\[.1cm]
Facultad de Ciencias Exactas\\[.1cm]
Universidad Nacional de La Plata}

\vskip 2 cm { \large April 2004.}
\end{center}




\chapter*{Abstract}
In this thesis we present contributions in the field of the
applications of quantum field theories techniques to condensed
matter models. In chapter 3 we investigate on the non covariant
fermionic determinant and its connection to Luttinger liquids. We
address the problem of the regularization of the theory. In
chapter 4 we treat spin flipping interactions in the non local
Thirring model and we obtain an effective bosonic actions that
describe separated spin and charge degrees of freedom. In chapter
4 we apply the self consistent harmonic approximation to
previously derived bosonic action and we obtain potential
depending equations for the spectrum gap. In chapter 5 we include
spin-orbit couplings and compute correlations functions. We show
that the spin orbit interactions modify the exponents and the
phase diagram of the system and makes new susceptibilities diverge
for low temperature. Finally in chapter 6 we summarize the main
results and the conclusions.

\chapter*{Agradecimientos}

En primer lugar me gustar\'{\i}a agradecer el apoyo del Consejo
Nacional de Investigaciones Cient\'{\i}ficas y T\'ecnicas
(CONICET), de la Universidad Nacional de La Plata (UNLP), de la
Fundaci\'on Antorchas y del Departamento de F\'{\i}sica de la
UNLP.\\

Quiero agradecer muy especialmente a Carlos por haberme guiado
aconsejado y ense\~nado un mont\'on de cosas en estos cuatro
a\~nos, y haber estado siempre dispuesto a las m\'as variadas
discusiones. A los compa\~neros del Grupo Vicky, Mariano, Marta,
Kang, y otros compa\~neros y amigos del Departamento,
Daniel Cabra, Ale, Juli\'an, Gaby, Virginia, Cecilia, Juan.\\

Por \'ultimo, nada de todo esto hubiera sido posible sin el afecto
y el apoyo de los amigos, Nico, la gente del Coro del Nacio, de mi
familia, Carlos, Alicia, Emilia, Mat\'{\i}as, Leti y Fede, los
Cardoso, los gatos Luisa y Cultr\'um, y muy especialmente de vos,
Yami.


\tableofcontents

\chapter{Introducci\'on}\label{part:introducción}

La descripci\'on de sistemas de fermiones altamente
correlacionados es un problema central de la f\'{\i}sica de
materia condensada. Durante las d\'ecadas pasadas, diversos
experimentos llevados a cabo en diferentes clases de materiales
mostraron que la alta correlaci\'on es un ingrediente fundamental
a considerar para la comprensi\'on de sus propiedades
f\'{\i}sicas. Entre otros podemos mencionar sistemas de efecto
Hall, superconductores de alta $T_c$, y diversos metales,
superconductores y aisladores org\'anicos. Por otro lado, el
tratamiento te\'orico de estos sistemas constituye una tarea
formidable. Las ideas b\'asicas acerca del comportamiento de los
electrones en materiales son conocidas, al menos intuitivamente,
desde hace muchos a\~nos. En aparente contradicci\'on con lo
expresado m\'as arriba, Sommerfeld \cite{sommerfeld28} mostr\'o
que el comportamiento lineal del calor espec\'{\i}fico de los
metales a bajas temperaturas, al igual que el comportamiento
asint\'otico a bajas temperaturas de la resistividad y de la
conductividad \'optica pod\'{\i}an ser entendidos suponiendo que
los electrones en el metal se comportaban como un gas de fermiones
no interactuantes. Simultaneamente, Pauli \cite{pauli26} calcul\'o
la susceptibilidad paramagn\'etica de electrones \emph{libres} y
hall\'o que es independiente de la temperatura, en perfecto
acuerdo con los experimentos. Al mismo tiempo, a partir de los
trabajos de Bloch \cite{bloch29} y Wigner \cite{wigner34}, se
encontr\'o que las energ\'{\i}as de interacci\'on de los
electrones en el rango met\'alico de densidades era comparable a
la energ\'{\i}a cin\'etica.

La resoluci\'on de esta paradoja surgi\'o con los trabajos de
Landau\cite{landau56,landau57} donde se introdujeron las ideas
fundamentales que dominar\'{\i}an la visi\'on de los sistemas
interactuantes en materia condensada hasta nuestros d\'{\i}as.
Landau postul\'o que los sistemas en interacci\'on evolucionan a
partir de los sistemas libres al conectar la interacci\'on de
manera adiab\'atica. Y que los estados de part\'{\i}culas en el
sistema no interactuante se corresponden uno a uno con estados de
\emph{cuasipart\'{\i}culas} o excitaciones elementales en el
sistema en interacci\'on, es decir que poseen los mismos n\'umeros
cu\'anticos. Existen sin embargo algunas restricciones: la m\'as
importante impone que s\'olo pueden considerarse excitaciones en
una escala de energ\'{\i}a peque\~na comparada con la energ\'ia de
Fermi. Esta restricci\'on sin embargo nada dice acerca de la
intensidad de las interacciones que ocurren entre los electrones;
\'estas pueden ser arbitrariamente fuertes. De all\'{\i} que los
sistemas fuertemente interactuantes que pueden tratarse con esta
teor\'{\i}a se comportan cualitativamente como sistemas libres,
con \emph{par\'ametros renormalizados por las interacciones}. Sin
embargo, las propiedades de baja temperatura de muchos materiales
que exhiben este tipo de comportamiento, poseen coeficientes que
difieren hasta un factor de $10^3$ respecto de los valores para
electrones libres. Por otro lado los estados en el sistema libre e
interactuante deben tener la misma simetr\'{\i}a, y adem\'as, al
conectar la interacci\'on no deben formarse estados ligados. Estas
restricciones impiden atacar con este formalismo problemas tales
como el ferromagnetismo o la superconductividad, que se
caracterizan justamente por esos efectos.

La teor\'{\i}a microsc\'opica que respald\'o esta teor\'{\i}a
fenomenol\'ogica pronto se desarroll\'o a partir de un
Hamiltoniano fermi\'onico con interacciones de dos
cuerpos\cite{nozieres64}. En general los objetos de inter\'es son
las funciones de Green $G(\mathbf{k},\omega)$; si se las conoce
para todos los valores de $\mathbf{k}$ y $\omega$, pueden
obtenerse en principio todas las propiedades term\'odinamicas del
sistema. Su comportamiento a bajas energ\'{\i}as y grandes
longitudes de onda est\'a relacionado con el estado fundamental y
los estados excitados m\'as bajos, y como el espectro de baja
energ\'{\i}a est\'a determinado cualitativamente por unos pocos
par\'ametros universales como la dimensi\'on, simetr\'{\i}as y
leyes de conservaci\'on, el comportamiento infrarojo de las
funciones de Green permite efectuar una clasificaci\'on de los
sistemas de muchos cuerpos en interacci\'on.

En la mayor\'{\i}a de las situaciones de inter\'es es imposible
calcular las funciones de Green en forma exacta, de modo que es
necesario recurrir a m\'etodos aproximados. El enfoque usual
consiste en hacer un desarrollo perturbativo de
$G(\textbf{k},\omega)$ en potencias de la interacci\'on. En los
llamados \emph{l\'{\i}quidos de Landau Fermi}, este enfoque
perturbativo es posible, y aunque para interacciones fuertes deben
sumarse infinitos ordenes del desarrollo, las integrales generadas
en la expansi\'on perturbativa est\'an libres de divergencias.
Este desarrollo arroja como resultado aproximado

\begin{equation}\label{eq:greenFunction}
G(\mathbf{k},\omega+i0^+)\approx
\frac{Z_\mathbf{k}}{\omega-\tilde{\xi}_\mathbf{k}+i\gamma_\mathbf{k}},
\end{equation}
para la funci\'on de Green retardada con $\mathbf{k}$ en la
vecindad de la superficie de Fermi. El n\'umero $Z_\mathbf{k}$ es
el llamado \emph{residuo de la cuasipart\'{\i}cula}, y la
energ\'{\i}a $\tilde{\xi}_\mathbf{k}$ es la energ\'{\i}a de
excitaci\'on de una cuasipart\'{\i}cula. Como en general los
l\'{\i}quidos de Landau Fermi son metales,
$\tilde{\xi}_\mathbf{k}$ no debe tener gap. Esto significa que
existe una superficie en el espacio $\mathbf{k}$ en la que
$\tilde{\xi}_\mathbf{k}=0$, lo que define la superficie de Fermi,
y $Z_{\mathbf{k}\f}$ representa la magnitud del salto de la
distribuci\'on de momentos en dicha superficie. La energ\'{\i}a
$\gamma_\mathbf{k}$ puede identificarse con el amortiguamiento de
la cuasi-part\'{\i}cula (o lo que es lo mismo,
$\tau_\mathbf{k}=1/\gamma_\mathbf{k}$ con su tiempo de vida).
N\'otese que en el plano complejo $\omega$,
$G(\mathbf{k},\omega+i0^+)$ posee un polo simple en
$\omega=\tilde{\xi}_\mathbf{k}-i\gamma_\mathbf{k}$ con residuo
$Z_\mathbf{k}$. La funci\'on de Green de los electrones no
interactuantes, denotada $G_0(\mathbf{k},\omega)$, puede obtenerse
como caso especial de la Ec. (\ref{eq:greenFunction}) tomando
$Z_\mathbf{k}=1$, $\gamma_\mathbf{k}=0$ e identificando
$\tilde{\xi}_\mathbf{k}$ con la dispersi\'on del sistema sin
interacciones. En este caso el polo simple en
$\omega=\tilde{\xi}_\mathbf{k}-i0^+$ con residuo unidad representa
la propagaci\'on no amortiguada de una part\'{\i}cula con
energ\'{\i}a $\tilde{\xi}_\mathbf{k}$. El correspondiente polo en
la funci\'on de Green del l\'iquido de Fermi en interacci\'on se
asocia con el llamado \emph{polo de la cuasipart\'{\i}cula}. El
punto importante es que en la vecindad del polo de la
cuasipart\'{\i}cula, las funciones de Green del sistema
interactuante poseen la misma estructura que las del sistema
libre. Si definimos la autoenerg\'{\i}a
$\Sigma(\mathbf{k},\omega)$ como

\begin{equation}
\Sigma(\mathbf{k},\omega)=\left[G_0(\mathbf{k},\omega)\right]^{-1}
-\left[G(\mathbf{k},\omega)\right]^{-1},
\end{equation}
entonces las cantidades $Z_\mathbf{k}$, $\tilde{\xi}_\mathbf{k}$ y
$\gamma_\mathbf{k}$ pueden calcularse de las derivadas de la
autoenerg\'{\i}a.

En algunos casos, sin embargo, la aplicaci\'on de esta maquinaria
conduce a integrales divergentes en la expansi\'on perturbativa de
$\Sigma(\mathbf{k},\omega)$. El colapso del desarrollo
perturbativo es un indicador de que la funci\'on de Green de la
teor\'{\i}a en interacci\'on no se relaciona m\'as en forma simple
con la funci\'on de Green de la teor\'{\i}a libre, por ejemplo por
la existencia de polos m\'ultiples o de singularidades no
algebraicas, lo que impide definir las cuasipart\'{\i}culas. En
este caso el sistema no puede ser un l\'{\i}quido de Fermi.

En a\~nos m\'as recientes, el inter\'es en sistemas en una
dimensi\'on espacial se vio incrementado debido a la realizaci\'on
experimental de materiales en los cuales el movimiento
electr\'onico correlacionado se encuentra efectivamente confinado
a una dimensi\'on. Podemos citar como ejemplos nanotubos de
carbono\cite{bockrath99}, estados de borde en sistemas de efecto
Hall\cite{wen90a,wen90b,kang00}, conductores
org\'anicos\cite{jerome82}, heteroestructuras semiconductoras
(alambres cu\'anticos)\cite{tarucha95}, etc. Al intentar aplicar
la teor\'{\i}a del l\'{\i}quido de Fermi a tales sistemas, se
arriba a los problemas antes mencionados. El modelo m\'as sencillo
considerado con el objeto de describir el estado met\'alico normal
de estos sistemas es el modelo de
Tomonaga-Luttinger\cite{tomonaga50,luttinger63,mattis63}. Como
veremos en el cap\'{\i}tulo siguiente, este modelo es exactamente
soluble, sus funciones de correlaci\'on pueden ser calculadas, y
todas sus propiedades se vuelven accesibles, por ejemplo el
espectro de bajas energ\'{\i}as resulta lineal y sin gap. Las
excitaciones fundamentales no son m\'as las cuasipart\'{\i}culas,
sino fluctuaciones bos\'onicas colectivas \emph{independientes} de
grados de libertad de carga y spin. Esta independencia entre ambos
se denomina separaci\'on spin-carga\cite{voit95}.

El colapso de la teor\'{\i}a de Landau del l\'{\i}quido de Fermi
en este modelo puede entenderse al observar el comportamiento de
su funci\'on de Green, que presenta un decaimiento algebraico no
universal, es decir, el exponente del decaimiento depende de las
interacciones. Adem\'as el c\'alculo de la distribuci\'on de
momentos tambi\'en arroja un comportamiento algebraico en las
proximidades de la superficie de Fermi, y se vuelve continuo en
dicha superficie, a\'un a temperatura cero.

Haldane\cite{haldane81} conjetur\'o que este conjunto de
propiedades no es exclusivo del modelo de Tomonaga-Luttinger, sino
que son propiedades gen\'ericas del estado met\'alico normal de
sistemas de electrones interactuantes en una dimensi\'on. M\'as
a\'un, llam\'o \emph{L\'{\i}quidos de Luttinger} a estos sistemas,
y propuso que el modelo de Tomonaga-Luttinger es su l\'{\i}mite a
bajas energ\'{\i}as, en el mismo sentido en el que el gas de Fermi
es el modelo libre sobre el que se construye el l\'{\i}quido de
Fermi. La inclusi\'on de interacciones que sacan al sistema del
punto fijo del l\'{\i}quido de Luttinger, como dispersi\'on hacia
atr\'as o umklapp generan gaps en los espectros de carga o spin, y
dependiendo de los valores de los acoplamientos pueden ocurrir
transiciones de fase entre un estado sin gap de tipo L\'{\i}quido
de Luttinger y otros con gaps en los espectros de carga o spin,
por ejemplo la llamada transici\'on metal aislador de
Mott\cite{voit95}. La importancia de los l\'{\i}quidos de
Luttinger recobr\'o \'{\i}mpetu hace pocos a\~nos desde la
propuesta de Anderson de que poseen propiedades que son semejantes
a la de los superconductores de alta temperatura
cr\'{\i}tica\cite{anderson90a,anderson90b}.

La correspondencia entre teor\'{\i}a y experimento para estos
sistemas unidimensionales se ha visto puesta a prueba en diversos
materiales. Las leyes de potencia de funciones de correlaci\'on se
han verificado en muchos de ellos, por ejemplo en las Ref.
\citen{postma00,gao03,hilke03} mediante propiedades de transporte.
Sin embargo la separaci\'on spin-carga ha sido m\'as elusiva, y
las pruebas m\'as convincentes son muy
recientes\cite{tserkovnyak03}.

El marco te\'orico en el cual se da tratamiento a estos sistemas
se encuentra \'{\i}ntimamente vinculado a los formalismos
utilizados en teor\'{\i}a cu\'antica de campos y en el tratamiento
de las interacciones fundamentales de la naturaleza. La
bosonizaci\'on abeliana, o trasmutaci\'on de fermiones en bosones,
es la t\'ecnica construida para la resoluci\'on original del
modelo de Tomonaga-Luttinger en materia
condensada\cite{mattis63,mattis74,luther74a}, y fue desarrollada
en paralelo con la t\'ecnica del mismo nombre en teor\'{\i}a de
campos \cite{klaiber68,coleman75,mandelstam75}. \'Esta se aplic\'o
a la resoluci\'on del modelo de Thirring, y el modelo de Thirring
masivo, que son versiones en el lenguaje de la teor\'{\i}a de
campos del modelo de Tomonaga-Luttinger. M\'as tarde, a partir del
trabajo de Fujikawa\cite{fujikawa79}, la t\'ecnica de
bosonizaci\'on en teor\'ia de campos se extendi\'o para ser
aplicada en el contexto de las integrales
funcionales\cite{furuya82,naon85}, y un proceso equivalente
ocurri\'o en materia condensada\cite{fogebdy76,lee88}. M\'as
espec\'{\i}ficamente, la bosonizaci\'on funcional se aplic\'o en
la Ref. \citen{naon95} al estudio de una versi\'on del modelo de
Thirring, en la que se modific\'o el acoplamiento entre las
corrientes fermi\'onicas para dar lugar a la posibilidad de una
interacci\'on no local. En la Ref. \citen{iucci00} se aplic\'o
dicha formulaci\'on al c\'alculo de las funciones de Green del
modelo con interacciones de largo alcance, m\'as
espec\'{\i}ficamente de tipo columbiano.

En esta tesis presentamos contribuciones originales en el campo de
las aplicaciones de las teor\'{\i}as cu\'anticas de campos a la
formulaci\'on de modelos de materia condensada en los que el spin
electr\'onico juega un rol crucial. En particular nos concentramos
en el modelo de Tomonaga-Luttinger con spin y dos extensiones
posibles: interacciones de inversi\'on de spin, y acoplamiento
spin-\'orbita\cite{moroz00b,moroz00c}. Adem\'as estudiamos
problemas que surgen al realizar la bosonizaci\'on de teor\'{\i}as
de materia condensada en el marco de la integral funcional. El
plan es el siguiente: en el cap\'{\i}tulo
\ref{part:bosonizaciónOp} presentamos la bosonizaci\'on en el
marco operacional en forma detallada, y la aplicamos al estudio de
modelos concretos de teor\'{\i}as de muchos cuerpos en baja
dimensi\'on. Para esto seguimos la bibliograf\'{\i}a
est\'andar\cite{voit95,vondelft98,varma02}. En el cap\'{\i}tulo
\ref{part:jacobiano} analizamos las ambig\"{u}edades que se
presentan en la bosonizaci\'on debido a los necesarios mecanismos
de regularizaci\'on que deben implementarse, particularmente
cuando la teor\'{\i}a bajo estudio no posee la invarianza de
Lorentz de las teor\'{\i}as de campos usuales, y allanamos el
camino para estudiar teor\'{\i}as m\'as complejas de materia
condensada mediante bosonizaci\'on funcional\cite{iucci04}. En el
cap\'{\i}tulo \ref{part:spinflip} atacamos uno de esos modelos: el
modelo de Thirring no local con dos especies de fermiones, que es
una versi\'on de teor\'{\i}a de campos del modelo de
Tomonaga-Luttinger con spin. Consideramos adem\'as el efecto de
a\~nadir al Lagrangiano t\'erminos de inversi\'on de spin.
Obtenemos una acci\'on bos\'onica efectiva que representa
oscilaciones de densidad de carga y de spin de forma
independiente; la funci\'on de partici\'on resulta factorizada,
dando lugar a la separaci\'on spin-carga\cite{iucci01}. En el
cap\'{\i}tulo \ref{part:aaac} examinamos la aproximaci\'on
arm\'onica autoconsistente y su utilizaci\'on en el marco de la
integral funcional y las teor\'{\i}as de materia condensada.
Hallamos una f\'ormula para el gap de las excitaciones del sector
de spin del modelo estudiado en el cap\'{\i}tulo anterior, como
funci\'on de potenciales arbitrarios de interacciones
electr\'on-electr\'on de tipo dispersi\'on hacia adelante
\cite{iucci01}. En este cap\'{\i}tulo proponemos adem\'as un nuevo
m\'etodo para determinar el par\'ametro inc\'ognita asociado a la
aproximaci\'on y lo aplicamos al estudio del r\'egimen de escala
del modelo de Ising en 2D fuera del punto cr\'{\i}tico y en
presencia de un campo magn\'etico $h$\cite{iucci02}. En el
cap\'{\i}tulo \ref{part:spinorbit} calculamos funciones de
correlaci\'on en sistemas unidimensionales de electrones en
interacci\'on en los que los grados de libertad de carga y spin se
encuentran acoplados a trav\'es de la interacci\'on spin-\'orbita.
Este acoplamiento est\'a representado por una asimetr\'{\i}a en el
espectro libre de los electrones. Estudiamos fluctuaciones de tipo
ondas de densidad de carga y spin, y de tipo superconductor
singulete y triplete. Mostramos que la interacci\'on spin-\'orbita
modifica los exponentes del decaimiento de las funciones de
correlaci\'on y el diagrama de fases del sistema. Adem\'as
encontramos que susceptibilidades que eran finitas a bajas
temperaturas, se vuelven divergentes cuando la interacci\'on
spin-\'orbita es suficientemente intensa\cite{iucci03}. Para
concluir, en el cap\'{\i}tulo \ref{part:conclusionesFinales}
reunimos los resultados m\'as destacados y las conclusiones.

Finalmente queremos mencionar que las investigaciones realizadas
en esta tesis se complementan con los estudios del modelo de
Tomonaga Luttinger sin spin realizados en las Ref.
\citen{fernandez02a} y \citen{fernandez02b}. En la primera se
estudi\'o el efecto de interacciones de tipo dispersi\'on hacia
adelante y umklapp no locales, y la generalizaci\'on de la
ecuaci\'on del gap a estos casos, y en la segunda se consider\'o
la asimetr\'{\i}a en el espectro libre, pero en el caso
simplificado de electrones sin spin.


\chapter{Bosonizaci\'on}\label{part:bosonizaciónOp}

\begin{center}
\begin{minipage}{5.6in}
\textsl{Presentaremos una derivaci\'on exhaustiva de la t\'ecnica
de bosonizaci\'on en el marco operacional. Seguiremos las
referencias usuales para mostrar la equivalencia entre operadores
fermi\'onicos y operadores bos\'onicos. Este es el punto de
partida para el estudio de teor\'{\i}as de materia condensada en
una dimensi\'on.}
\end{minipage}
\end{center}

\section{Equivalencia entre operadores fermi\'onicos y
bos\'o\-nicos}

\subsection{Campos fermi\'onicos}

Tomemos una teor\'{\i}a que puede formularse en t\'erminos de un
conjunto de operadores de creaci\'on y de aniquilaci\'on
fermi\'onicos en una dimensi\'on espacial, que satisfacen
relaciones can\'onicas de anticonmutaci\'on

\begin{equation}\label{eq:operadoresAnticonmutan}
\left\{c_{krs},\cd_{k'r's'}\right\}=\delta_{kk'}\delta_{rr'}\delta_{ss'}.
\end{equation}
Estos operadores est\'an etiquetados por los \'{\i}ndices $r$, que
distingue part\'{\i}culas que se mueven a la derecha ($r=+1$) o a
la izquierda ($r=-1$), $s$ que en general puede referirse a $M$
especies de fermiones, por ejemplo en problemas de m\'ultiples
cadenas, pero que usualmente se utilizar\'a para indicar el spin
electr\'onico ($s=+1$ para spin para arriba y $s=-1$ para spin
para abajo), y un \'{\i}ndice discreto y no acotado $k$ que denota
el momento (o n\'umero de onda), de la forma

\begin{equation}\label{eq:operadoresIndice}
k=\frac{2\pi}{L}\left(n_k-\frac{1}{2}\delta_b\right),\qquad\text{con}\;n_k\in\mathbb{Z}\;
\text{y}\;\delta_b\in[0,2).
\end{equation}
Aqu\'{\i} $L$ es la longitud asociada al tama\~no del sistema y
$\delta_b$ es un par\'ametro que determina las condiciones de
contorno del problema. $k$ usualmente etiqueta las
autoenerg\'{\i}as $\epsilon_k$ del sistema libre (con $\epsilon_0$
correspondiente a la energ\'{\i}a de Fermi $\epsilon\f$). Que este
\'{\i}ndice sea discreto y no acotado es un requisito
indispensable para realizar una derivaci\'on rigurosa de las
identidades de bosonizaci\'on. Estas identidades son
independientes de un problema espec\'{\i}fico como puede serlo el
modelo de Tomonaga-Luttinger, o el problema de Kondo; y de la
relaci\'on de dispersi\'on $\epsilon_k$. Esto es posible porque
dichas identidades son igualdades entre operadores, es decir,
v\'alidas cuando act\'uan sobre cualquier estado del espacio de
Fock. Son independientes entonces del Hamiltoniano, cuya forma
detallada s\'olo se vuelve relevante al calcular funciones de
correlaci\'on. Su aplicaci\'on a modelos m\'as concretos ser\'a
analizada m\'as adelante. En esta secci\'on seguiremos en detalle
la exposici\'on hecha en la Ref. \citen{vondelft98} con dos
diferencias m\'{\i}nimas: un cambio en la normalizaci\'on, para
adecuarla a las aplicaciones de materia condensada m\'as usuales,
y la inclusi\'on expl\'{\i}cita del \'{\i}ndice $r$, debido a que
ciertos conmutadores dependen expl\'{\i}citamente de \'el (en la
mencionada referencia se lo incluy\'o junto con el \'{\i}ndice $s$
en un \'unico \'{\i}ndice $\eta=1,...,M$).

Comenzando por el dado conjunto de operadores de destrucci\'on
$c_{krs}$ con las propiedades (\ref{eq:operadoresAnticonmutan}) y
(\ref{eq:operadoresIndice}) definimos un conjunto de campos
fermi\'onicos de la siguiente manera:

\begin{equation}\label{eq:camposFermionicos}
\psi_{rs}(x)=\frac{1}{\sqrt{L}}\sum_{k=-\Infinity}^\Infinity
e^{irkx}c_{krs},\qquad\psid_{rs}(x)=\frac{1}{\sqrt{L}}\sum_{k=-\Infinity}^\Infinity
e^{-irkx}\cd_{krs},
\end{equation}
donde $x\in[-\Infinity,\Infinity]$ es la variable espacial. Sus
inversas son

\begin{equation}
c_{krs}=\frac{1}{\sqrt{L}}\int_0^L
dx\,e^{-irkx}\psi_{rs}(x),\qquad
\cd_{krs}=\frac{1}{\sqrt{L}}\int_0^L dx\,e^{irkx}\psid_{rs}(x).
\end{equation}
Los operadores $\psi_{rs}$ satisfacen las condiciones de contorno

\begin{equation}
\psi_{rs}(x+L)=e^{i\pi\delta_b}\psi_{rs}(x)\qquad\text{y}\qquad
\psid_{rs}(x+L)=e^{-i\pi\delta_b}\psid_{rs}(x),
\end{equation}
peri\'odicas para $\delta_b=0$ y antiperi\'odicas para
$\delta_b=1$. Las ecuaciones (\ref{eq:operadoresAnticonmutan}) y
(\ref{eq:operadoresIndice}), junto con la identidad

\begin{equation}\label{eq:deltaPeriodica}
\sum_{n\in\mathbb{Z}}e^{iny}=2\pi\sum_{m\in\mathbb{Z}}\delta(y-2\pi
m)
\end{equation}
implican de inmediato las relaciones de anticonmutaci\'on

\begin{align}
\left\{\psi_{rs}(x),\psid_{r's'}(x')\right\}=&
\delta_{rr'}\delta_{ss'}\sum_{n\in\mathbb{Z}}\delta(x-x'-nL)e^{in\pi\delta_b}\\
\Big\{\psi_{rs}(x),\psi_{r's'}(x')\Big\}=&\left\{\psid_{rs}(x),\psid_{r's'}(x')\right\}=0.
\end{align}
Para $x,x'\in[0,L]$ o $L\rightarrow\Infinity$, y condiciones de
contorno peri\'odicas se reducen a las relaciones usuales para
campos fermionicos.

El vac\'{\i}o fermi\'onico $\vacio_0$ (llamado a veces \emph{mar
de Fermi}) se define en la forma

\begin{align}
c_{krs}\vacio_0&\equiv 0\qquad\text{para}\qquad k>0\,(n>0)\label{eq:defVacio1}\\
\cd_{krs}\vacio_0&\equiv 0\qquad\text{para}\qquad
k\leq0\,(n\leq0),\label{eq:defVacio2}
\end{align}
es decir que es un estado tal que para todos los valores de $r$ y
$s$ los niveles llenos m\'as altos corresponden a $n_k=0$ y los
vac\'{\i}os m\'a bajos a $n_k=1$. Respecto a este vac\'{\i}o se
define la operaci\'on de orden normal del producto de operadores
$ABC\dots$ como

\begin{equation}\label{eq:ordenNormalFerm}
:ABC\dots:\;=\;ABC\dots\;-\;{}_0\langle\ceroS|ABC\dots\vacio_0,
\end{equation}
para $A,\,B,\,C,\,\dots\,\in\{c_{krs};\cd_{krs}\}$. Esta
definici\'on es equivalente a agrupar todos los operadores
$c_{krs}$ con $k>0$ y todos los $\cd_{krs}$ con $k\leq0$ a la
derecha de los dem\'as.

El operador n\'umero de part\'{\i}culas de tipo $rs$ se define
como

\begin{equation}
\hat{N}_{rs}\;\equiv\;\sum_{k=-\Infinity}^\Infinity :\cd_{krs}
c_{krs} :\; = \;\sum_{k=-\Infinity}^\Infinity\left[\cd_{krs}
c_{krs}-{}_0\langle\ceroS|\cd_{krs} c_{krs}\vacio_0\right].
\end{equation}
Mediante la aplicaci\'on de operadores de creaci\'on y
aniquilaci\'on sobre el vac\'{\i}o construimos estados con
distintas configuraciones de part\'{\i}culas y agujeros, $\npart$,
autoestados de los operadores $\hat{N}_{rs}$, con autovalores
$N_{rs}$

\begin{equation}
\hat{N}_{rs}\npart=N_{rs}\npart.
\end{equation}

Designamos con la letra $\NS$ al conjunto de los autovalores
$N_{rs}$ para los diferentes $r,s$, y por abuso de lenguaje
diremos que un estado de $\NS$ part\'{\i}culas es un estado en el
que hay $N_{rs}$ part\'{\i}culas de tipo $rs$. N\'otese que es
posible aniquilar part\'{\i}culas con $k<0$ (ya que justamente el
mar de Fermi est\'a lleno hasta el nivel $k=0$). Alternativamente
en este caso decimos que creamos un agujero con impulso $k$. Esto
disminuye el autovalor $N_{rs}$, que puede tomar as\'{\i} valores
negativos.

El conjunto de autoestados con un dado $\NS$ conforman el espacio
de Hilbert de $\NS$ part\'{\i}culas $\H_\NS$. El espacio de Fock
$\F$ se define como suma directa de los espacios de Hilbert con
n\'umero fijo de part\'{\i}culas $\F=\sum_{\oplus\NS}\H_\NS$.

Entre todos los estados con el mismo $\NS$ hay uno que posee menor
energ\'{\i}a, es aquel que est\'a lleno hasta un determinado
nivel, y vac\'{\i}o de all\'{\i} en m\'as. \'Este es el estado
fundamental de $\H_\NS$, $\npart_0$. Podemos dar una definici\'on
m\'as precisa de este estado:

\begin{equation}
\npart_0\equiv\prod_{r,s}C_{rs}^{N_{rs}}\vacio_0,
\end{equation}
donde

\begin{equation}
C_{rs}^{N_{rs}}\equiv
\begin{cases}
\cd_{N_{rs}rs}\cd_{(N_{rs}-1)rs}\dots\cd_{1rs}&\text{para}\,N_{rs}>0,\\
1&\text{para}\,N_{rs}=0,\\
c_{(N_{rs}+1)rs}c_{(N_{rs}+2)rs}\dots
c_{0rs}&\text{para}\,N_{rs}<0.
\end{cases}
\end{equation}

\subsection{Campos bos\'onicos}

A partir del estado $\npart_0$ pueden construirse el resto de las
excitaciones de $\NS$ part\'{\i}culas. Definimos los siguientes
operadores de \emph{creaci\'on y aniquilaci\'on bos\'onicos} que
cumplen dicha tarea,

\begin{equation}
\bd_{qrs}\equiv\frac{i}{\sqrt{n_q}}\sum_{k=-\Infinity}^\Infinity\cd_{k+q\;rs}
c_{krs},\qquad
b_{qrs}\equiv\frac{-i}{\sqrt{n_q}}\sum_{k=-\Infinity}^\Infinity\cd_{k-q\;rs}
c_{krs},\label{eq:operadoresBosonicos}
\end{equation}
donde $n_q\in\mathbb{Z}^+$ es un entero positivo, y $q=2\pi
n_q/L>0$. Estos operadores, al actuar sobre cualquier estado
$\npart$ crean una combinaci\'on de excitaciones de
part\'{\i}cula-agujero sobre ese estado con $q$ unidades de
momento m\'as (o menos), pero sin salirse de $\H_\NS$. En este
sentido son operadores que aumentan y disminuyen el momento. Su
normalizaci\'on se eligi\'o de modo que satisfagan relaciones de
conmutaci\'on bos\'onicas

\begin{equation}
\Big[b_{qrs},b_{q'r's'}\Big]=\,\Big[\bd_{qrs},\bd_{q'r's'}\Big]=0,\qquad
\Big[\hat{N}_{rs},b_{q'r's'}\Big]=\,\Big[\hat{N}_{rs},\bd_{q'r's'}\Big]=0,\label{eq:conmBos1}
\end{equation}

\begin{align}
\Big[b_{qrs},\bd_{q'r's'}\Big]=&\;\delta_{rr'}\delta_{ss'}\frac{1}{\sqrt{n_q
n_q'}} \sum_{k=-\Infinity}^\Infinity\left(\cd_{k+q'-q\,rs}
c_{krs}-\cd_{k+q'\,rs} c_{k+q\,rs}\right)\notag\\
=&\;\delta_{rr'}\delta_{ss'}\delta_{qq'}\sum_k\frac{1}{n_q}\bigg\{\left[:\cd_{krs}
c_{krs}:-:\cd_{k+q\,rs} c_{k+q\,rs}:\right]\notag\\
&\qquad\qquad\qquad\qquad+\left({}_0\langle\ceroS|\cd_{krs}
c_{krs}\vacio_0 - {}_0\langle\ceroS|\cd_{k+q\,rs}
c_{k+q\,rs}\vacio_0\right)\bigg\}\notag\\
=&\;\delta_{rr'}\delta_{ss'}\delta_{qq'}\label{eq:conmBos2}.
\end{align}

Las ecuaciones (\ref{eq:conmBos1}) se pueden verificar
f\'acilmente, pero la derivaci\'on de (\ref{eq:conmBos2}) requiere
cierto cuidado, como notaron por primera vez Mattis y
Lieb\cite{mattis63}: para $q\neq q'$ los dos t\'erminos en la
primera l\'{\i}nea ya est\'an ordenados normalmente (esto es
porque sus valores medios de vac\'{\i}o son nulos) y pueden
restarse trivialmente mediante un cambio $k\rightarrow k-q'$ en el
segundo t\'ermino, dando cero como resultado. Sin embargo, para
$q=q'$ antes de hacer la sustracci\'on debemos construir
expresiones ordenadas normalmente, de otro modo estar\'{\i}amos
restando expresiones infinitas de un modo no controlado. Los
t\'erminos en la segunda l\'{\i}nea se cancelan, reemplazando en
el segundo t\'ermino $k\rightarrow k-q$ (esto ahora si se puede
hacer porque est\'an ordenados normalmente). La definici\'on del
vac\'{\i}o (Ecs. (\ref{eq:defVacio1}) y (\ref{eq:defVacio2}))
implica que la diferencia en los valores de expectaci\'on de la
tercera l\'{\i}nea arroja como resultado

\begin{equation}
\frac{1}{n_q}\left(\;\sum_{n_k=-\Infinity}^0 -
\sum_{n_k=-\Infinity}^{-n_q}\;\right)=\frac{1}{n_q}\;n_q=1.
\end{equation}

N\'otese que la construcci\'on de los operadores $b_{qrs}$
(\ref{eq:operadoresBosonicos}) y la derivaci\'on de los
conmutadores (\ref{eq:conmBos2}) descansa fuertemente en el hecho
de que el conjunto de $k$s es infinito y no acotado por debajo.

Es f\'acil verificar que dentro de $\H_\NS$, $\npart_0$ act\'ua
como estado fundamental para las excitaciones bos\'onicas:

\begin{equation}\label{eq:estadoFunBos}
b_{qrs}|\textsf{N}\rangle_0=0,\qquad\text{para todo }q,r,s.
\end{equation}
Intuitivamente esto es claro: si $\npart_0$ es el estado
fundamental entre todos aquellos que contienen $\NS$
part\'{\i}culas, entonces no se le pueden quitar unidades de
momento sin quitar part\'{\i}culas, es decir, sin salir de
$\H_\NS$.

Es obvio que los estados excitados $\npart$ que conforman el
espacio de Hilbert de $\NS$ part\'{\i}culas se pueden obtener
actuando sobre $\npart_0$ con alguna funci\'on de los operadores
fermi\'onicos: $\npart=\bar{f}(\cd_{krs},c_{k'rs})\npart_0$.
Haldane\cite{haldane81} mostr\'o que tambi\'en existe una
representaci\'on en t\'erminos de los $\bd_{qrs}$. M\'as
espec\'{\i}ficamente, mostr\'o lo siguiente:

\begin{teorema}
Para cualquier estado $\npart$, existe una funci\'on $f(\bd)$ tal
que

\begin{equation}\label{eq:factorizacionNpart}
\npart=f(\bd)\npart_0.
\end{equation}
\end{teorema}
\noindent Esta es una afirmaci\'on para nada trivial ya que los
operadores $\bd$ crean complejas combinaciones de excitaciones
part\'{\i}cula-agujero; y constituye el coraz\'on de la
bosonizaci\'on debido a que implica una igualdad entre espacios de
Fock bos\'onicos y fermi\'onicos. Omitiremos aqu\'{\i} la
demostraci\'on, y remitiremos al lector a la mencionada
referencia, y tambi\'en a la Ref. \citen{vondelft98}.

El estado fundamental $\npart_0$ sirve para definir una
operaci\'on de orden normal bos\'onica de un producto de
operadores de tipo $b_{qrs}$ y $\bd_{qrs}$ de manera an\'aloga al
orden normal fermi\'onico (\ref{eq:ordenNormalFerm}). M\'as a\'un,
ambos son equivalentes, es decir que si un producto de operadores
bos\'onicos est\'a ordenado normalmente de acuerdo al orden
bos\'onico, entonces tambi\'en lo est\'a de acuerdo al orden
fermi\'onico, y viceversa. Por este motivo se utiliza la misma
notaci\'on para ambos.

Con los operadores bos\'onicos definidos en la Ec
(\ref{eq:operadoresBosonicos}) podemos definir campos bos\'onicos:

\begin{equation}\label{eq:camposBosonicos}
\varphi_{rs}(x)=-\sum_{q>0}\frac{1}{\sqrt{n_q}}e^{irqx}b_{qrs}e^{-aq/2},\qquad
\varphid_{rs}(x)=-\sum_{q>0}\frac{1}{\sqrt{n_q}}e^{-irqx}\bd_{qrs}e^{-aq/2},
\end{equation}
y su combinaci\'on herm\'{\i}tica

\begin{equation}
\phi_{rs}(x)=\varphi_{rs}(x)+\varphid_{rs}(x)=
-\sum_{q>0}\frac{1}{\sqrt{n_q}}\left(e^{irqx}b_{qrs}+e^{-irqx}\bd_{qrs}\right)e^{-aq/2}.
\end{equation}
Aqu\'{\i} $a>0$ es un par\'ametro infinitesimal que regulariza
divergencias ultravioletas que ocurren en ciertas expresiones y
conmutadores no ordenados normalmente. Usualmente se toma del
orden del espaciado de red $a\sim1/k\f$. Los campos as\'i
definidos satisfacen las relaciones de conmutaci\'on

\begin{align}
\Big[\varphi_{rs}(x),\varphi_{r's'}(x')\Big]
&=\;\Big[\varphid_{rs}(x),\varphid_{r's'}(x')\Big]=0,\\
\Big[\varphi_{rs}(x),\varphid_{r's'}(x')\Big]
&=\;\delta_{rr'}\delta_{ss'}\sum_{q>0}\frac{1}{n_q}e^{q[ir(x-x')-a]}\\
&=\;-\delta_{rr'}\delta_{ss'}\ln\left[1-e^{\frac{2\pi}{L}[ir(x-x')-a]}\right]\label{eq:conmBosLog}\\
& \xrightarrow{L\rightarrow\Infinity} -\delta_{rr'}\delta_{ss'}
\ln\left[\frac{2\pi}{L}[a-ir(x-x')]\right].
\end{align}
La Ec. (\ref{eq:conmBosLog}) se obtuvo utilizando la expansi\'on
en serie de $\log(1-y)$. Aqu\'i se ve claramente que $a$ act\'ua
como cut-off de la divergencia ultravioleta para $x=x'$. Estos
conmutadores son \'utiles en la evaluaci\'on del producto de
operadores de v\'ertice (exponenciales de campos bos\'onicos).
Utilizando la identidad

\begin{equation}
e^A e^B=e^{A+B}e^{[A,B]/2},
\end{equation}
para operadores $A$ y $B$ que conmutan con $[A,B]$, obtenemos

\begin{align}
e^{i\varphid_{rs}(x)}e^{i\varphi_{rs}(x)}=&
\;e^{i(\varphid_{rs}+\varphi_{rs})(x)}e^{[i\varphid_{rs}(x),\,i\varphi_{rs}(x)]/2}=
\left(\frac{L}{2\pi a}\right)^{1/2}e^{i\phi_{rs}(x)}\label{eq:opVertice1}\\[10pt]
e^{-i\varphi_{rs}(x)}e^{-i\varphid_{rs}(x)}=&
\;e^{-i(\varphi_{rs}+\varphid_{rs})(x)}e^{[-i\varphi_{rs}(x),\,-i\varphid_{rs}(x)]/2}=
\left(\frac{2\pi
a}{L}\right)^{1/2}e^{-i\phi_{rs}(x)}\label{eq:opVertice2}
\end{align}
N\'otese que estas f\'ormulas son v\'alidas para cualquier valor
de $L$ siempre que $a$ sea suficientemente chico (esto es as\'{\i}
porque para $x=x'$ el l\'{\i}mite $L\rightarrow\Infinity$ en
(\ref{eq:conmBosLog}) es equivalente a $a\rightarrow 0$). Resulta
interesante tambi\'en la evaluaci\'on del conmutador del campo
$\phi_{rs}(x)$ con su derivada:

\begin{equation}\label{eq:conmBosDeriv}
\Big[\phi_{rs}(x),\partial_{x'}\phi_{r's'}(x')\Big]=
-\delta_{rr'}\delta_{ss'}ir\frac{2\pi}{L} \sum_{n_q=1}^\Infinity
\left[e^{\frac{2\pi}{L}[ir(x-x')-a]n_q}+e^{\frac{2\pi}{L}[-ir(x-x')-a]n_q}\right]
\end{equation}
A partir de aqu\'i podemos obtener dos expresiones diferentes de
acuerdo a c\'omo se tomen los l\'{\i}mites para
$L\rightarrow\Infinity$ y $a$ infinitesimal. Si queremos una
expresi\'on no peri\'odica, para $L$ grande, es conveniente hacer
la suma geom\'etrica, y posteriormente tomar los l\'{\i}mites
dejando el l\'{\i}mite $a\rightarrow0$ para el final:

\begin{multline}
\Big[\phi_{rs}(x),\partial_{x'}\phi_{r's'}(x')\Big]=
-\delta_{rr'}\delta_{ss'}\frac{2\pi}{L}ir
\left[\frac{1}{e^{\frac{2\pi}{L}[a-ir(x-x')]}-1}+
\frac{1}{e^{\frac{2\pi}{L}[a+ir(x-x')]}-1}\right]\\[10pt]
\xrightarrow{L\rightarrow\Infinity}-\delta_{rr'}\delta_{ss'}2\pi
ir\frac{a/\pi}{(x-x')^2+a^2}
\xrightarrow{a\rightarrow0}-\delta_{rr'}\delta_{ss'}2\pi
ir\delta(x-x').
\end{multline}
N\'otese que para tomar correctamente el l\'{\i}mite
$L\rightarrow\Infinity$ en la primera l\'{\i}nea de las
expresiones precedentes, se deben desarrollar los exponenciales
hasta orden cuadr\'atico en $1/L$. Para $L$ finito, en cambio,
tomamos primero el l\'{\i}mite $a\rightarrow 0$ en
(\ref{eq:conmBosDeriv}), y utilizamos la identidad
(\ref{eq:deltaPeriodica}):

\begin{equation}
\Big[\phi_{rs}(x),\partial_{x'}\phi_{r's'}(x')\Big]=-\delta_{rr'}\delta_{ss'}2\pi
ir\left[\sum_{n\in\mathbb{Z}}\delta(x-x'-nL)-\frac{1}{L}\right],
\end{equation}
donde el t\'ermino $1/L$ en esta \'ultima ecuaci\'on aparece
debido a la ausencia del t\'ermino $n_q=0\,(q=0)$ en la Ec.
(\ref{eq:conmBosDeriv}). Finalmente podemos calcular el conmutador
del campo $\phi_{rs}$ con si mismo, obteniendo

\begin{equation}\label{eq:conmCampoPhi}
\Big[\phi_{rs}(x),\phi_{r's'}(x')\Big]
\xrightarrow{L\rightarrow\Infinity, a\rightarrow0} 2\pi i
r\delta_{rr'}\delta_{ss'}\epsilon(x-x')\qquad\text{donde}\;\epsilon(x)=
\begin{cases}
    \pm1 & \text{si}\;x\gtrless 0,\\
    0    & \text{si}\;x=0.
\end{cases}
\end{equation}

\subsection{Factores de Klein}

Los operadores $b$ y $\bd$ crean excitaciones dentro del espacio
de Hilbert de $\NS$ part\'{\i}culas. Debemos definir entonces
operadores que conecten espacios de Hilbert con diferente n\'umero
de part\'{\i}culas, es decir, operadores escalera que aumenten o
disminuyan el n\'umero fermi\'onico total, cosa que no pueden
hacer los operadores bos\'onicos.

Definimos los \emph{factores de Klein} $F_{rs}$ y $\Fd_{rs}$ como
operadores con las siguientes propiedades: i) conmutan con todos
los operadores bos\'onicos:

\begin{equation}\label{eq:defFactoresKlein1}
\Big[b_{qrs},\Fd_{r's'}\Big]=\Big[b_{qrs},F_{r's'}\Big]=
\Big[\bd_{qrs},\Fd_{r's'}\Big]=\Big[\bd_{qrs},F_{r's'}\Big]=0\quad\text{para
todo}\;q,r,r',s,s',
\end{equation}
y ii) su acci\'on sobre un estado $\npart_0$, es la de agregar una
part\'{\i}cula en el nivel m\'as bajo posible, y la de quitar una
en el m\'as alto respectivamente:

\begin{align}\label{eq:defFactoresKlein2}
\Fd_{rs}\npart_0 &\equiv \cd_{(N_{rs}+1)rs}\npart_0,\\
F_{rs}\npart_0 &\equiv c_{N_{rs}rs}\npart_0.
\end{align}
de este modo queda definida su acci\'on sobre cualquier estado
$\npart$. En efecto, el estado $\npart$ puede descomponerse de
acuerdo a la Ec. (\ref{eq:factorizacionNpart}), y por lo tanto,

\begin{align}
\Fd_{rs}\npart &= f(\bd)\cd_{(N_{rs}+1)rs}\npart_0,\\
F_{rs}\npart &= f(\bd)c_{N_{rs}rs}\npart_0.
\end{align}
Es decir, que el estado $\Fd_{rs}\npart$ (o $F_{rs}\npart$) posee
el mismo conjunto de excitaciones bos\'onicas que el estado
$\npart$, pero creadas sobre un estado con una part\'{\i}cula
m\'as (o menos). As\'{\i} definidos, los factores de Klein poseen
las siguientes propiedades:

\begin{align}
&F_{rs}\Fd_{rs}=\Fd_{rs}F_{rs}=1\qquad(\text{unitariedad});\\[10pt]
&\Big\{\Fd_{rs},F_{r's'}\Big\}=
2\delta_{rr'}\delta_{ss'}\qquad\text{para todo}\;r,r',s,s';\\[10pt]
&\Big\{\Fd_{rs},\Fd_{r's'}\Big\}=\Big\{F_{rs},F_{r's'}\Big\}=0\qquad\text{para}\;r\neq r',s\neq s';\\[10pt]
&\Big[\Nh_{rs},\Fd_{r's'}\Big]=\delta_{rr'}\delta_{ss'}\Fd_{rs},\qquad
\Big[\Nh_{rs},F_{r's'}\Big]=-\delta_{rr'}\delta_{ss'}F_{rs}.
\end{align}
Para probar la unitariedad es fundamental que el espectro del
operador $\Nh_{rs}$ sea no acotado.

\subsection{Identidades de bosonizaci\'on}

Con todas las definiciones y propiedades estudiadas estamos en
condiciones de establecer igualdades entre operadores de campos
bos\'onicos y fermi\'onicos. La primera de ellas, la m\'as simple
de derivar, establece una igualdad entre la densidad electr\'onica
ordenada normalmente, y la derivada del campo bos\'onico
$\partial_x\phi_{rs}(x)$:

\begin{align}
\rho_{rs}(x)\equiv & :\psid_{rs}(x)\psi_{rs}(x):=
\frac{1}{L}\sum_{q}e^{irqx}\sum_k:\cd_{k-q\,rs}c_{krs}:\label{eq:defDensidad}\\
=&\;\frac{1}{L}\sum_{q>0}i\sqrt{n_q}\left(e^{irqx}b_{qrs}-e^{-irqx}\bd_{qrs}\right)
+ \frac{1}{L}\sum_k:\cd_{krs}c_{krs}:\\
=&\;-r\frac{1}{2\pi}\partial_x\phi_{rs}(x)+\frac{1}{L}\Nh_{rs}\qquad(\text{para}\;a\rightarrow
0)\label{eq:bosDensidad}.
\end{align}

La segunda, relaciona el campo fermi\'onico con el operador de
v\'ertice bos\'onico. Para derivarla debemos mostrar previamente
la siguiente propiedad:

\begin{propiedad}
$\psi_{rs}(x)\npart_0$ es un estado coherente bos\'onico
\end{propiedad}
Mostraremos que dicho estado es un autoestado de $b_{qrs}$ y por
lo tanto posee una representaci\'on como estado coherente. Para
ello basta con calcular los conmutadores de $b$ y $\bd$ con
$\psi$:

\begin{align}
\Big[b_{qr's'},\psi_{rs}(x)\Big]&=\delta_{rr'}\delta_{ss'}\alpha_{qr}(x)\psi_{rs}(x)\label{eq:conmPsib}\\
\Big[\bd_{qr's'},\psi_{rs}(x)\Big]&=\delta_{rr'}\delta_{ss'}\alpha^*_{qr}(x)\psi_{rs}(x),\label{eq:conmPsibd}
\end{align}
donde $\alpha_{qr}(x)=\frac{i}{\sqrt{n_q}}e^{-irqx}$. Estos
conmutadores y la ecuaci\'on (\ref{eq:estadoFunBos}) implican
inmediatamente que

\begin{equation}
b_{qr's'}\psi_{rs}(x)\npart_0=\delta_{rr'}\delta_{ss'}\alpha_{qr}(x)\psi_{rs}(x)\npart_0.
\end{equation}
Y por lo tanto, este estado posee una representaci\'on como estado
coherente bos\'onico\cite{negele88}:

\begin{equation}\label{eq:repCoherente}
\psi_{rs}(x)\npart_0=\exp\left[\sum_{q>0}\alpha_{qr}(x)
\bd_{qrs}\right]F_{rs}\lambdah_{rs}(x)\npart_0=
e^{-i\varphid_{rs}(x)}F_{rs}\lambdah_{rs}(x)\npart_0
\end{equation}
Aqu\'{\i} utilizamos la definici\'on del campo $\varphid$
(\ref{eq:camposBosonicos}) en la segunda igualdad. Hemos agregado
el operador de fase $\lambdah_{rs}$ que derivaremos en lo
sucesivo; y el factor de Klein, que es necesario porque
$\psi_{rs}$ remueve una part\'{\i}cula del estado $\npart_0$, cosa
que los campos bos\'onicos $\bd$ no pueden hacer. Para obtener el
operador $\lambdah$ calculamos el siguiente valor medio de dos
formas diferentes: por un lado,

\begin{equation}
{}_0\langle\NS|\Fd_{rs}\psi_{rs}(x)|\NS\rangle_0=
{}_0\langle\NS|\lambdah_{rs}(x)|\NS\rangle_0\equiv\lambda_{rs}(x)
\end{equation}
donde hemos pasado adelante el factor de Klein $F_{rs}$ en
(\ref{eq:repCoherente}), ya que seg\'un su definici\'on
(\ref{eq:defFactoresKlein1}) conmuta con todos los $\bd$;
utilizamos la unitariedad de los $F$'s, y expandimos en serie el
exponencial, qued\'andonos con el t\'ermino de orden $0$, ya que
${}_0\langle\NS|\bd_{qrs}=0$.

Por otro lado, insertamos la descomposici\'on de Fourier
(\ref{eq:camposFermionicos}) para $\psi(x)$ y la definici\'on del
factor de Klein (\ref{eq:defFactoresKlein2}), y nos quedamos
s\'olo con el t\'ermino $n_k=N_{rs}$ (o bien
$k=\frac{2\pi}{L}(N_{rs}-\frac{1}{2}\delta_b)$):

\begin{equation}
{}_0\langle\NS|\Fd_{rs}\psi_{rs}(x)|\NS\rangle_0=
\frac{1}{\sqrt{L}}e^{irkx}{}_0\langle\NS|\cd_{N_{rs}rs}c_{krs}|\NS\rangle_0=
\frac{1}{\sqrt{L}}e^{ir\frac{2\pi}{L}(N_{rs}-\frac{1}{2}\delta_b)x}.
\end{equation}
Concluimos entonces que el operador $\lambdah_{rs}(x)$ est\'a dado
por

\begin{equation}\label{eq:defOpLambda}
\lambdah_{rs}(x)=\frac{1}{\sqrt{L}}e^{ir\frac{2\pi}{L}(\Nh_{rs}-\frac{1}{2}\delta_b)x}.
\end{equation}

Para derivar las identidades de bosonizaci\'on debemos estudiar la
acci\'on del campo $\psi_{rs}(x)$ sobre un estado $\npart$
arbitrario (que seg\'un (\ref{eq:factorizacionNpart}) puede
escribirse como $\npart=f(\{\bd_{qr's'}\})\npart_0$). Para ello
utilizaremos las siguientes identidades \cite{vondelft98},

\begin{equation}\label{eq:idf1}
\psi_{rs}(x)f(\{\bd_{qr's'}\})=
f(\{\bd_{qr's'}-\delta_{rr'}\delta_{ss'}\alpha^*_{qr}(x)\})\psi_{rs}(x),
\end{equation}

\begin{equation}\label{eq:idf2}
f(\{\bd_{qr's'}-\delta_{rr'}\delta_{ss'}\alpha^*_{qr}(x)\})=
e^{-i\varphi_{rs}(x)}f(\{\bd_{qrs}\})e^{i\varphi_{rs}(x)},
\end{equation}
que se pueden mostrar facilmente a partir de la f\'ormula de
Baker-Hausdorff

\begin{equation}
e^{-B}Ae^B=A+[A,B]+\frac{1}{2!}[[A,B]B]+\dots,
\end{equation}
expandiendo en serie de Taylor la funci\'on $f$ y empleando los
conmutadores (\ref{eq:conmPsib}) y (\ref{eq:conmPsibd}). Podemos
evaluar entonces $\psi_{rs}(x)\npart$ conmutando $\psi_{rs}(x)$
con $f(\{\bd_{qrs}\})$, insertando la representaci\'on
(\ref{eq:repCoherente}) y reordenando los factores:

\begin{align}
\psi_{rs}(x)\npart=&\;\psi_{rs}(x)f(\{\bd_{qrs}\})\npart_0\notag& \\[5pt]
=&\;f(\{\bd_{qr's'}-\delta_{rr'}\delta_{ss'}\alpha^*_{qr}(x)\})\psi_{rs}(x)\npart_0&{\text{[por la Ec. \ref{eq:idf1}]}}\notag\\[5pt]
=&\;f(\{\bd_{qr's'}-\delta_{rr'}\delta_{ss'}\alpha^*_{qr}(x)\})e^{-i\varphid_{rs}(x)}F_{rs}\lambdah_{rs}(x)\npart_0&{\text{[por la Ec. \ref{eq:repCoherente}]}}\notag\\[5pt]
=&\;F_{rs}\lambdah_{rs}(x)e^{-i\varphid_{rs}(x)}f(\{\bd_{qr's'}-\delta_{rr'}\delta_{ss'}\alpha^*_{qr}(x)\})\npart_0&{\text{[por la Ec. \ref{eq:defFactoresKlein1}]}}\notag\\[5pt]
=&\;F_{rs}\lambdah_{rs}(x)e^{-i\varphid_{rs}(x)}\left[e^{-i\varphi_{rs}(x)}f(\{\bd_{qrs}\})e^{i\varphi_{rs}(x)}\right]\npart_0&{\text{[por la Ec. \ref{eq:idf2}]}}\notag\\[5pt]
=&\;F_{rs}\lambdah_{rs}(x)e^{-i\varphid_{rs}(x)}e^{-i\varphi_{rs}(x)}f(\{\bd_{qrs}\})\npart_0&{\text{[por la Ec. \ref{eq:estadoFunBos}]}}\notag\\[5pt]
=&\;F_{rs}\lambdah_{rs}(x)e^{-i\varphid_{rs}(x)}e^{-i\varphi_{rs}(x)}\npart.&{\text{[por
la Ec. \ref{eq:factorizacionNpart}]}}
\end{align}
Dado que $\npart$ es arbitrario, y que todo estado del espacio de
Fock es de esta forma, concluimos que las siguientes
\emph{f\'ormulas de bosonizaci\'on} valen como identidades entre
operadores en el espacio de Fock, y para todo $L$:

\begin{align}
\psi_{rs}(x)=&\;F_{rs}\lambdah_{rs}(x)e^{-i\varphid_{rs}(x)}e^{-i\varphi_{rs}(x)}& \\
=&\;F_{rs}\frac{1}{\sqrt{L}}e^{ir\frac{2\pi}{L}(\Nh_{rs}-\frac{1}{2}\delta_b)x}e^{-i\varphid_{rs}(x)}e^{-i\varphi_{rs}(x)}&{\text{[por la Ec. \ref{eq:defOpLambda}]}}\\
=&\;\frac{1}{\sqrt{2\pi a}}
F_{rs}e^{ir\frac{2\pi}{L}(\Nh_{rs}-\frac{1}{2}\delta_b)x}e^{-i\phi_{rs}(x)}.&{\text{[por
la Ec. \ref{eq:opVertice1}]}}\label{eq:bosPsi}
\end{align}

Por \'ultimo estudiaremos como representar un Hamiltoniano
fermi\'onico libre con una relaci\'on de dispersi\'on lineal.
M\'as espec\'{\i}ficamente tomemos un Hamiltoniano de la forma
(con $\hbar=1$)

\begin{equation}\label{eq:hamLibreFerm1}
H_0=\sum_{r,s}H_{0rs}
\end{equation}
con

\begin{equation}\label{eq:hamLibreFerm2}
H_{0rs}\equiv-irv_{rs}\int_0^L
dx\,:\psid_{rs}(x)\partial_x\psi_{rs}(x): =v_{rs}\sum_k
k:\cd_{krs}c_{krs}:
\end{equation}
La segunda forma se obtiene de la primera insertando el desarrollo
de Fourier del campo $\psi$ (\ref{eq:camposFermionicos}).

La ecuaci\'on (\ref{eq:factorizacionNpart}) implica que los
$\bd$'s actuando sobre $\npart_0$ generan todo el espacio de
Hilbert de $\NS$ part\'{\i}culas. Esto significa entonces, que
$H_{0rs}$ debe tener una representaci\'on en t\'erminos solamente
de variables bos\'onicas. Para hallar esa representaci\'on,
estudiemos el conmutador de $\bd_{qrs}$ con $H_{0rs}$:

\begin{equation}\label{eq:conmHb}
\Big[H_{0rs},\bd_{qr's'}\Big]=q\bd_{qrs}\delta_{rr'}\delta_{ss'}.
\end{equation}
Adem\'as, dado que $[H_{0rs},\Nh_{r's'}]=0$ para todo $r,r',s,s'$,
todo autoestado de $\Nh_{rs}$ lo es tambi\'en de $H_{0rs}$, en
particular el estado fundamental de $\NS$ part\'{\i}culas,
$\npart_0$. Su autovalor es

\begin{equation}\label{eq:enrgiaN0}
E_{0rs}^\NS={}_0\langle\NS|H_{0rs}\npart_0=
\frac{v_{rs}}{2}\left(\frac{2\pi}{L}\right)N_{rs}\left(N_{rs}+1-\delta_b\right).
\end{equation}

Se comprueba que la \'unica forma bos\'onica para $H_{0rs}$ que
reproduce las Ecs. (\ref{eq:conmHb}) y (\ref{eq:enrgiaN0}) es:

\begin{align}
H_{0rs}=&\sum_{q>0}q\bd_{qrs}
b_{qrs}+\frac{v_{rs}}{2}\left(\frac{2\pi}{L}\right)\Nh_{rs}\left(\Nh_{rs}+1-\delta_b\right)\\
=&\frac{v_{rs}}{2}\int_0^L \frac{dx}{2\pi}\,
:\left(\partial_x\phi_{rs}(x)\right)^2:
+\frac{v_{rs}}{2}\left(\frac{2\pi}{L}\right)\Nh_{rs}\left(\Nh_{rs}+1-\delta_b\right)\label{eq:hamLibreBos}.
\end{align}

Con esta \'ultima ecuaci\'on completamos la derivaci\'on de las
identidades de bosonizaci\'on, que valen para $L$ finito. Para
obtener expresiones con $L\rightarrow\Infinity$ basta con
despreciar los t\'erminos $\sim 1/L$. En este cap\'{\i}tulo
seguimos un enfoque constructivo, de modo que no es necesario
verificar los conmutadores de los campos fermi\'onicos o igualdad
entre funci\'ones de Green. A continuaci\'on veremos las
aplicaciones del proceso de bosonizaci\'on, y como se vuelve
extremadamente \'util para el estudio de complicadas teor\'{\i}as
fermi\'onicas en una dimensi\'on espacial.

\section{Aplicaciones de la bosonizaci\'on: el modelo g-ology}

Veremos a continuaci\'on c\'omo aplicar la t\'ecnica a una
teor\'{\i}a de electrones interactuantes en una dimensi\'on. El
punto de comienzo de las teor\'{\i}as de muchos cuerpos en materia
condensada es un Hamiltoniano de la forma $H=H_0+H\Tint$, con

\begin{align}
H_0=&\sum_\mathbf{k}\sum_s\epsilon_\mathbf{k}\cdh_{\mathbf{k}s}\ch_{\mathbf{k}s},\label{eq:hamLibre}\\
H\Tint=&\frac{1}{2V}\sum_{\mathbf{q}\mathbf{k}\mathbf{k}'}\sum_{ss'}f_\mathbf{q}^{\mathbf{k}s\mathbf{k}'s'}\cdh_{\mathbf{k}+\mathbf{q}\,s}\cdh_{\mathbf{k}'-\mathbf{q}\,s'}\ch_{\mathbf{k}'s'}\ch_{\mathbf{k}s},\label{eq:hamInt}
\end{align}
donde $V$ es el vol\'umen del sistema, y $\cdh_{\mathbf{k}s}$ y
$\ch_{\mathbf{k}s}$ son los operadores de creaci\'on y
aniquilaci\'on de electrones, que satisfacen las relaciones de
anticonmutaci\'on can\'onicas

\begin{equation}
\Big\{\ch_{\mathbf{k}s},\cdh_{\mathbf{k}'s'}\Big\}=\delta_{\mathbf{k}\mathbf{k}'}\delta_{ss'}.
\end{equation}
Las cantidades $f_\mathbf{q}^{\mathbf{k}s\mathbf{k}'s'}$ son los
llamados \emph{par\'ametros de Landau}, que describen la
dispersi\'on de dos part\'{\i}culas desde un estado inicial con
n\'umeros cu\'anticos $(\mathbf{k},s)$ y $(\mathbf{k'},s)$ a un
estado final con n\'umeros cu\'anticos $(\mathbf{k}+\mathbf{q},s)$
y $(\mathbf{k}-\mathbf{q},s)$. En general $\mathbf{k}$ es el
momento o pseudomomento de la part\'{\i}cula, $s$ su spin y
$\mathbf{q}$ es el momento transferido. $\epsilon_\mathbf{k}$ es
la energ\'{\i}a cin\'etica de los electrones medida desde el nivel
de Fermi.

\begin{figure}
\begin{center}
\includegraphics{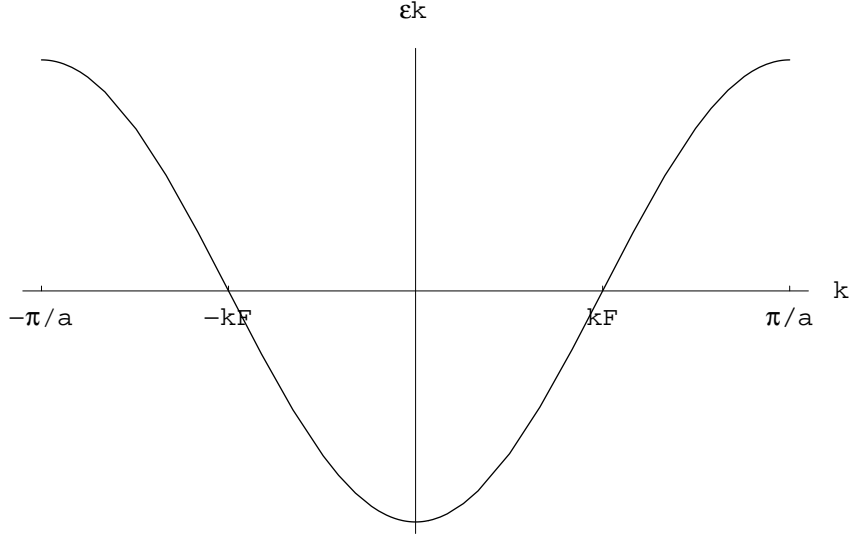}
\caption{\label{fig:dispCos} Relaci\'on de dispersi\'on de un gas
de electrones unidimensionales no interactuantes.}
\end{center}
\end{figure}

Como ya hemos comentado, en esta tesis estamos interesados
principalmente en problemas de electrones en una dimensi\'on
espacial, y de aqu\'{\i} en m\'as nos restrigiremos a este caso.
En la aproximaci\'on de electrones casi libres, o de ligadura
fuerte, la energ\'{\i}a cin\'etica o relaci\'on de dispersi\'on se
ilustra esquem\'aticamente en la Fig. \ref{fig:dispCos}. En
particular, en el modelo de Hubbard\cite{emery79}, el modelo
realista m\'as simple que puede plantearse en una dimensi\'on,
tenemos una expresi\'on expl\'{\i}cita para $\epsilon_k$:

\begin{equation}\label{eq:dispersiónHubbard}
\epsilon_k=-2t\cos ka
\end{equation}
donde $t$ es la constante de intercambio (acoplamiento entre
sitios vecinos) y $a$ es el espaciado entre los \'atomos de la
red.

\begin{figure}
\begin{center}
\includegraphics{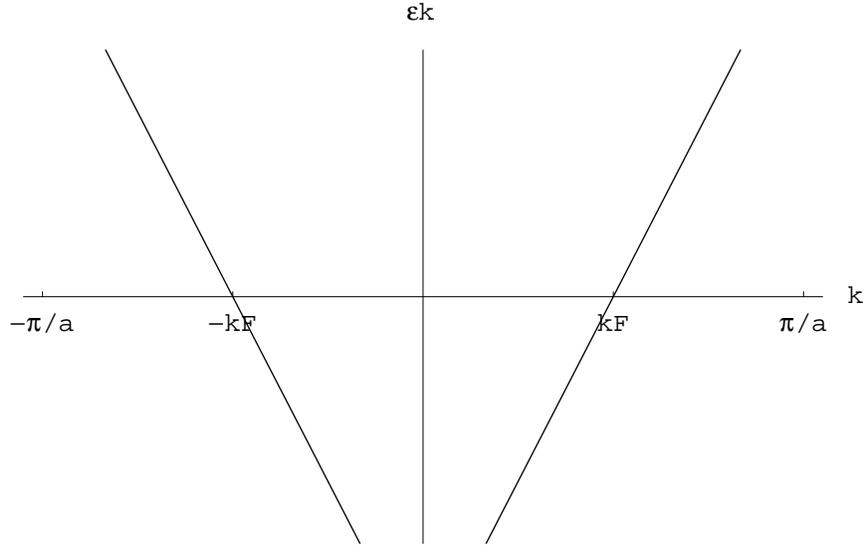}
\caption{\label{fig:dispLineal}Relaci\'on de dispersi\'on
linealizada alrededor de los puntos de Fermi $\pm k\f$.}
\end{center}
\end{figure}

El espectro del modelo de Hubbard puede hallarse mediante la
t\'ecnica del ansatz de Bethe\cite{bethe31} (lo que a menudo en la
literatura se denomina ``resolver" el modelo). Sin embargo, esta
t\'ecnica no brinda resultados para las funciones de Green. Por
este motivo se vuelve necesaria la aplicaci\'on de otros m\'etodos
que permitan obtener una descripci\'on de los tipos de
fluctuaciones que tienen lugar. Una de las t\'ecnicas m\'as
utilizadas para el c\'alculo de funciones de correlaci\'on es la
bosonizaci\'on. Como se ha mostrado en las secciones previas, en
su formulaci\'on m\'as usual (operacional) este procedimiento se
basa en identidades entre los operadores fermi\'onicos originales
de la teor\'{\i}a y ciertos operadores bos\'onicos que se pueden
construir a partir de los primeros. En particular, el Hamiltoniano
de los diferentes modelos resulta escrito completamente en
t\'erminos de fluctuaciones bos\'onicas. Sin embargo debemos hacer
una aclaraci\'on: el modelo de Hubbard originalmente se formula en
la red, mientras que la bosonizaci\'on se aplica a modelos
continuos. Por ello, al estudiar este modelo, en alg\'un punto
debe tomarse dicho l\'{\i}mite. Al hacerlo las propiedades que se
derivan tienen validez para grandes distancias (ej. los
decaimientos de las funciones de Green), o lo que es equivalente,
para las excitaciones de baja energ\'{\i}a.

La superficie de Fermi en un metal estrictamente unidimensional
consiste en dos puntos, $+k\f$ y $-k\f$; en su vecindad podemos
linealizar la relaci\'on de dispersi\'on
(\ref{eq:dispersiónHubbard}) \cite{solyom79}:

\begin{equation}
\epsilon_k=v\f(|k|-k\f)
\end{equation}
Esta aproximaci\'on, en principio, es razonable en un rango finito
alrededor de los puntos de Fermi. Sin embargo, los tratamientos
matem\'aticos se simplifican enormemente si tomamos esta versi\'on
linealizada para todos los valores de $k$ entre $-\Infinity$ y
$+\Infinity$, es decir, si reemplazamos el espectro libre de la
Fig. \ref{fig:dispCos} por el de la Fig. \ref{fig:dispLineal}. Por
otro lado, de acuerdo a lo dicho m\'as arriba, s\'olo estamos
interesados en excitaciones de baja energ\'{\i}a, a las que
contribuyen estados pr\'oximos a la superficie de Fermi; de modo
que la inclusi\'on de los estados adicionales por efecto de la
linealizaci\'on es despreciable en \'este r\'egimen. Esto \'ultimo
se verifica a posteriori al estudiar los efectos producidos por la
curvatura de banda, es decir incluyendo t\'erminos cuadr\'aticos y
c\'ubicos en la relaci\'on de dispesi\'on. Se puede mostrar que
las contribuciones de estos t\'erminos son irrelevantes frente a
las del t\'ermino lineal.

La linealizaci\'on genera dos ramas bien definidas en la
relaci\'on de dispersi\'on. Los electrones que pertenecen a la
rama que contiene al punto $+k\f$ ($-k\f$) se mueven hacia la
derecha (izquierda), a los operadores que los representan los
denotaremos $\cd_{(k-k\f)Rs}$ y $c_{(k-k\f)Rs}$
($\cd_{(-k-k\f)Ls}$ y $c_{(-k-k\f)Ls}$). Este conjunto de
operadores as\'i definidos satisface los requisitos de ser un
conjunto infinito y no acotado, y los identificamos inmediatamente
con los descriptos en las Ecs. (\ref{eq:operadoresAnticonmutan}) y
(\ref{eq:operadoresIndice}). En t\'ermino de estos operadores, el
Hamiltoniano libre se escribe

\begin{equation}\label{eq:hamLibreUnaDim}
H_0=\sum_{ks}v\f
k\left(:\cd_{kRs}c_{kRs}+\cd_{kLs}c_{kLs}:\right)=-iv\f\sum_s\int_0^L:\left(\psid_{Rs}\partial_x\psi_{Rs}
-\psid_{Ls}\partial_x\psi_{Ls}\right):.
\end{equation}
Para obtener la primera igualdad cambiamos $k\rightarrow-k$ en el
segundo t\'ermino, y posteriormente efectuamos la traslaci\'on
$k\rightarrow k+k\f$. Para obtener la segunda, utilizamos la
definici\'on de los operadores de campo fermi\'onicos
(\ref{eq:camposFermionicos}).

\begin{figure}
\begin{center}
\includegraphics{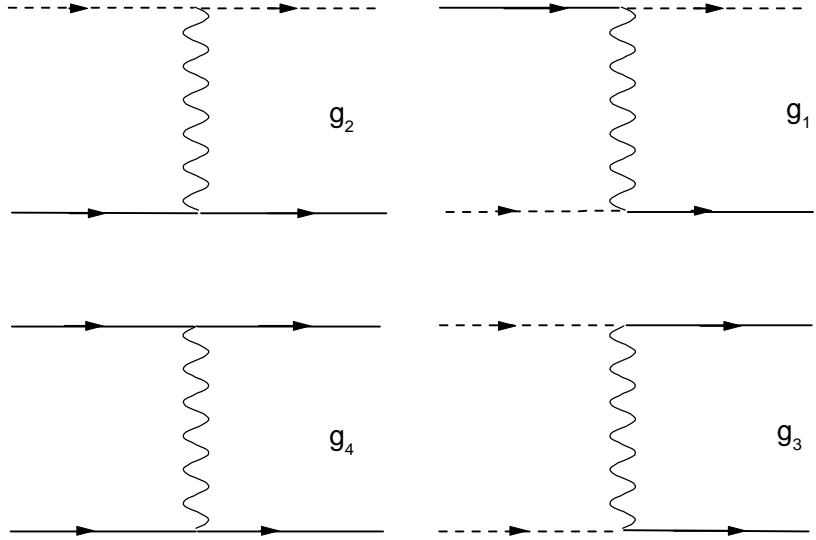}
\caption{\label{fig:scattering} Procesos que intervienen en la
dispesi\'on del gas de electrones unidimensional: $g_1$
dispersi\'on hacia atr\'as; $g_2$ y $g_4$ dispersi\'on hacia
adelante; $g_3$ umklapp.}
\end{center}
\end{figure}

Los t\'erminos de interacci\'on se pueden clasificar en cuatro
tipos diferentes mostrados en la Fig. \ref{fig:scattering}. Los
electrones pertenecientes a ambas ramas se distinguen mediante
l\'{\i}neas punteadas y s\'olidas. El proceso con constante de
acoplamiento $g_1$ corresponde a dispersi\'on hacia atr\'as, y
posee una transferencia de momento de $2k\f$. Los procesos con
constantes $g_2$ y $g_4$ son de dispersi\'on hacia adelante, su
transferencia de momento es nula. Por \'ultimo el proceso con
constante $g_3$ es de tipo umklapp, y su transferencia de momento
es de $4k\f$. Este \'ultimo proceso s\'olo es importante cuando
estamos con un llenado medio, es decir, pensando en una
situaci\'on del tipo del modelo de Hubbard, cuando tenemos un
electr\'on por sitio. En ese caso, $4k\f$ es igual al vector de
red rec\'{\i}proco.

Al introducir el spin electr\'onico, para cada uno de estos
procesos aparecen dos variantes, de acuerdo a la orientaci\'on
relativa de los electrones incidentes. Si ambos poseen spines
alineados, le agregamos un sub\'{\i}ndice $\parallel$ a la
constante de acoplamiento; en cambio, si los spines son
antiparalelos, agregamos el sub\'{\i}ndice $\perp$. N\'otese que
estos procesos no invierten el spin, es decir, el spin de las
part\'{\i}culas finales es id\'entico al de las part\'{\i}culas
iniciales. La generalizací\'on de este modelo al caso en que las
interacciones pueden cambiar el estado de spin es uno de los
objetivos de esta tesis (ver Cap\'itulo 4). Volviendo al caso
presente, la expresi\'on para el Hamiltoniano de interacci\'on que
describe estos procesos, en espacio de coordenadas es

\begin{align}
\H\Tint=&\sum_{ss'}\int_0^L dx\,
\left(g_{1\parallel}\delta_{ss'}+g_{1\perp}\delta_{s,-s'}\right)
\psid_{Ls}\psi_{Rs}\psid_{Rs'}\psi_{Ls'}\nonumber\\
+&\sum_{ss'}\int_0^L dx\,
\left(g_{2\parallel}\delta_{ss'}+g_{2\perp}\delta_{s,-s'}\right)
\rho_{Ls}\rho_{Rs'}\nonumber\\
+\frac{1}{2}&\sum_{rss'}\int_0^L dx\,
\left(g_{3\parallel}\delta_{ss'}+g_{3\perp}\delta_{s,-s'}\right)
\psid_{-rs}\psi_{rs}\psid_{-rs'}\psi_{rs'}\nonumber\\
+\frac{1}{2}&\sum_{rss'}\int_0^L dx\,
\left(g_{4\parallel}\delta_{ss'}+g_{4\perp}\delta_{s,-s'}\right)
\rho_{rs}\rho_{rs'}.
\end{align}
$\rho_{rs}$ est\'a definido en la Ec. (\ref{eq:defDensidad}). La
teor\'ia definida por las dos ecuaciones anteriores, versi\'on
cont\'inua del modelo de bajas energ\'ias de un ensemble de
electrones, se conoce popularmente como modelo de
``geolog\'{\i}a", o ``g-ology" \cite{solyom79}. La bosonizaci\'on
del Hamiltoniano es inmediata aplicando las identidades ya
mostradas en la secci\'on previa. Por simplicidad tomaremos
$L\rightarrow\Infinity$, y $g_{3\parallel}=g_{1\parallel}=0$,
$g_{3\perp}=g_3$, $g_{1\perp}=g_1$. Para $H_0$ usamos las ecs.
(\ref{eq:hamLibreFerm1}), (\ref{eq:hamLibreFerm2}) y
(\ref{eq:hamLibreBos}):

\begin{equation}
H_0=\frac{v\f}{2}\sum_{rs}\int \frac{dx}{2\pi}
\left(\partial_x\phi_{rs}\right)^2,
\end{equation}
mientras que para $H_\text{int}$ utilizamos
(\ref{eq:defDensidad})-(\ref{eq:bosDensidad}) para los t\'erminos
de dispersi\'on hacia adelante, y (\ref{eq:bosPsi}) para los de
dispersi\'on hacia atr\'as y umklapp. El resultado es

\begin{align}
\H\Tint=&\sum_s \int dx\, g_1\,
e^{i\phi_{Ls}}e^{-i\phi_{Rs}}e^{i\phi_{R,-s}}e^{-i\phi_{L,-s}}\nonumber\\
+&\sum_{ss'}\int dx\,
\left(g_{2\parallel}\delta_{ss'}+g_{2\perp}\delta_{s,-s'}\right)
\frac{-1}{(2\pi)^2}\partial_x\phi_{Ls}\partial_x\phi_{Rs'}\nonumber\\
+\frac{1}{2}&\sum_{rs}\int dx\, g_3\,
e^{i\phi_{-rs}}e^{-i\phi_{rs}}e^{i\phi_{-r,-s}}e^{-i\phi_{r,-s}}\nonumber\\
+\frac{1}{2}&\sum_{rss'}\int dx\,
\left(g_{4\parallel}\delta_{ss'}+g_{4\perp}\delta_{s,-s'}\right)
\frac{1}{(2\pi)^2}\partial_x\phi_{rs}\partial_x\phi_{rs'}.
\end{align}
En los t\'erminos de dispersi\'on hacia atr\'as y umklapp no
tuvimos en cuenta los factores de Klein porque sus contribuciones
pueden despreciarse en el l\'{\i}mite
$L\rightarrow\Infinity$\cite{voit95}. Introducimos a
continuaci\'on los campos $\theta_\nu$ y $\phi_\nu$ con
$\nu=\rho,\sigma$.

\begin{equation}\label{eq:camposCargaSpin}
\phi_{rs}=\sqrt{\frac{\pi}{2}}
\left[\theta_\rho-r\phi_\rho+s\left(\theta_\sigma-r\phi_\sigma\right)\right].
\end{equation}
Para entender el significado f\'{\i}sico de estos campos,
estudiamos los operadores de densidad de carga y de spin,
definidos del siguiente modo:

\begin{equation}
\rhoh=\sum_{rs}\rho_{rs}\qquad\sigmah=\sum_{rs}s\rho_{rs}.
\end{equation}
Utilizando la equivalencia de bosonizaci\'on
(\ref{eq:bosDensidad}), podemos encontrar expresiones para estos
operadores en t\'erminos de los campos recientemente definidos:

\begin{equation}
\rhoh=\sqrt{\frac{2}{\pi}}\partial_x\phi_\rho;
\qquad\sigmah=\sqrt{\frac{2}{\pi}}\partial_x\phi_\sigma.
\end{equation}
Entonces los campos $\phi_\rho$ y $\phi_\sigma$ est\'an
relacionados con la densidad de carga y de spin respectivamente.
M\'as a\'un, utilizando los conmutadores del campo $\phi_{rs}$ con
si mismo (\ref{eq:conmCampoPhi}) y con su derivada
(\ref{eq:conmBosDeriv}) es posible mostrar que $\phi_\rho$ y
$\theta_\rho$, y sus derivadas conmutan con $\phi_\sigma$ y
$\theta_\sigma$ y sus derivadas. Adem\'as si definimos
$\Pi_\nu=\partial_x\theta_\nu$, hallamos que

\begin{equation}
\Big[\phi_\nu(x),\Pi_\mu(x')\Big]=i\delta_{\mu\nu}\delta(x-x'),
\end{equation}
que constituyen las relaciones de conmutaci\'on can\'onicas para
campos bos\'onicos. Introducimos adem\'as las constantes

\begin{equation}
g_2^\nu=\frac{1}{2}\left(g_{2\parallel}\pm g_{2\perp}\right),
\end{equation}

\begin{equation}
g_4^\nu=\frac{1}{2}\left(g_{4\parallel}\pm g_{4\perp}\right),
\end{equation}
las constantes de dureza

\begin{equation}
K_\nu=\sqrt{\frac{\pi v\f+g_4^\nu-g_2^\nu}{\pi
v\f+g_4^\nu+g_2^\nu}},
\end{equation}
y las velocidades renormalizadas

\begin{equation}
v_\nu=\sqrt{\left(\pi v\f+g_4^\nu-g_2^\nu\right)\left(\pi
v\f+g_4^\nu+g_2^\nu\right)}.
\end{equation}

Con estas definiciones el Hamiltoniano total se escribe como
$H=H_\rho+H_\sigma$, donde

\begin{equation}
H_\sigma=\int
dx\frac{v_\sigma}{2}\left[K_\sigma\left(\partial_x\theta_\sigma\right)^2+\frac{1}{K_\sigma}
\left(\partial_x\phi_\sigma\right)^2\right]+\frac{2g_1}{(2\pi
a)^2}\cos\left(\sqrt{8\pi}\phi_\sigma\right)
\end{equation}

\begin{equation}
H_\rho=\int
dx\frac{v_\rho}{2}\left[K_\rho\left(\partial_x\theta_\rho\right)^2+\frac{1}{K_\rho}
\left(\partial_x\phi_\rho\right)^2\right]+\frac{2g_3}{(2\pi
a)^2}\cos\left(\sqrt{8\pi}\phi_\rho\right).
\end{equation}

De este modo el Hamiltoniano fermi\'onico original queda escrito
completamente en t\'ermino de dos campos bos\'onicos $\phi_\rho$ y
$\phi_\sigma$ que representan oscilaciones independientes de
densidad de carga y spin respectivamente, y que se propagan con
velocidades diferentes. Adem\'as los campos de spin conmutan con
los campos de carga, y lo mismo ocurre con los Hamiltonianos de
cada sector. Estas caracter\'{\i}sticas dan lugar a una propiedad
fundamental de este tipo de modelos en una dimensi\'on: la
separaci\'on spin-carga. Por \'ultimo mencionemos que las
funciones de Green del sistema admiten una factorizaci\'on
semejante, en t\'erminos de un factor de carga y un factor de
spin. Uno de los objetivos de esta tesis es analizar de qu\'e
manera este fen\'omeno emerge en un estudio de la teor\'{\i}a en
el marco de la integral funcional, y c\'omo se manifiesta al nivel
de las funciones de Green y del diagrama de fases la ruptura de la
separaci\'on spin carga mediante la presencia de interacciones
spin-\'orbita (ver Cap\'itulo 6).


\chapter{El determinante fermi\'onico no covariante y su relaci\'on con los l\'{\i}quidos de
Luttinger}\label{part:jacobiano}

\begin{center}
\begin{minipage}{5.6in}
\textsl{En este cap\'{\i}tulo consideramos el procedimiento de
bosonizaci\'on en el marco de la integral funcional. Esto nos
conduce naturalmente al estudio del determinante fermi\'onico
asociado a una Teor\'{\i}a Cu\'antica de Campos no covariante,
utilizada para describir un sistema no relativista en $(1+1)$
dimensiones. Explotando la libertad que brinda no mantener la
invarianza de Lorentz, determinamos el operador regulador
correspondiente al m\'etodo del n\'ucleo del calor (heat-kernel)
que permite reproducir la relaci\'on de dispersi\'on de las
excitaciones bos\'onicas y los exponentes cr\'{\i}ticos correctos
del modelo de Tomonaga-Luttinger. Adem\'as derivamos el
Hamiltoniano del modelo bosonizado funcionalmente y las
correspondientes corrientes. De este modo establecemos la
regularizaci\'on precisa mediante heat-kernel, que conduce a un
completo acuerdo entre el abordaje operacional a la bosonizaci\'on
de modelos de materia condensada, y su alternativa mediante
integrales funcionales. Estos resultados son parte de las
contribuciones originales a esta tesis\cite{iucci04}.}
\end{minipage}
\end{center}

\section{Introducci\'on}

Los determinantes fermi\'onicos juegan un rol central en las
formulaciones modernas de las Teor\'{\i}as Cu\'anticas de Campos
(QFT's). Como es bien sabido, emergen naturalmente al considerar
funcionales generatrices asociadas a campos fermi\'onicos en el
marco de la integral funcional\cite{ramond96}. En los \'ultimos
veinte a\~nos ha sido especialmente fruct\'{\i}fero el estudio de
determinantes fermi\'onicos en $(1+1)$ dimensiones. La
observaci\'on de Fujikawa concerniente a la no trivialidad del
jacobiano asociado a las transformaciones quirales en las
variables fermi\'onicas\cite{fujikawa79,fujikawa80a,fujikawa80b}
fue aplicada al caso en $(1+1)$ dimensiones, y condujo a avances
significativos en nuestra comprensi\'on de los llamados
\emph{modelos de juguete}\@ paradigm\'aticos, tales como la
electrodin\'amica cu\'antica en dos dimensiones ($\text{QED}_2$),
el modelo de Thirring, y sus versiones no
abelianas\cite{roskies81,gamboa81,furuya82}. De hecho, basado en
un tratamiento adecuado del determinante fermi\'onico, se
desarroll\'o una t\'ecnica de bosonizaci\'on mediante integrales
funcionales\cite{stone94}. El punto crucial se encuentra en el
c\'alculo del mencionado jacobiano. Un c\'alculo ingenuo arroja un
resultado mal definido y se vuelve necesario implementar un
procedimiento de regularizaci\'on. En teor\'{\i}as de gauge con
fermiones de Dirac, es natural considerar un esquema de
regularizaci\'on que preserve la invarianza de gauge. Por otro
lado, cuando los campos vectoriales que se encuentran presentes en
la teor\'{\i}a son s\'olo campos auxiliares (usualmente
introducidos a trav\'es de una transformaci\'on de
Hubbard-Stratonovich), se puede elegir un regulador m\'as
general\cite{rubin86,cabra89}. El modelo de
Thirring\cite{klaiber68} y el modelo de Schwinger
quiral\cite{jackiw85a,jackiw85b} constituyen ejemplos en los que
tienen lugar este tipo de ambig\"uedades en la regularizaci\'on.

El tema de la regularizaci\'on del jacobiano de Fujikawa, su
relaci\'on con contrat\'erminos locales, y su rol en el an\'alisis
de anomal\'{\i}as cu\'anticas ha sido extensivamente examinado en
la literatura\cite{fujikawa03}. En todos los casos, los modelos
bajo estudio son QFT's relativistas, es decir, teor\'{\i}as
covariantes de Lorentz. Sin embargo, en ciertas situaciones
relevantes, el inter\'es recae en teor\'{\i}as de campos no
covariantes. Es el caso del an\'alisis de sistemas de electrones
unidimensionales que pueden estudiarse mediante el modelo g-ology.
En este contexto, la bosonizaci\'on funcional, alternativa al
enfoque operacional usual, fue sugerida por primera vez por
Fogebdy\cite{fogebdy76} y sucesivamente elaborada por Lee y
Chen\cite{lee88}. La conexi\'on expl\'{\i}cita entre la
bosonizaci\'on funcional que condujo a una acci\'on efectiva para
la din\'amica de las excitaciones bos\'onicas colectivas y el
jacobiano de Fujikawa fue establecida por primera vez en la Ref.
\citen{naon95}. Pero a\'un en este caso se emple\'o una
regularizaci\'on covariante, tomada de la teor\'{\i}a de campos
relativista. Como resultado las expresiones generales para las
relaciones de dispersi\'on de los modos bos\'onicos y los
exponentes de decaimiento de las funciones de correlaci\'on en
t\'erminos de las constantes de acoplamiento iniciales del modelo
fermi\'onico no acordaron con las obtenidas mediante
bosonizaci\'on operacional usual. En este cap\'{\i}tulo mostramos
que el or\'{\i}gen de este desacuerdo se encuentra en el tipo de
regularizaci\'on escogido para calcular el jacobiano de Fujikawa.
Dado que se viola la invarianza de Lorentz subyacente, parecen
posibles un n\'umero arbitrario de esquemas de regularizaci\'on.
S\'olo uno de ellos conduce al resultado usual para los modelos
g-ology y Tomonaga-Luttinger. En este punto quisi\'eramos
enfatizar que no fuimos capaces de hallar un principio f\'{\i}sico
que sirviese de gu\'{\i}a para elegir a priori entre diferentes
esquemas de regularizaci\'on, al estilo del principio de
preservaci\'on a nivel cu\'antico de simetr\'{\i}as que se hallan
en la teor\'{\i}a bajo estudio a nivel cl\'asico, como lo son las
ya mencionadas simetr\'{\i}as de gauge o de Lorentz. Sin embargo,
hasta donde sabemos, tal principio tampoco ha sido identificado en
el marco de la bosonizaci\'on operacional de teor\'{\i}as de
materia condensada. Por supuesto, este es un aspecto importante
que merece futuras investigaciones.

El plan de este cap\'{\i}tulo es el siguiente. En la secci\'on
\ref{part:model} presentamos el modelo y expresamos su funcional
generatriz en t\'erminos de un determinante fermi\'onico. En la
secci\'on \ref{part:decoupling}, con el objeto de clarificar la
discusi\'on comenzamos con un trazado de los pasos principales del
enfoque desacoplante a la bosonizaci\'on, y los resultados que se
obtienen al emplear una regularizaci\'on est\'andar invariante de
Lorentz. Aqu\'{\i} incluimos dos subsecciones donde presentamos
dos tipos de regularizaci\'on diferentes: el m\'etodo
point-splitting, y el m\'etodo heat-kernel. En este \'ultimo caso
determinamos la forma precisa del operador necesario para obtener
la respuesta correcta para las relaciones de dispersi\'on y los
exponentes. En la secci\'on \ref{part:hamiltonianAndCurrents}
mostramos como derivar, en nuestro marco de bosonizaci\'on
funcional, el Hamiltoniano bos\'onico y las correspondientes
corrientes bosonizadas. Finalmente discutimos brevemente la
conservaci\'on de la corriente. En la secci\'on
\ref{part:conclusions} reunimos los resultados y las conclusiones.

\section{El modelo y el determinante fermi\'onico}\label{part:model}

Consideraremos una versi\'on no covariante del modelo de Thirring
definido por el Lagrangiano eucl\'{\i}deo

\begin{equation}
\L=\psib i \partial\!\!\!\slash \psi - \frac{g^2}{2}
V_{(\mu)}j_\mu j_\mu,
\end{equation}
donde $V_0$ and $V_1$ son las constantes de acoplamiento y las
derivadas est\'an redefinidas para incluir a la velocidad de
Fermi:

\begin{align}
\partial_0=&\frac{\partial}{\partial x_0}\\
\partial_1=& v\f\frac{\partial}{\partial x_1}.
\end{align}
N\'otese que $ v\f$ juega el rol de la velocidad de la luz en QFT
que usualmente se toma unitaria. Para $ v\f=1$ y $V_0=V_1=1$
tenemos el modelo de Thirring usual (la constante $g^2$ se incluye
para facilitar la comparaci\'on con los resultados invariantes de
Lorentz). La corriente fermi\'onica se define como

\begin{equation}
j_\mu=\psib\gamma_\mu\psi,
\end{equation}
la cual satisface la ley de conservaci\'on cl\'asica

\begin{equation}
\partial_\mu j_\mu=0.
\end{equation}

La funcional generatriz es

\begin{equation}
\Z[S]=\int\D\psib\D\psi\,\exp\left[-\int d^2x(\L+j_\mu
S_\mu)\right].
\end{equation}

Por medio de una transformaci\'on de Hubbard-Stratonovich, puede
ser escrita en la forma

\begin{equation}\label{generatingFunctional}
\Z[S]=\N\int\D A_\mu \det D\!\!\!\!\slash\;[A]
\,\exp\left[-\frac{1}{2g^2}\int
d^2x\,d^2y\,V^{-1}_{(\mu)}(x-y)(gA_\mu
-S_\mu)(x)(gA_\mu-S_\mu)(y)\right],
\end{equation}
donde

\begin{equation}
 D\!\!\!\!\slash\;[A]=i \partial\!\!\!\slash + g A\!\!\!\slash ,
\end{equation}
y

\begin{equation}
\det  D\!\!\!\!\slash\;[A]=\int\D\psib\D\psi\,\exp\left[-\int
d^2x\,\psib  D\!\!\!\!\slash\;[A]\psi\right].
\end{equation}

\section{Enfoque desacoplante de la bosonizaci\'on}\label{part:decoupling}

Habiendo expresado la funcional generatriz en t\'erminos de un
determinante fermi\'onico, trazaremos ahora un esquema del
m\'etodo desacoplante, que se encuentra en la base del enfoque
funcional de la
bosonizaci\'on\cite{roskies81,gamboa81,furuya82,stone94}. En
$(1+1)$ dimensiones espacio-temporales, el campo vectorial $A_\mu$
puede descomponerse en sus partes transversal y longitudinal del
siguiente modo:

\begin{equation}\label{AphietaRelation}
A_\mu = -(1/g) (\epsilon_{\mu\nu}
\partial_\nu\phi-\partial_\mu\eta),
\end{equation}
donde $\eta$ ($\phi$) es un campo escalar (pseudoescalar). Notemos
que si realizamos la siguiente transformaci\'on en los campos
fermi\'onicos

\begin{align}
\psi=&e^{t[\gamma_5\phi+i\eta]}\chi\\
\psib=&e^{t[\gamma_5\phi-i\eta]}\chib,
\end{align}
con $t$ un par\'ametro real, entonces la densidad lagrangiana
fermi\'onica cambia como

\begin{equation}
\psib D\!\!\!\!\slash\;[A]\psi=\chib D\!\!\!\!\slash_t[A]\chi
\end{equation}
donde

\begin{equation}
 D\!\!\!\!\slash_t[A]= D\!\!\!\!\slash\;[(1-t)A].
\end{equation}

Como fue observado por primera vez por Fujikawa \cite{fujikawa79},
el jacobiano asociado al mencionado cambio en las variables
fermi\'onicas es no trivial, y depende de los campos $\phi$ y
$\eta$:

\begin{equation}
\det(i
\partial\!\!\!\slash +gA\!\!\!\slash )=J[\phi,\eta;t]\det(i \partial\!\!\!\slash +g(1-t)A\!\!\!\slash ).
\end{equation}
Debe notarse que para $t=1$ los grados de libertad bos\'onicos y
fermi\'onicos se desacoplan completamente. Puede mostrarse que

\begin{equation}\label{jacobian}
J[\phi,\eta;1]\equiv J =\exp\left[-\int_0^1\omega(t)\,dt\right]
\end{equation}
con

\begin{equation}\label{omega}
\omega(t)=-\tr D\!\!\!\!\slash_t[A]^{-1}\,g\,A\!\!\!\slash
=-\lim_{y\rightarrow x}\tr^D\int d^2x\,
D\!\!\!\!\slash_t[A]^{-1}(x,y)\, g\,A\!\!\!\slash (x),
\end{equation}
donde $\tr^D$ indica la operaci\'on de traza en el espacio de
Dirac. Todas estas f\'ormulas guardan una gran analog\'{\i}a con
las correspondientes a una QFT covariante. M\'as a\'un, la \'unica
diferencia entre ellas es la presencia de $ v\f$ en lugar de la
velocidad de la luz, aunque esto tiene consecuencias no triviales.
La \'ultima ecuaci\'on debe ser regularizada, de otro modo
aparecen divergencias, como es obvio al tomar el l\'{\i}mite
$y\rightarrow x$. En QFT, cualquier regularizaci\'on aceptable
tiene que ser invariante de Lorentz. En el presente caso, no
tenemos esa limitaci\'on por dos razones: i) la invarianza de
Lorentz est\'a rota desde el principio dado que estamos en una
teor\'{\i}a no relativista; ii) Hay una covarianza remanente: la
teor\'{\i}a sin interacciones es invariante con respecto al grupo
de Lorentz, donde la velocidad de la luz se ha reemplazado por la
de Fermi, pero \'esta es una simetr\'{\i}a artificial, y no existe
raz\'on para respetarla. M\'as a\'un, al tomar $V_0\neq V_1$
($g_2\neq g_4$) queda expl\'{\i}citamente rota. Antes de utilizar
la libertad que proviene de la ausencia de covarianza, ser\'{\i}a
instructivo rever los resultados obtenidos previamente eligiendo
una regularizaci\'on invariante de Lorentz\cite{naon95}.
Utilizando un regulador de la forma

\begin{equation}
\left( D\!\!\!\!\slash_t[A]\, D\!\!\!\!\slash_t[A]^{\dagger} +
 D\!\!\!\!\slash_t[A]^{\dagger}\, D\!\!\!\!\slash_t[A]\right)/2,
\end{equation}
que fue propuesto por primera vez por Fujikawa en su an\'alisis de
las anomal\'{i}as covariante y consistente\cite{fujikawa03}, se
obtiene

\begin{equation}
J_\text{cov}=\exp\left\{-\frac{a}{2\pi v\f}\int
d^2x\,\left[(\partial_1\phi)^2+(\partial_0\phi)^2\right]\right\},
\end{equation}
donde $a$ es un par\'ametro vinculado a posibles ambig\"uedades en
la regularizaci\'on. Para $a=1$ se obtiene una regularizaci\'on
invariante de gauge. Aunque el modelo de Thirring no posee
invarianza de gauge local, en el presente contexto estamos
interesados fundamentalmente en la invarianza de Lorentz y podemos
fijar $a=1$ sin perder generalidad. Insertando el jacobiano
anterior en la funcional generatriz, absorbiendo el determinante
fermi\'onico libre, lo que resulta del procedimiento de desacople
en un factor de normalizaci\'on, y expresando $A_\mu$ en
t\'erminos de $\phi$ y $\eta$ seg\'un la Ec.
(\ref{AphietaRelation}), se obtiene una acci\'on bosonizada. En el
contexto de la materia condensada estos grados de libertad
bos\'onicos se interpretan como campos asociados a oscilaciones de
densidad de carga. De esta acci\'on bos\'onica derivada a trav\'es
de una regularizaci\'on que preserva la covarianza se puede
facilmente calcular la relaci\'on de dispersi\'on correspondiente

\begin{equation}
p_0^2 + v_\text{cov}^2\,p_1^2=0
\end{equation}
donde

\begin{equation}
v_\text{cov}^2= v\f^2\frac{\left( v\f+\frac{g^2 V_0}{\pi}\right)}
{\left( v\f+\frac{g^2 V_1}{\pi}\right)}.
\end{equation}
Debemos enfatizar aqu\'{\i}, que s\'olo para $V_1=0$ esta
velocidad, llamada velocidad renormalizada, acuerda con el valor
obtenido usando bosonizaci\'on operacional en materia condensada,
que es

\begin{equation}\label{velocity}
v^2=\left( v\f-\frac{V_1 g^2}{\pi}\right)\left( v\f+\frac{V_0
g^2}{\pi}\right).
\end{equation}

A continuaci\'on describiremos dos m\'etodos diferentes para
regularizar el jacobiano, que no preservan la invarianza de
Lorentz y que permiten obtener la acci\'on bos\'onica efectiva que
conduce a la respuesta correcta para las relaciones de
dispersi\'on.

\subsection{M\'etodo Point-splitting}

Como es bien sabido, el m\'etodo de regularizaci\'on
point-splitting rompe la invarianza de Lorentz
expl\'{\i}citamente. Consiste en una prescripci\'on para tomar el
l\'{\i}mite $y\rightarrow x$ antes mencionado al definir

\begin{equation}\label{defPointSplitting}
\lim_{y\rightarrow x}
D\!\!\!\!\slash_t[A]^{-1}(x,y)=\unmedio(\lim_{\epsilon\rightarrow
0^+}+\lim_{\epsilon\rightarrow 0^-})
D\!\!\!\!\slash_t[A]^{-1}(x_0,x_1;x_0,x_1+\epsilon),
\end{equation}
es decir, tomando un l\'{\i}mite sim\'etrico en la variable
espacial. Necesitamos entonces la funci\'on de Green del operador
de Dirac, que satisface

\begin{equation}
 D\!\!\!\!\slash_t[A]_x D\!\!\!\!\slash_t[A]^{-1}(x,y)=\delta^2(x-y)
\end{equation}

Como es usual, proponemos el ansatz

\begin{equation}
 D\!\!\!\!\slash
_t[A]^{-1}(x,y)=e^{(1-t)\left[\gamma_5\phi(x)+i\eta(x)\right]}\,
G_0(x,y)\, e^{(1-t)\left[\gamma_5\phi(y)-i\eta(y)\right]},
\end{equation}
donde $G_0$ es la funci\'on de Green del operador de Dirac libre:

\begin{equation}
i \partial\!\!\!\slash _x G_0(x,y)=\delta^2(x-y).
\end{equation}

Con esta receta, hallamos el siguiente resultado para la
ecuaci\'on (\ref{defPointSplitting}):

\begin{equation}
\lim_{y\rightarrow x} D\!\!\!\!\slash
_t[A]^{-1}(x,y)=-\frac{i}{2\pi v\f}(1-t)\,\gamma_1
\partial_1 \left[\gamma_5\phi(x)-i\eta(x)\right]
\end{equation}
y entonces, el Jacobiano (Ecs. (\ref{jacobian}) y (\ref{omega}))
est\'a dado por

\begin{equation}\label{resultJacobian}
J=\exp\left\{-\frac{1}{2\pi v\f}\int
d^2x\,\left[(\partial_1\phi)^2-(\partial_1\eta)^2-2\partial_1\phi\partial_0\eta\right]\right\}.
\end{equation}
La funcional de vac\'{\i}o puede entonces escribirse como

\begin{equation}
\Z[S=0]= \N \int\D\phi\D\eta\,e^{-S\eff}
\end{equation}
donde $\N$ es un factor de normalizaci\'on que incluye el
determinante fermi\'onico libre (independiente de las
interacciones). Tambi\'en hemos definido $S\eff$, que en el
espacio de momentos toma la forma

\begin{equation}
S\eff=\int
\frac{d^2p}{(2\pi)^2}\,\left[\phi(p)A\phi(-p)+\eta(p)B\eta(-p)+2\phi(p)C\eta(-p)\right].
\end{equation}
con

\begin{align}
A=& v\f^2p_1^2\left(\frac{1}{2g^2V_0}+\frac{1}{2\pi v\f}\right)+\frac{p_0^2}{2g^2V_1}\\
B=& v\f^2p_1^2\left(\frac{1}{2g^2V_1}-\frac{1}{2\pi v\f}\right)+\frac{p_0^2}{2g^2V_0}\\
C=&p_1p_0
v\f\left(\frac{1}{2g^2V_1}-\frac{1}{2g^2V_0}-\frac{1}{2\pi
v\f}\right)
\end{align}

El contenido f\'{\i}sico del modelo puede extraerse de $S\eff$ que
describe la din\'amica de los modos colectivos del sistema. Cuando
el modelo fermi\'onico original est\'a relacionado al modelo de
Tomonaga-Luttinger utilizado para el estudio de sistemas
electr\'onicos en una dimensi\'on
\cite{voit95,schulz00,rao01,varma02}, estas excitaciones
colectivas corresponden a oscilaciones de densidad de carga
(plasmones). Su relaci\'on de dispersi\'on puede obtenerse de los
ceros del determinante de la matriz

\begin{equation}\label{Mmatrix}
\begin{pmatrix}
      A & C \\
      C & B \\
\end{pmatrix}.
\end{equation}
El resultado es

\begin{equation}
p_0^2 + v^2\,p_1^2=0
\end{equation}
donde $v$ es la velocidad renormalizada de los modos de densidad
de carga dada por la Ec. (\ref{velocity}).

\subsection{M\'etodo Heat-kernel}

Otro modo popular de tratar la regularizaci\'on de los
determinantes fermi\'onicos es el m\'etodo heat-kernel
\cite{fujikawa79,gamboa84}. En este esquema $J$ se regula
insertando un operador de la forma $e^{-R/M^2}$, $R$ es un
operador definido positivo, y $M$ es un par\'ametro que juega el
rol de una masa, y que se deja fijo en los c\'alculos intermedios.
El l\'{\i}mite $M^2\rightarrow\infty$ se toma al final. De nuevo,
debemos mencionar que en contextos de QFT estandar el operador $R$
puede elegirse entre aquellos compatibles con la invarianza de
Lorentz (dejemos de lado, por el momento, otras posibles
simetr\'{\i}as), por ejemplo $R= D\!\!\!\!\slash _t[A]^2$.
Aqu\'{\i} no tenemos esa limitaci\'on, y nuestro objetivo es
hallar la forma precisa de $R$ que conduce a una acci\'on efectiva
que contenga las relaciones de dispersi\'on deseadas.

Comenzamos por reescribir la ecuaci\'on (\ref{omega}) como

\begin{equation}\label{omega2}
\omega(t)=\tr\left\{ D\!\!\!\!\slash
_t[A]^{-1}\left[\left(\gamma_5\phi-i\eta\right) D\!\!\!\!\slash
_t[A]+  D\!\!\!\!\slash
_t[A]\left(\gamma_5\phi+i\eta\right)\right]\right\}.
\end{equation}

La operaci\'on de traza est\'a mal definida, y necesita ser
regularizada. Definimos nuestra $\omega$ regularizada como

\begin{equation}
\omega(t)_R=\lim_{M\rightarrow\infty}\tr\left\{ D\!\!\!\!\slash
_t[A]^{-1}\left[\left(\gamma_5\phi-i\eta\right) D\!\!\!\!\slash
_t[A]+  D\!\!\!\!\slash
_t[A]\left(\gamma_5\phi+i\eta\right)\right]e^{-R/M^2}\right\}.
\end{equation}
La elecci\'on de $R$ siempre se encuentra dictada por
consideraciones f\'{\i}sicas, por ejemplo, si consideramos una
teor\'{\i}a de gauge, debemos tener en cuenta regularizaciones que
no destruyan la invarianza de gauge a nivel cu\'antico. Esto es
usualmente realizado tomando $R= D\!\!\!\!\slash _t[A]^2$, donde
$A_\mu$ es un campo de gauge. Aqu\'{\i} el modelo bajo estudio no
es una teor\'{\i}a de gauge y por lo tanto tenemos m\'as libertad
de elegir el regulador. Emplearemos un operador de la forma $R=
D\!\!\!\!\slash _t[B]^2$, donde $B_\mu$ es un cierto campo
vectorial a ser determinado. Podemos escribir $\omega(t)_R$ como
$\omega(t)_R=\omega_0(t)+\omega_\text{nc}(t)$ donde

\begin{align}
\omega_0(t)=&\tr\left(2\gamma_5\phi e^{-R/M^2}\right)\\
\omega_\text{nc}(t)=&\tr\left\{ D\!\!\!\!\slash
_t[A]^{-1}\left(\gamma_5\phi-i\eta\right) \left[ D\!\!\!\!\slash
_t[A],e^{-R/M^2}\right]\right\}.
\end{align}
Aqu\'{\i} el sub\'{\i}ndice $0$ indica el t\'ermino que
habr\'{\i}amos obtenido si hubi\'eramos empleado la propiedad
c\'{\i}clica de la traza en la ecuaci\'on (\ref{omega2}). El
sub\'{\i}ndice $\text{nc}$ se refiere a un t\'ermino ``no
c\'{\i}clico" (esta cuesti\'on se discute en detalle en la Ref.
\citen{cabra89}). Las expresiones finales para estos dos
t\'erminos son

\begin{align}
\omega_0(t)&=-(1-t)\frac{g}{\pi}\int
d^2x\,\phi\epsilon_{\mu\nu}\partial_\mu B_\nu\\
\omega_\text{nc}(t)&=-(1-t)\frac{g}{2\pi}\int
d^2x\,\partial_\mu(B_\nu-A_\nu)(\epsilon_{\nu\mu}\phi+\delta_{\nu\mu}\eta).
\end{align}

En este punto, un c\'alculo directo nos permite comprobar que
tomando

\begin{align}
B_0=&A_0\\
B_1=&-A_1,
\end{align}
se arriba al mismo resultado obtenido en la subsecci\'on previa
Ecs. (\ref{resultJacobian})-(\ref{Mmatrix})). De este modo, hemos
hallado una forma expl\'{\i}cita para el operador que regula el
jacobiano asociado a un determinante fermi\'onico no covariante. A
su vez, esta forma conduce a las relaciones de dispersi\'on
correctas para el modelo de Tomonaga-Luttinger. La derivaci\'on de
este jacobiano, y del operador regulador era una de las
principales motivaciones de este cap\'{\i}tulo, y constituye uno
de los aportes originales de esta tesis.

\section{El Hamiltoniano bosonizado y las corrientes}\label{part:hamiltonianAndCurrents}

Hasta este punto hemos trabajado en la formulaci\'on lagrangiana.
Sin embargo en aplicaciones de materia condensada, el marco
Hamiltoniano es usualmente el preferido. Es deseable entonces
mostrar la derivaci\'on del Hamiltoniano usual para sistemas
electr\'onicos unidimensionales en el marco de la bosonizaci\'on
funcional discutido en este cap\'{\i}tulo, es decir, la forma
bos\'onica del modelo de
Tomonaga-Luttinger\cite{voit95,schulz00,rao01,varma02}. El otro
punto que tratamos en esta secci\'on es la forma bos\'onica de las
corrientes fermi\'onicas originales (densidad de carga y corriente
el\'ectrica), y sus leyes de conservaci\'on.

Teniendo en cuenta la expresi\'on para $\det D\!\!\!\!\slash\;[A]$
calculada en las secci\'ones precedentes, y la relaci\'on entre
los campos $\phi$ y $\eta$ y el campo $A_\mu$
(Ec.(\ref{AphietaRelation})), podemos expresar la funcional
generatriz (\ref{generatingFunctional}) en t\'erminos del campo
$A_\mu$:

\begin{multline}
\Z[S]=\N\int\D A_\mu\exp\left(-\frac{1}{2}\int
d^2x\,d^2y\,A_\mu(x)D_{\mu\nu}(x-y)A_\nu(y)\right)\times\\\exp\left(
-\frac{1}{2 g^2}\int d^2x\,d^2y\, S_\mu(x) V_{(\mu)}^{-1}(x-y)
S_\mu(y)\right)\exp\left(-\frac{1}{g}\int d^2x\,d^2y\,A_\mu(x)
V_{(\mu)}^{-1}(x-y) S_\mu(y)\right),
\end{multline}
donde $D_{\mu\nu}$ est\'a dado en el espacio de Fourier por

\begin{equation}
D_{\mu\nu}(p)=\frac{g^2}{\pi(p_0^2+ v\f^2 p_1^2)}
\begin{pmatrix}
     v\f p_1^2 & p_0p_1\\
    p_0 p_1  & - v\f p_1^2 \\
\end{pmatrix} +
\begin{pmatrix}
    \frac{1}{V_0} & 0 \\
    0 & \frac{1}{V_1} \\
\end{pmatrix}.
\end{equation}
El campo $A_\mu$ se desacopla de la fuente $S_\mu$ a trav\'es del
procedimiento usual de efectuar la traslaci\'on

\begin{equation}
A_\mu\rightarrow A_\mu + \frac{D_{\mu\nu}^{-1}S_\nu}{gV_{(\nu)}},
\end{equation}
obteniendo

\begin{multline}
\Z[S]=\N\int\D A_\mu\exp\left(-\frac{1}{2}\int
d^2x\,d^2y\,A_\mu(x)D_{\mu\nu}(x-y)A_\nu(y)\right)\times\\\exp\left[\frac{1}{2}\int
d^2x\,d^2y\,S_\mu(x)\Delta_{\mu\nu}^{-1}(x-y)S_\nu(y)\right],
\end{multline}
con $\Delta_{\mu\nu}^{-1}$ y la constante de dureza $K$ dados por
(en el espacio de Fourier)

\begin{equation}
\Delta_{\mu\nu}^{-1}(p)=\frac{1}{\pi(p_0^2+v^2 p_1^2)}
\begin{pmatrix}
    -Kvp_1^2 & p_0p_1\\
    p_0 p_1  & \frac{v}{K}p_1^2\\
\end{pmatrix},
\end{equation}

\begin{equation}
K=\sqrt{\frac{ v\f-g^2V_1/\pi}{ v\f+g^2V_0/\pi}}.
\end{equation}
Esta constante gobierna los exponentes del decaimiento de las
funciones de correlaci\'on. A continuaci\'on podemos multiplicar y
dividir por

\begin{equation}
\int\D A_\mu\exp\left(-\frac{1}{2}\int
d^2x\,d^2y\,A_\mu(x)\Delta_{\mu\nu}(x-y)A_\nu(y)\right),
\end{equation}
y efectuar una traslaci\'on adicional en el campo $A_\mu$

\begin{equation}
A_\mu\rightarrow A_\mu + \Delta_{\mu\nu}^{-1} S_\nu,
\end{equation}
para obtener

\begin{equation}
\Z[S]=\tilde{\N}\int\D A_\mu \exp\left(-\frac{1}{2}\int
d^2x\,d^2y\,A_\mu(x) \Delta_{\mu\nu}(x-y) A_\nu(y)+\int
d^2x\,S_\mu(x)A_\mu(x)\right).
\end{equation}
Finalmente, definiendo los campos $\varphi$ y $\theta$ del
siguiente modo:

\begin{align}
A_0&=\frac{-1}{\sqrt{\pi}}\,\partial_x\varphi\\
A_1&=\frac{i}{\sqrt{\pi}}\,\partial_x \theta,
\end{align}
llegamos a la funcional generatriz

\begin{multline}
\Z[S]=\bar{\N}\int\D\varphi\D\theta\exp\left(-\frac{1}{2}\int dx
d\tau\left[\frac{v}{K}(\partial_x\varphi)^2+v K
(\partial_x\theta)^2+2i\partial_x\theta\partial_\tau\varphi\right]\right)\times\\
\exp\left(\int dx d\tau\,\left[-S_0\partial_x\varphi/\sqrt{\pi} +
i S_1\partial_x\theta/\sqrt{\pi}\right]\right).
\end{multline}

Naturalmente identificamos entonces el campo $\varphi$ con el modo
de densidad de carga del sistema y $\Pi=\partial_x\theta$ como su
campo can\'onico conjugado. M\'as a\'un, los primeros dos
t\'erminos en la acci\'on cuadr\'atica de la expresi\'on previa
pueden ser identificados con el Hamiltoniano del sistema:

\begin{equation}
\H=\frac{1}{2}\int dx\,\left[\frac{v}{K}(\partial_x\varphi)^2+v K
(\partial_x\theta)^2\right],
\end{equation}
el cual coincide exactamente con el Hamiltoniano obtenido
utilizando bosonizaci\'on operacional
estandar\cite{voit95,schulz00,rao01,varma02}. Ahora, por
derivaci\'on funcional obtenemos la forma bos\'onica de las
corrientes

\begin{align}
j_0&=\frac{-1}{\sqrt{\pi}}\,\partial_x\varphi\\
j_1&=\frac{i}{\sqrt{\pi}}\,\Pi,
\end{align}
que son id\'enticas, por supuesto, a las halladas en el enfoque
operacional. Es importante enfatizar que estas corrientes no
obedecen la ecuaci\'on de continuidad. Siguiendo la Ref.
\cite{giamarchi91}, se introduce una corriente el\'ectrica
f\'{\i}sica $j$, que es en general diferente de $j_1$. La densidad
de carga es identificada con $j_0$ ($j_0=\rho$). La corriente
f\'{\i}sica se determina al imponer la ecuaci\'on de continuidad:

\begin{equation}
\frac{\partial\rho}{\partial\tau} + \frac{\partial j}{\partial
x}=0.
\end{equation}
Obtenemos

\begin{equation}
j=\frac{i}{\sqrt{\pi}}\,v\,K\,\Pi.
\end{equation}
N\'otese que s\'olo para $V_1=0$ ($g_2=g_4$ en el lenguaje de
Tomonaga-Luttinger) se llega a $v\,K=1$ y $j=j_1$.

\section{Conclusiones}\label{part:conclusions}

En este cap\'{\i}tulo consideramos un determinante fermi\'onico
asociado a teor\'{\i}as cu\'anticas de campos no covariantes. En
particular estudiamos el determinante que aparece al implementar
la bosonizaci\'on mediante integrales funcionales basada en el
desacople del determinante fermi\'onico a trav\'es de cambios
apropiados en las variables de integraci\'on. El modelo analizado
(una versi\'on no covariante del modelo de Thirring) se ha
utilizado previamente para describir sistemas de electrones
altamente correlacionados en una dimensi\'on (l\'{\i}quidos de
Luttinger).

En el contexto del m\'etodo de regularizaci\'on del n\'ucleo del
calor, explotando la libertad originada en la no covarianza,
determinamos el operador que act\'ua como regulador y que da lugar
a una acci\'on bos\'onica, a una relaci\'on de dispersi\'on, y a
un decaimiento de funciones de correlaci\'on correctos en
t\'erminos de las constantes de acoplamiento, es decir, en
completo acuerdo con aquellas que se obtienen, y son bien
conocidas en el marco operacional. C\'alculos previos mediante
integrales funcionales hab\'{\i}an hecho uso de reguladores
covariantes tomados de teor\'{\i}as de campos relativistas, y
daban un espectro y exponentes correctos s\'olo para valores
particulares de las constantes de acoplamiento.

Finalmente mostramos como derivar el Hamiltoniano bosonizado y las
corrientes , que coinciden con los obtenidos mediante
bosonizaci\'on operacional estandar. De este modo fuimos capaces
de establecer el regulador heat-kernel que brinda completo acuerdo
entre los enfoques operacionales y mediante integrales funcionales
de la bosonizaci\'on del modelo de Tomonaga-Luttinger.


\chapter{Interacciones de inversi\'on de spin en el modelo de Thirring no
local}\label{part:spinflip}
\begin{center}
\begin{minipage}{5.6in}
\textsl{Extendemos una versi\'on no local y no covariante del
modelo de Thirring con el objeto de describir sistemas de muchos
cuerpos que poseen interacciones de inversi\'on de spin.
Introduciendo un modelo con dos especies de fermiones evitamos el
uso de bosonizaci\'on no abeliana, necesario en un enfoque previo.
Obtenemos una expresi\'on bosonizada para la funci\'on de
partici\'on, que describe la din\'amica de los modos colectivos
del sistema. Los resultados de este cap\'{\i}tulo constituyen un
aporte original de esta tesis\cite{iucci01}.}
\end{minipage}
\end{center}

\section{Introducci\'on}

En el cap\'{\i}tulo anterior estudiamos el modelo de
Tomonaga-Luttinger mediante bosonizaci\'on funcional. Como ya se
ha mencionado \'este es un modelo en el que se desprecia el spin
electr\'onico y que s\'olo contempla interacciones de dispersi\'on
hacia adelante, que poseen momento transferido nulo; sus
caracter\'{\i}sticas m\'as notables son la existencia de funciones
de correlaci\'on no anal\'{\i}ticas y no universales, y de una
singularidad algebraica en la distribuci\'on de momentos sobre la
superficie de Fermi; los modos colectivos bos\'onicos resultantes
poseen una relaci\'on de dispersi\'on sin
gap\cite{voit95,schulz00,varma02}, de modo que una cantidad
arbitrariamente peque\~na de energ\'{\i}a es capaz de producir
excitaciones a partir del estado fundamental. Sin embargo, en
situaciones m\'as realistas se deben tener en cuenta el spin
electr\'onico y procesos de dispersi\'on electr\'onica m\'as
complejos como dispersi\'on hacia atr\'as y umklapp. En estos
procesos el momento transferido es $2k\f$ y $4k\f$
respectivamente. Estos problemas han sido tratados en el contexto
de la bosonizaci\'on operacional usual en el cap\'{\i}tulo 2,
siguiendo las referencias \cite{voit95,vondelft98}. La inclusi\'on
del spin electr\'onico da lugar a la separaci\'on spin-carga, y
los nuevos t\'erminos de dispersi\'on hacia atr\'as y umklapp
generan gaps en los espectros de ambos sectores respectivamente, y
por supuesto cambian los decaimientos algebraicos de las funciones
de correlaci\'on por decaimientos exponenciales.

En trabajos recientes\cite{naon95,naon97,manias98} se trataron
estos modelos m\'as realistas mediante el enfoque de la
teor\'{\i}a de campos y la bosonizaci\'on funcional. Se introdujo
una versi\'on no local y no covariante del modelo de Thirring en
el que las densidades y corrientes fermi\'onicas se acoplan
mediante potenciales bilocales generales, y una generalizaci\'on a
un modelo de Gross-Neveu SU(N) no local con N=2 para tener en
cuenta el spin electr\'onico. Estos modelos contienen al modelo de
TL como caso particular (para N=1). Aunque constituyen un marco
elegante para tratar problemas de muchos cuerpos en 1D, presentan
dos inconvenientes: i) la t\'ecnica de bosonizaci\'on funcional
empleada descansa en la regularizaci\'on del jacobiano tomada de
la teor\'{\i}a de campos covariante, con las consiguientes
diferencias halladas con las Ref. \citen{voit95,schulz00,varma02}
mencionadas en el cap\'{\i}tulo anterior, y ii) la descripci\'on
de los modelos con simetr\'{\i}a SU(N) se realiz\'o mediante
bosonizaci\'on no abeliana, lo que resulta poco pr\'actico para el
an\'alisis final.

En el presente cap\'{\i}tulo salvamos estos problemas
introduciendo un modelo de tipo Thirring con dos especies de
fermiones, generalizando el modelo de dos fermiones de la Ref.
\citen{zinn-justin97}, construido originalmente como una
teor\'{\i}a local y covariante, al caso en que las interacciones
entre corrientes fermi\'onicas se hallan mediadas por funciones
bilocales. Este modelo no incluye interacciones de tipo umklapp.
S\'olo posee interacciones de tipo inversi\'on de spin (IS), que
en el caso local se reducen a las de dispersi\'on hacia atr\'as.
Adicionalmente utilizamos la t\'ecnica de bosonizaci\'on abeliana
en materia condensada, tal como se la formul\'o en el
cap\'{\i}tulo anterior.

El resultado final puede considerarse una extensi\'on del trabajo
de Grinstein, Minnhagen y Rosengren \cite{grinstein79} donde se
estudi\'o una versi\'on simplificada del problema de inversi\'on
de spin (contenido en nuestro modelo). Extendiendo formulaciones
previas de modelos masivos\cite{coleman75,naon85,li98}, nuestro
procedimiento permite obtener una funcional de vac\'{\i}o
bosonizada. El cap\'{\i}tulo est\'a organizado de la siguiente
manera. En la secci\'on \ref{part:modelo} introducimos el
mencionado modelo no local de dos fermiones y explicamos su
relaci\'on con la descripci\'on no abeliana previa. En la
secci\'on \ref{part:bosonización} establecemos la equivalencia
entre la funci\'on de partici\'on fermi\'onica inicial y la que
corresponde a una extensi\'on no local de dos bosones del modelo
seno-Gordon. Finalmente, en la secci\'on
\ref{part:conclusionesSpinFlip} reunimos los puntos principales de
nuestras investigaciones.

\section{El modelo y su relaci\'on con una descripci\'on no abeliana
previa.}\label{part:modelo}

En esta secci\'on introducimos una versi\'on no local del modelo
de Thirring que incorpora el spin electr\'onico considerando dos
especies de fermiones, donde cada especie representa un estado de
spin. La acci\'on eucl\'{\i}dea inicial es

\begin{multline}\label{eq:model}
S = \int d^2x\,\Psib^a i \slp \Psi^a - \frac{g^2}{2}\int
d^2x\,d^2y\, j_\mu^a(x) V^{ab}_{(\mu)}(x-y)j_\mu^b(y)\\ - g_s\int
d^2x\,d^2y\,\Psib^1 (x)\gamma_\mu\Psi^2
(x)U_{(\mu)}(x-y)\Psib^2(y)\gamma_\mu\Psi^1(y)
\end{multline}
donde $a,b=1,2$, con $1=\uparrow ,2=\downarrow$, las corrientes
$j^a_\mu$ son las corrientes fermi\'onicas usuales

\begin{equation}
j_\mu^1=\Psib^1\gamma_\mu\Psi^1,\;\;\;\;
j_\mu^2=\Psib^2\gamma_\mu\Psi^2,
\end{equation}
y las matrices $V_{(\mu)}^{ab}$ tienen la forma

\begin{equation}
V_{(\mu)} = \frac{1}{2}
\begin{pmatrix}
      V^\rho_{(\mu)}+V^\sigma_{(\mu)} & V^\rho_{(\mu)}-V^\sigma_{(\mu)} \\
      V^\rho_{(\mu)}-V^\sigma_{(\mu)} & V^\rho_{(\mu)}+V^\sigma_{(\mu)} \\
\end{pmatrix},
\end{equation}
donde $V^\rho_{(\mu)}$ y $V^\sigma_{(\mu)}$ son los potenciales de
acoplamiento relacionados a las constantes $g_2$ y $g_4$ de
dispersi\'on hacia adelante de Solyom\cite{solyom79}. Estas
funciones se vinculan a interacciones del tipo corriente-corriente
que no invierten el spin. Las funcions $U_{(\mu)}(x-y)$ son los
acoplamientos que gobiernan los procesos que cambian el estado del
spin electr\'onico. Hemos dejado las constantes $g$ y $g_s$ para
facilitar la comparaci\'on con versiones locales del modelo. Por
ejemplo, el caso $g_s=0$ y
$V^\rho_{(\mu)}=V^\sigma_{(\mu)}=\delta^2(x-y)$ corresponde a dos
modelos de Thirring usuales desacoplados. La acci\'on
(\ref{eq:model}) tiene simetr\'{\i}a quiral U(1) manifiesta, es
decir, es invariante bajo la transformaci\'on $\Psi^a\rightarrow
e^{i\gamma_5 \theta}\Psi^a$, $\Psib^a\rightarrow \Psib^a
e^{i\gamma_5 \theta}$. La conservaci\'on del spin fermi\'onico
tambi\'en se preserva en esta teor\'{\i}a.

La teor\'{\i}a definida m\'as arriba es similar a la descripta por
Zinn-Justin(ZJ) \cite{zinn-justin97}. Existen, sin embargo dos
diferencias importantes. En primer lugar nuestro modelo tiene en
cuenta la posible naturaleza de largo alcance de los potenciales,
mientras que el modelo de ZJ es local. Por otro lado, el modelo de
ZJ no incluye t\'erminos del tipo $g_4$ asociados a procesos de
dispersi\'on fermi\'onica que involucran s\'olo una rama (derecha
o izquierda), ni para diagramas de IS ni para diagramas
ordinarios.

En lo concerniente a la relaci\'on entre nuestra acci\'on y
modelos previos inspirados en materia condensada, debemos
mencionar los primeros trabajos de Luther y Emery\cite{luther74b},
y Grinstein, Minnhagen y Rosengren \cite{grinstein79}. Los
primeros introdujeron el llamado modelo de dispersi\'on hacia
atr\'as. Aunque este sistema, en principio, no posee IS, en el
l\'{\i}mite local que ellos consideraron, los diagramas de
dispersi\'on hacia atr\'as coinciden con aquellos que cambian el
spin. Grinstein y col. incluyeron desde el comienzo interacciones
de tipo IS teniendo en cuenta un potencial coulombiano. Aunque
dicho modelo es no local, considera el mismo potencial para todas
las clases de diagramas (IS y los ordinarios). Por otro lado, con
el objeto de establecer una relaci\'on entre su teor\'{\i}a y un
sistema de tipo gas de Coulomb, los autores consideraron de nuevo
el l\'{\i}mite local. Y, nuevamente, no incluyeron t\'erminos de
tipo $g_4$.

Mostremos ahora que la acci\'on(\ref{eq:model}) puede escribirse
de un modo alternativo. Consideremos las corrientes U(N)

\begin{equation}
J^\alpha_\mu=\Psib\gamma_\mu\lambda^\alpha\Psi
\end{equation}
con

\begin{align}
\lambda^0&=\unmedio I \\ \lambda^j&=t^j,
\end{align}
siendo $t^j$ los generadores de SU(N) normalizados de acuerdo a

\begin{equation}
\tr{t^it^j}=\unmedio\delta^{ij}.
\end{equation}

Con estas corrientes, podemos definir un modelo de Gross-Neveu,
invariante quiral y no local, con una acci\'on dada por

\begin{equation}\label{eq:grossneveu}
S=\int d^2x\, \Psib i\slp \Psi - \int
d^2x\,d^2y\,J^\alpha_\mu(x)\V^{\alpha\beta}_{(\mu)}(x-y)J^\beta_\mu(y),\qquad
\alpha,\beta=0,1,...,N^2-1
\end{equation}
donde $\V_{(\mu)}$ son $N^2\times N^2$ matrices sim\'etricas que
pesan la interacci\'on . Tomando $N=2$, las acciones
(\ref{eq:model}) y (\ref{eq:grossneveu}) son iguales si las
matrices $\V_{(\mu)}$ se escriben como

\begin{equation}
\V_{(\mu)} =
\begin{pmatrix}
      g^2 V^\rho_{(\mu)} & 0 & 0 & 0 \\
      0 & g_s U_{(\mu)} & 0 & 0  \\
      0 & 0 & g_s U_{(\mu)} & 0  \\
      0 & 0 & 0 & g^2 V^\sigma_{(\mu)} \\
\end{pmatrix}.
\end{equation}

Este modelo no abeliano fue considerado en la Ref. \citen{naon95}.
La acci\'on efectiva bosonizada obtenida mediante bosonizaci\'on
no abeliana dio lugar a una funcional de Wess-Zumino-Witten (WZW),
que resulta dificil de tratar a la hora de obtener el espectro
f\'{\i}sico. En la pr\'oxima secci\'on mostramos como esta tarea
se ve simplificada comenzando por (\ref{eq:model}) en lugar de
(\ref{eq:grossneveu}) y combinando bosonizaci\'on {\em abeliana}
en el marco de la integral funcional y la aproximaci\'on
arm\'onica autoconsistente.

\section{La acci\'on bos\'onica
equivalente}\label{part:bosonización}

Comenzamos considerando la funci\'on de partici\'on

\begin{equation}
\Z=\N\int\D\Psib^a\D\Psi^a\,e^{-S},
\end{equation}
donde $\N$ es una constante de normalizaci\'on. Es conveniente
escribir

\begin{equation}
S=S_0 + S\flip,
\end{equation}
donde

\begin{equation}
S_0=\int d^2x\,\Psib^a i \slp \Psi^a - \frac{g^2}{2}\int
d^2x\,d^2y\, j_\mu^a(x) V^{ab}_{(\mu)}(x-y)j_\mu^b(y)
\end{equation}
y

\begin{equation}\label{eq:flip}
S\flip=-g_s\int d^2x\,d^2y\,\Psib^1(x)\gamma_\mu\Psi^2(x)
U_{(\mu)}(x-y)\Psib^2(y)\gamma_\mu\Psi^1(y).
\end{equation}
La raz\'on para esta separaci\'on se halla en el hecho de que
$S_0$ contiene todos los t\'erminos de interacci\'on que poseen
invarianza quiral separada para cada especie fermi\'onica (estados
de spin). Son interacciones de tipo Thirring, es decir que se
transforman en t\'erminos libres de la acci\'on. El segundo
t\'ermino no posee invarianza quiral separada, y ser\'a expandido
en una serie perturbativa, en analog\'ia con lo hecho con el
t\'ermino de masa en la bosonizaci\'on funcional del modelo de
Thirring masivo \cite{naon85}.

De igual forma que en el cap\'{\i}tulo anterior, la introducci\'on
de campos vectoriales auxiliares $A_\mu^a$ mediante la
transformaci\'on de Hubbard-Stratonovich permite escribir

\begin{equation}
\Z=\N'\int\D\Psib^a\D\Psi^a\D A_\mu^a\exp\left[-\int
d^2x\,\Psib^a\left(i\slp + g\slA^a\right)\Psi^a - S[A] -
S\flip\right],
\end{equation}
donde $\N'$ es una nueva constante de normalizaci\'on que incluye
al determinante fermi\'onico libre, y

\begin{equation}
S[A]=\frac{1}{2}\int
d^2x\,d^2y\,\left(V^{-1}_{(\mu)}\right)^{ab}(x-y)A_\mu^a(x)A_\mu^b(y),
\end{equation}
con $\left(V^{-1}_{(\mu)}\right)^{ab}$ definido a trav\'es de la
ecuaci\'on

\begin{equation}\label{eq:bV}
\int
d^2y\,\left(V^{-1}_{(\mu)}\right)^{ab}(x-y)V_{(\mu)}^{bc}(y-z)=\delta^{(2)}(x-z)\delta^{ac}.
\end{equation}
Descomponemos ahora $A_\mu^a$ en sus partes transversal y
longitudinal

\begin{equation}
A_\mu^a(x)=\epsilon_{\mu\nu}\partial_{\nu}\phi^a(x)+
\partial_\mu\eta^a(x),
\end{equation}
donde $\phi^a$ y $\eta^a$ son campos escalares. Adem\'as
realizamos un cambio en los campos fermi\'onicos

\begin{equation} \label{f1}
\Psi^a(x) = e^{-g[\gamma_5 \phi^a(x) - i \eta^a(x)]}\chi^a(x)
\end{equation}

\begin{equation} \label{f2}
\Psib^a(x) = \chib^a(x) e^{-g[\gamma_5 \phi^a(x) + i\eta^a(x)]},
\end{equation}
cuyo Jacobiano fue calculado en el cap\'{\i}tulo anterior.
Obtenemos entonces

\begin{equation}
\Z=\N\int\D\chib^a\D\chi^a\D\phi^a\D\eta^a e^{-S\eff},
\end{equation}
siendo $S\eff$ una suma de tres partes:

\begin{equation}
S\eff=S\ferm+S\bos+S\flip
\end{equation}
donde

\begin{equation}
S\ferm=\int d^2x\,\left(\chib^1 i\slp \chi^1 + \chib^2 i\slp
\chi^2\right),
\end{equation}

\begin{multline}
S\bos=\frac{g^2}{2\pi}\int
d^2x\,\left[(\partial_1\phi^a)^2-(\partial_1\eta^a)^2-2\partial_1\phi^a\partial_0\eta^a\right]
+\frac{1}{2}\int d^2x\,d^2y\,\left(V^{-1}_{(\mu)}\right)^{ab}(x-y)\\
\times\left[\epsilon_{\mu\nu}\epsilon_{\mu\rho}\partial_\nu\phi^a(x)\partial_\rho\phi^b(y)
+\partial_\mu\eta^a(x)\partial_\mu\eta^b(y) +
2\epsilon_{\mu\nu}\partial_\nu\phi^a(x)\partial_\mu\eta^b(y)\right],
\end{multline}
y $S\flip$ es el mismo t\'ermino de interacci\'on IS ya definido
en la Ec. (\ref{eq:flip}). Respecto de este \'ultimo t\'ermino,
desde ahora nos restringiremos al caso de interacciones IS de
contacto:

\begin{equation}
U(x-y)_{(0)}=U(x-y)_{(1)}=\delta^{(2)}(x-y),
\end{equation}
y por medio de una transformaci\'on de Fierz seguida del cambio
quiral definido en las Ecs. (\ref{f1}) y (\ref{f2}), podemos
escribirlo en la forma

\begin{equation}
S\flip=2 g_s\int d^2x\left[
e^{-2g(\phi^1-\phi^2)}\chib^1\frac{1+\gamma_5}{2}\chi^1\cdot
\chib^2\frac{1-\gamma_5}{2}\chi^2 +
e^{2g(\phi^1-\phi^2)}\chib^1\frac{1-\gamma_5}{2}\chi^1\cdot
\chib^2\frac{1+\gamma_5}{2}\chi^2\right].
\end{equation}

Ahora estamos listos para hacer la expansi\'on de la funci\'on de
partici\'on tomando $g_s$ como par\'ametro perturbativo:

\begin{multline}
\Z=\N\int\D\phi^a\D\eta^a e^{-S\bos}
\sum_{n=0}^\Infinity\frac{(-2g_s)^n}{n!}\int\prod_{i=1}^n d^2x_i\\
\times\bra\prod_{i=1}^n\bigg[e^{-2g(\phi^1-\phi^2)}\chib^1\frac{1+\gamma_5}{2}\chi^1\cdot
\chib^2\frac{1-\gamma_5}{2}\chi^2 +
e^{2g(\phi^1-\phi^2)}\chib^1\frac{1-\gamma_5}{2}\chi^1\cdot
\chib^2\frac{1+\gamma_5}{2}\chi^2\bigg]\ket\ferm
\end{multline}
donde $\bracket\ferm$ significa valor medio en una teor\'{\i}a con
acci\'on $S\ferm$. S\'olo valores medios que involucran un mismo
n\'umero de factores de la forma
$\unmedio\chib^a(1+\gamma_5)\chi^a$ y
$\unmedio\chib^a(1-\gamma_5)\chi^a$ (en esta expresi\'on no debe
entenderse suma sobre \'{\i}ndices repetidos) son no nulos, por lo
tanto la funci\'on de partici\'on puede escribirse en la forma

\begin{multline}
\Z=\N\int\D\phi^a\D\eta^a\,
e^{-S\bos}\sum_{n=0}^\Infinity\frac{(2g_s)^{2n}}{(n!)^2}\int\left(\prod_{i=1}^n
d^2x_i\,d^2y_i\,
e^{-2g\left[\phi^1(x_i)-\phi^2(x_i)-\phi^1(y_i)+\phi^2(y_i)\right]}\right)
\\ \times \bra\prod_{i=1}^n
\chib^1(x_i)\frac{1+\gamma_5}{2}\chi^1(x_i)
\chib^1(y_i)\frac{1-\gamma_5}{2}\chi^1(y_i)\ket\ferm\\
\times\bra\prod_{i=1}^n
\chib^2(x_i)\frac{1-\gamma_5}{2}\chi^2(x_i)
\chib^2(y_i)\frac{1+\gamma_5}{2}\chi^2(y_i)\ket\ferm.
\end{multline}
El siguiente paso consiste en introducir dos campos escalares
locales sin masa $\vartheta^a$ que ser\'an asociados a los
fermiones libres $\chib^a$ y $\chi^a$. Este truco permite
reemplazar los valores medios fermi\'onicos en la expresi\'on de
arriba por sus contrapartes bos\'onicos
regularizados\cite{zinn-justin97,naon85}, lo que conduce a

\begin{multline}\label{eq:bosExpSerie}
\Z=\N\int\D\phi^a\D\eta^a\,
e^{-S\bos}\sum_{n=0}^\Infinity\frac{(2g_s)^{2n}}{(n!)^2}\int\left(\prod_{i=1}^n
d^2x_i\,d^2y_i\,
e^{-2g\left[\phi^1(x_i)-\phi^2(x_i)-\phi^1(y_i)+\phi^2(y_i)\right]}\right)
\\ \times \left(\frac{i\Lambda}{2\pi}\right)^{2n}
\bra\prod_{i=1}^n
e^{i\sqrt{4\pi}\left[\vartheta^1(x_i)-\vartheta^1(y_i)\right]}\ket_{0,\vartheta^1}
\\ \times \left(\frac{i\Lambda}{2\pi}\right)^{2n}
\bra\prod_{i=1}^n
e^{i\sqrt{4\pi}\left[\vartheta^2(x_i)-\vartheta^2(y_i)\right]}\ket_{0,\vartheta^2}.
\end{multline}

Ahora usamos el hecho de que las divergencias infrarojas del
propagador del campo $\vartheta^a$ proveen una condici\'on de
neutralidad para los valores medios de los operadores de
v\'ertice. Esto significa que el valor medio $\bra \prod_{i=1}^n
e^{i\beta_i\vartheta(x_i)}\ket_{0,\vartheta}$ es no nulo solamente
si $\sum_i\beta_i=0$\cite{zinn-justin97}. La condici\'on de
neutralidad permite rearmar la serie perturbativa de un modo no
trivial. Llegamos entonces a la acci\'on completamente bosonizada
$S_{\text{bos}}$

\begin{equation}
S\bosonic=S\bos + \int d^2x\,
\left\{\frac{1}{2}(\partial_\mu\vartheta^a)^2
-\frac{g_s\Lambda^2}{\pi^2}
\cos\left[2ig(\phi^1-\phi^2)+\sqrt{4\pi}(\vartheta^1+\vartheta^2)\right]\right\}.
\end{equation}

En este punto es interesante observar que existe un cambio de
variables que permite expresar la acci\'on en un modo muy
sugestivo. Efectivamente, escribiendo

\begin{equation}
\theta=\frac{1}{\sqrt{2}}(\vartheta^1+\vartheta^2)
\end{equation}

\begin{equation}
\tilde{\theta}=\frac{1}{\sqrt{2}}(\vartheta^1-\vartheta^2)
\end{equation}

\begin{equation}\label{scSep1}
\phi^{1,2}=\frac{1}{\sqrt{2}}(\phi_\rho\pm\phi_\sigma)
\end{equation}

\begin{equation}\label{scSep2}
\eta^{1,2}=\frac{1}{\sqrt{2}}(\eta_\rho\pm\eta_\sigma)
\end{equation}
donde el signo m\'as (menos) corresponde al par
${\phi^1,\eta^1}$(${\phi^2,\eta^2}$), se ve que el campo
$\tilde{\theta}$ se desacopla completamente de los otros y por lo
tanto puede integrarse, y la acci\'on bos\'onica queda

\begin{equation}
S_{\text{bos}}=S_\rho+S_\sigma,
\end{equation}
donde

\begin{multline}
S_\rho=\frac{g^2}{2\pi}\int
d^2x\,\left[(\partial_1\phi_\rho)^2-(\partial_1\eta_\rho)^2-2\partial_1\phi_\rho\partial_0\eta_\rho\right]
+\frac{1}{2}\int d^2x\,d^2y\,\left(V^\rho_{(\mu)}\right)^{-1}(x-y)\\
\times\left[\epsilon_{\mu\nu}\epsilon_{\mu\rho}\partial_\nu\phi_\rho(x)\partial_\rho\phi_\rho(y)
+\partial_\mu\eta_\rho(x)\partial_\mu\eta_\rho(y) +
2\epsilon_{\mu\nu}\partial_\nu\phi_\rho(x)\partial_\mu\eta_\rho(y)\right],
\end{multline}
y

\begin{multline}\label{eq:SpinAction}
S_\sigma=\frac{g^2}{2\pi}\int
d^2x\,\left[(\partial_1\phi_\sigma)^2-(\partial_1\eta_\sigma)^2-2\partial_1\phi_\sigma\partial_0\eta_\sigma\right]
+\frac{1}{2}\int d^2x\,d^2y\,\left(V^\sigma_{(\mu)}\right)^{-1}(x-y)\\
\times\left[\epsilon_{\mu\nu}\epsilon_{\mu\rho}\partial_\nu\phi_\sigma(x)\partial_\rho\phi_\sigma(y)
+\partial_\mu\eta_\sigma(x)\partial_\mu\eta_\sigma(y) +
2\epsilon_{\mu\nu}\partial_\nu\phi_\sigma(x)\partial_\mu\eta_\sigma(y)\right]\\
+\int d^2x\left[\frac{1}{2}(\partial_\mu\theta)^2
-\frac{g_s\Lambda^2}{\pi^2}
\cos\left(\sqrt{8}ig\phi_\sigma+\sqrt{8\pi}\theta\right)\right]
\end{multline}
con las funciones $\left(V^{c,s}_{(\mu)}\right)^{-1}$ definidas
como

\begin{equation}
\int d^2y\,\left(V^{c,s}_{(\mu)}\right)^{-1}(x-y)
V^{c,s}_{(\mu)}(y-z) = \delta^{(2)}(x-z).
\end{equation}

Esto, a su vez, conduce a una factorizaci\'on de la funci\'on de
partici\'on en la forma $\Z=\Z_\rho \Z_\sigma$. Este resultado es
una clara manifestaci\'on de la separaci\'on
spin-carga\cite{luther74b,grinstein79}. $\Z_\rho$ es la funci\'on
de partici\'on asociada a las excitaciones de densidad de carga.
Coincide con el modelo de Tomonaga-Luttinger sin spin estudiado en
el cap\'{\i}tulo anterior y en la Ref. \citen{iucci04}.
$\Z_\sigma$ describe excitaciones de densidad de spin. La acci\'on
$S_\sigma$ corresponde a un modelo seno-Gordon con t\'ermino
cin\'etico no local, similar al considerado previamente en la Ref.
\citen{li98}. En el cap\'{\i}tulo siguiente derivaremos una
expresi\'on para el gap de su espectro en funci\'on de los
potenciales $V^\sigma_{(\mu)}$.

Dado que nuestro objetivo principal es analizar el sector de spin,
desde ahora enfocaremos nuestra atenci\'on en $S_\sigma$.
Tranformando Fourier la acci\'on (con la excepci\'on del t\'ermino
del coseno, cuya transformada de Fourier no es muy esclarecedora)
se obtiene

\begin{multline}
S_\sigma=\int\difp\left[\phih(p)\phih(-p)A(p)+\etah(p)\etah(-p)B(p)
+
\phih(p)\etah(-p)C(p)+\frac{p^2}{2}\theta(p)\theta(-p)\right]\\
-\frac{g_s\Lambda^2}{\pi^2}\int d^2x\,
\cos\left(\sqrt{8}ig\phi_\sigma+\sqrt{8\pi}\theta\right)
\end{multline}
con

\begin{align}
A=&p_1^2\left(\frac{1}{2V_0}+\frac{g^2}{2\pi}\right)+\frac{p_0^2}{2V_1}\\
B=&p_1^2\left(\frac{1}{2V_1}-\frac{g^2}{2\pi}\right)+\frac{p_0^2}{2V_0}\\
C=&p_1p_0\left(\frac{1}{V_1}-\frac{1}{V_0}-\frac{g^2}{\pi}\right)
\end{align}

En la expresi\'on de arriba, $\phih(p)$, $\etah(p)$ y $\theta(p)$
son las transformadas de Fourier de $\phi_\sigma(x)$,
$\eta_\sigma(x)$ y $\theta(x)$ respectivamente. Es conveniente
considerar todav\'{\i}a otro cambio de variables que diagonaliza
la parte cuadr\'atica de $S_\sigma$. Este cambio est\'a dado por

\begin{equation}\label{eq:changeVariables}
\theta=\frac{\Delta\pi}{\Delta\pi+2Bg^2p^2}\xi-\frac{ig}{\sqrt{\pi}}\zeta
\end{equation}

\begin{equation}
\eta=\frac{i\sqrt{\pi}gCp^2}
{\Delta\pi+2Bg^2p^2}\,\xi-\varphi-\frac{C}{2B}\,\zeta
\end{equation}

\begin{equation}
\phi=\zeta-\frac{2i\sqrt{\pi}gp^2B}{\Delta\pi+2Bg^2p^2}\,\xi
\end{equation}
donde $\xi$, $\zeta$ y $\chi$ son los nuevos campos bos\'onicos, y
$\Delta=C^2-4AB$. La acci\'on resultante se lee

\begin{align}
S_\sigma=&\int\difp\frac{1}{2K_\sigma v_\sigma}(p_0^2+v_\sigma
p_1^2)\xi(p)\xi(-p)
-\frac{g_s\Lambda^2}{\pi^2}\int d^2x\,\cos(\sqrt{8\pi}\xi)\nonumber\\
+&\int\difp\frac{p^4(\pi-g^2V^\sigma_{(1)})} {2(p_0^2\pi
V^\sigma_{(1)}+p_1^2
V_{(0)} (\pi-g^2 V_{(1)})}\zeta(p)\zeta(-p)\nonumber\\
+&\int\difp\frac{1}{2}\left[\frac{p_0^2}{V^\sigma_{(0)}}+
p_1^2\left(\frac{-g^2}{\pi}+\frac{1}{V^\sigma_{(1)}}\right)\right]\varphi(p)\varphi(-p)
\end{align}
donde

\begin{equation}
v_\sigma=\sqrt{\left(1+\frac{g^2}{\pi}V^\sigma_{(0)}\right)\left(1-\frac{g^2}{\pi}V^\sigma_{(1)}\right)},
\end{equation}

\begin{equation}
K_\sigma=\sqrt{\frac{1+\frac{g^2}{\pi}V^\sigma_{(0)}}{1-\frac{g^2}{\pi}V^\sigma_{(1)}}}.
\end{equation}

Se observa que los campos $\zeta$ y $\varphi$ se desacoplaron
completamente de $\xi$. La parte de la acci\'on que depende de
$\xi$ coincide con la obtenida mediante bosonizaci\'on operacional
en la secci\'on \ref{part:bosonización} y es la parte relevante a
considerar. M\'as generalmente la reescribimos como

\begin{equation}\label{eq:accionSpin}
S_\sigma=\int\difp\xi(p)\frac{F(p)}{2}\xi(-p) -
\frac{g_s\Lambda^2}{\pi^2}\int d^2x\,\cos(\sqrt{8\pi}\xi),
\end{equation}
con

\begin{equation}
F(p)=\frac{1}{K_\sigma(p) v_\sigma(p)}\left[p_0^2+v_\sigma(p)
p_1^2\right].
\end{equation}
Esta acci\'on corresponde a un modelo seno-Gordon con t\'ermino
cin\'etico no local, introducido en la Ref. \citen{li98}.

\section{Conclusiones}\label{part:conclusionesSpinFlip}

En este cap\'{\i}tulo hemos mejorado una versi\'on no local del
modelo de Thirring que provee una descripci\'on tratable de los
l\'{\i}quidos de Luttinger con spin, basada en teor\'{\i}a de
campos. Efectivamente, en el contexto de teor\'{\i}as del tipo
Thirring no local, los tratamientos previos de las interaccciones
de inversi\'on de spin condujeron a un complicado modelo no
abeliano (ver por ejemplo\cite{naon95}). Espec\'{\i}ficamente
construimos una acci\'on basada en dos especies de fermiones que
permite tener en cuenta interacciones de inversi\'on de spin de un
modo elegante y simple. Aunque nuestro modelo se inspir\'o en el
considerado en la Ref. \cite{zinn-justin97}, incluye interacciones
no contenidas en ese trabajo previo (los llamados diagramas $g_4$
en la terminolog\'{\i}a de Solyom\cite{solyom79}). Adem\'as, la
teor\'{\i}a que presentamos posee potenciales bilocales generales
que gobiernan las interacciones que no invierten el spin.
Parametrizamos estos potenciales en t\'erminos de las funciones
$V^\rho$ y $V^\sigma$ que se asocian a la din\'amica de las
densidades de carga y spin respectivamente, una vez que la
separaci\'on spin-carga se hace manifiesta luego de un cambio de
variables apropiado (ver ecuaci\'on (\ref{scSep1}) y
(\ref{scSep2})). Aunque nuestro an\'alisis es \'unicamente
v\'alido para interacciones de inversi\'on de spin locales,
pudimos mantener la dependencia con la distancia en los
potenciales ordinarios (que no invierten el spin) hasta el final
de los c\'alculos. Bajo estas condiciones obtuvimos una acci\'on
bos\'onica efectiva cuyos grados de libertad de carga coinciden
con descripciones previamente encontradas en el cap\'{\i}tulo
anterior del problema de dispersi\'on hacia adelante sin spin.
Respecto del sector de spin, que es el de mayor inter\'es en el
presente contexto, hallamos que se corresponde con un modelo
seno-Gordon con t\'ermino cin\'etico no local. En el cap\'{\i}tulo
siguiente estudiaremos su espectro dentro del marco de la
aproximaci\'on arm\'onica autoconsistente, derivando una
expresi\'on para el gap de las excitaciones de baja energ\'{\i}a.


\chapter{La aproximaci\'on arm\'onica
autoconsistente}\label{part:aaac}

\begin{center}
\begin{minipage}{5.6in}
\textsl{Examinamos la aproximaci\'on arm\'onica autoconsistente y
su utilizaci\'on en el marco de la integral funcional y las
teor\'{\i}as de materia condensada. En particular hallamos una
f\'ormula para el gap de las excitaciones del sector de spin del
modelo estudiado en el cap\'{\i}tulo anterior, como funci\'on de
potenciales arbitrarios de interacci\'on electr\'on-electr\'on.
Proponemos adem\'as un nuevo m\'etodo para determinar el
par\'ametro inc\'ognita asociado a esta aproximaci\'on.
Comprobamos la validez de esta nueva t\'ecnica en el contexto del
modelo seno-Gordon y como aplicaci\'on no trivial consideramos el
r\'egimen de escala del modelo de Ising en 2D fuera del punto
cr\'{\i}tico y en presencia de un campo magn\'etico $h$. En este
caso derivamos una expresi\'on aproximada que relaciona la
longitud de correlaci\'on $\xi$, $T-T_c$ y $h$. Los resultados de
este cap\'{\i}tulo constituyen aportes originales de esta
tesis\cite{iucci01,iucci02}.}
\end{minipage}
\end{center}

\section{Introducci\'on}

La aproximac\'{\i}on arm\'onica autoconsistente (SCHA,
self-consistent harmonic approximation) es una t\'ecnica no
perturbativa que ha sido utilizado extensivamente en aplicaciones
de mec\'anica estad\'{\i}stica\cite{saito78,fisher85} y materia
condensada\cite{gogolin93,prokof'ev94,egger95,xu96,iucci02}.
Consiste en reemplazar una acci\'on verdadera $S\true$ por una
acci\'on de prueba $S\trial$ que hace que el problema resulte
tratable. Usualmente $S\trial$ es una acci\'on cuadr\'atica que
depende de cierto par\'ametro desconocido $\Omega$. Este
par\'ametro debe ser determinado mediante alg\'un criterio como
por ejemplo la minimizaci\'on de la energ\'{\i}a libre. Esta
aproximaci\'on est\'a \'{\i}ntimamente relacionada al
\emph{potencial efectivo
gausiano}\cite{stevenson84,stevenson85,ingermanson86} en
teor\'{\i}a cu\'antica de campos, una aproximaci\'on variacional
al potencial efectivo que utiliza como estado fundamental de
prueba una funcional de onda gaussiana dependiente de un
par\'ametro de masa. Tambi\'en est\'a basado en un \emph{principio
de m\'{\i}nima sensibilidad}\cite{stevenson81,kauffman84} para
determinar el par\'ametro adicional.

En el cap\'{\i}tulo anterior mostramos que la inclusi\'on de
t\'erminos que invierten el spin en la acci\'on de una teor\'{\i}a
de electrones fuertemente correlacionados en una dimensi\'on
espacial, condujo a una acci\'on bos\'onica de tipo seno-Gordon
para las fluctuaciones de los grados de libertad de spin. Aunque
el espectro del modelo seno-Gordon usual se conoce exactamente a
partir de los trabajos de Dashen, Hasslacher y Neveu (DHN)
\cite{dashen75}, en el caso que nos ocupa la presencia de
interacciones de largo alcance dan lugar a un t\'ermino cin\'etico
no local que destruye la solubilidad. Esta situaci\'on nos lleva
naturalmente a considerar m\'etodos aproximados. En el presente
cap\'{\i}tulo nos proponemos hallar el espectro de excitaciones
del mencionado sector mediante la SCHA. El presente c\'alculo
resulta de inter\'es por dos motivos adicionales: en primer lugar,
la SCHA es mucho m\'as sencilla de implementar que las t\'ecnicas
semicl\'asicas utilizadas en los mencionados trabajos. Y en
segundo lugar, la SCHA puede extenderse facilmente a modelos m\'as
complicados como por ejemplo el doble seno-Gordon\cite{delfino98}
en teor\'{\i}a de campos, o la versi\'on continua del modelo de
Hubbard extendido en llenado medio\cite{voit92} y l\'{\i}quidos de
Luttinger acoplados\cite{schulz96}.

Si bien nuestra motivaci\'on inicial para abordar el estudio de la
SCHA fue contar con un m\'etodo de aproximaci\'on para poder
estimar el valor del gap (en modelos en los que la no localidad
impide hallar una soluci\'on exacta) al familiarizarnos con la
t\'ecnica  pudimos realizar algunos aportes originales en lo
concerniente al m\'etodo en si mismo. Esto nos ha llevado a
presentar tanbi\'en en este cap\'itulo, a modo de leve
digresi\'on, los detalles de nuestra propuesta. En particular,
se\~nalamos que en problemas en dos dimensiones existe un modo
alternativo de determinar el par\'ametro $\Omega$. Este m\'etodo
est\'a basado en teor\'{\i}a de campos
conformes\cite{boyanovsky90,difrancesco99}. M\'as a\'un, mostramos
que nuestro m\'etodo conduce a mejoras en los resultados para el
modelo SG respecto del SCHA estandar y nos permite dar una nueva
descripci\'on del modelo de Ising bidimensional (MI2D) fuera del
punto cr\'{\i}tico. En el primer caso explotamos la existencia de
resultados exactos\cite{dashen75,zamolodchikov95} para verificar
la consistencia de nuestra propuesta, obteniendo una respuesta
cualitativamente buena para la masa del solit\'on. Aplicamos
entonces las mismas ideas al MI2D a $T\neq T_c$ y $h\neq0$, un
modelo no integrable para el cual se conocen pocos resultados
cuantitativos\cite{mccoy78,mccoy95,delfino96}. Utilizamos la
representaci\'on fermi\'onica del MI2D. Dado que la SCHA estandar
est\'a restringida a modelos bos\'onicos, el nuevo procedimiento
provee tambi\'en una extensi\'on de la aproximaci\'on gausiana a
teor\'{\i}as fermi\'onicas bidimensionales. Nuestro resultado
principal es una ecuaci\'on algebraica que permite obtener el
comportamiento de la longitud de correlaci\'on como funci\'on de
$T-T_c$ y $h$.

Debemos enfatizar que no estamos introduciendo una nueva
aproximaci\'on, sino s\'olo un m\'etodo para determinar el
par\'ametro. Como es bien sabido, la SCHA es una aproximaci\'on no
controlada, es decir, no hay ning\'un par\'ametro perturbativo
involucrado. Es claro entonces que la misma cr\'{\i}tica puede
hacerse a la presente propuesta.

\section{Detalles de la aproximaci\'on y el espectro del sector de spin en electrones unidimensionales}

Comenzaremos mostrando el desarrollo mediante integrales
funcionales de la SCHA. En general comenzamos con una funci\'on de
partici\'on

\begin{equation}
\Z\true=\int\D\mu\; e^{-S\true}
\end{equation}

\noindent donde $\D\mu$ es una medida de integraci\'on funcional
bos\'onica. Una manipulaci\'on elemental conduce a

\begin{equation}\label{eq:Z}
\Z\true=\frac{\int\D\mu\;
e^{-(S\true-S\trial)}\,e^{-S\trial}}{\int\D\mu\;
e^{-S\trial}}\int\D\mu\; e^{-S\trial}=\Z\trial\bra
e^{-(S\true-S\trial)} \ket\trial.
\end{equation}
para cualquier acci\'on \emph{de prueba} $S\trial$. Por medio de
la propiedad

\begin{equation}
\bra e^{-f} \ket\geq e^{-\bra f \ket},
\end{equation}
para $f$ real, y tomando logaritmo natural en la ecuaci\'on
(\ref{eq:Z}), obtenemos la desigualdad de Feynman\cite{feynman72}.

\begin{equation}\label{eq:Feynman}
\ln\Z\true\geq \ln\Z\trial - \bra S\true-S\trial\ \ket\trial
\end{equation}

La aproximaci\'on consiste en reemplazar la acci\'on verdadera,
dif\'{\i}cil de tratar, por una acci\'on de prueba m\'as simple,
que contenga alg\'un par\'ametro libre. Este se fija maximizando
el lado derecho de la desigualdad (\ref{eq:Feynman}).

Consideramos a continuaci\'on como acci\'on verdadera, la acci\'on
que describe la din\'amica del sector de spin en teor\'{\i}as de
electrones altamente correlacionados en una dimensi\'on
(\ref{eq:accionSpin}), obtenida en el cap\'{\i}tulo anterior.
Adem\'as como acci\'on de prueba tomamos una acci\'on cuadr\'atica
de la forma

\begin{equation}
S\trial=\int\difp\left[\xi(p)\frac{F(p)}{2}\xi(-p) +
\frac{\Omega^2}{2}\xi(p)\xi(-p)\right].
\end{equation}
Para la situaci\'on mencionada, $F$ se define como

\begin{equation}\label{eq:cineticoLuttinger}
F(p)=\frac{1}{K_s v_s}(p_0^2+v_s p_1^2),
\end{equation}
aunque para los desarrollos siguientes consideraremos una $F$
arbitraria, y obtendremos resultados generales. Una vez realizada
la sustituci\'on de acciones es inmediato obtener el espectro:

\begin{equation}
F(p)+\Omega^2=0.
\end{equation}
Volviendo a frecuencias reales, $p_0=i\omega$, $p_1=k$, y tomando
el t\'ermino cin\'etico (\ref{eq:cineticoLuttinger}) se obtiene la
siguiente ecuaci\'on

\begin{equation}\label{eq:spectrum}
v_s K_s \Omega^2 + v_s^2 k^2 -\omega^2=0.
\end{equation}

Como se dijo anteriormente, el par\'ametro $\Omega$ puede
determinarse maximizando el lado derecho de la ecuaci\'on
(\ref{eq:Feynman}). Para alcanzar este objetivo primero escribimos

\begin{align}
\ln \Z\trial=&\ln\int\D\xi\exp{\left[-\frac{1}{2}\int d^2x\,\xi(x)
(\hat{A}\xi)(x)\right]} = [\ln(\det\hat{A})^{-1/2}]+\text{const}
\\=&-\frac{1}{2}\tr\ln\hat{A} + \text{const},
\end{align}
donde el operador $\hat{A}$ est\'a definido, en espacio de
Fourier, por

\begin{equation}
(\hat{A}\xi)(p)=[F(p)+\Omega^2]\xi(p).
\end{equation}
Es f\'acil entonces obtener

\begin{equation}
\tr\ln\hat{A}=V\int\frac{d^2p}{(2\pi)^2} \ln[F(p)+\Omega^2],
\end{equation}
donde $V$ es el volumen (infinito) de todo el espacio $\int d^2x$.
Por otro lado, $\langle S\true-S\trial \rangle$ se puede calcular
directamente siguiendo, por ejemplo, los pasos explicados en la
Ref. \citen{li98} y la identidad del ap\'endice
\ref{part:identity}. El resultado es

\begin{align}
-\bra S\true-S\trial \ket\trial &= \int d^2x
\left[\frac{g_s\Lambda^2}{\pi^2}\bra
\cos(\sqrt{8\pi}\xi)\ket\trial + \frac{\Omega^2}{2}\bra \xi^2\ket\trial\right]\\
&=V\frac{g_s\Lambda^2}{\pi^2}
\exp\left[-4\pi\int\difp\frac{1}{F(p) + \Omega^2}\right] + V
\frac{\Omega^2}{2}\int\difp\frac{1}{F(p) + \Omega^2}.
\end{align}
Finalmente, podemos reunir todos los t\'erminos y escribir

\begin{equation}\label{eq:elegant}
\ln\Z\trial - \bra S\true-S\trial
\ket\trial=V\left[\frac{g_s\Lambda^2}{\pi^2} \exp\left(-4\pi
I_0(\Omega)\right) + \frac{\Omega^2}{2}I_0(\Omega) -
\frac{1}{2}I_1(\Omega)\right] + \text{const}.
\end{equation}
donde definimos las integrales

\begin{equation}
I_1(\Omega)=\int \frac{d^2k}{(2\pi)^2} \ln[F(p) + \Omega^2]
\end{equation}

\begin{equation}
I_{-n}(\Omega)=\int\frac{d^2k}{(2\pi)^2}\frac{1}{\left[F(p) +
\Omega^2\right]^{n+1}}
\end{equation}
con las propiedades formales

\begin{equation}
\frac{dI_1(\Omega)}{d\Omega}=2\Omega I_0
\end{equation}

\begin{equation}
\frac{dI_{-n}(\Omega)}{d\Omega}=-2(n+1)\Omega I_{-n-1}.
\end{equation}
Hallando el extremo de la expresi\'on (\ref{eq:elegant}) con
respecto a $\Omega$, y asumiendo que $I_{-1}(\Omega)$ es no nulo
(una condici\'on que vale para la mayor\'{\i}a de los potenciales
realistas), finalmente obtenemos la ecuaci\'on del gap.

\begin{equation}\label{eq:omegaeq}
\Omega^2-\frac{g_s\Lambda^2}{\pi^2} e^{-4\pi I_0(\Omega)}=0.
\end{equation}

La ecuaci\'on (\ref{eq:omegaeq}) es uno de los resultados
principales de este cap\'{\i}tulo. Dentro de la aproximaci\'on
arm\'onica autoconsistente, da una expresi\'on cerrada para el gap
como funcional de los potenciales $V^s_{(\mu)}(p)$. Resulta
interesante obtener un valor expl\'{\i}cito para $\Omega$ para el
potencial de contacto, dado por

\begin{equation}\label{eq:casoLocal}
V^s_{(0)}(p)=V^s_{(1)}(p)=1.
\end{equation}
En este caso $I_0(\Omega)$ es infinita. Si utilizamos la misma
regularizaci\'on empleada en el cap\'{\i}tulo anterior se obtiene

\begin{equation}
\frac{\Omega^2}{\Lambda^2}=\frac{g_s}{\pi^2}
\left[1+\frac{(1+g^2/\pi)\Lambda^2}{\Omega^2}\right]^{\frac{-1}{1+g^2/\pi}}.
\end{equation}
Esta ecuaci\'on puede f\'acilmente resolverse para
$\Lambda\gg\Omega$ y $\Omega\gg\Lambda$. Los resultados est\'an
dados respectivamente por

\begin{equation}
\Omega^2=\Lambda^2\left(\frac{g_s}{\pi}\right)^{\frac{1+g^2/\pi}{g^2/\pi}}
\left(1+\frac{g^2}{\pi}\right)^{-g^2/\pi},
\end{equation}
y

\begin{equation}
\Omega^2=\frac{g_s\Lambda^2}{\pi^2}.
\end{equation}

\subsection{El modelo seno-Gordon en teor\'ia de campos}

Si bien el modelo estudiado en la secci\'on precedente coincide
con el modelo seno-Gordon considerado tradicionalmente en la
literatura de la teor\'{\i}a de campos en el caso local dado por
(\ref{eq:casoLocal}), existe una diferencia en el tratamiento de
las divergencias en las integrales. Mientras que en teor\'{\i}a de
campos se busca implementar mecanismos que hagan finitos los
resultados finales, lo que se consigue al renormalizar los
par\'ametros de la teor\'{\i}a, en materia condensada las
divergencias que aparecen son artificiales, es decir son
introducidas por las t\'ecnicas mediante las cuales se trata la
teor\'{\i}a. Por ejemplo, en el modelo de Hubbard original, el
espectro es acotado, es decir, las sumas o integrales sobre $k$ se
realizan en una regi\'on finita. Al considerar la linealizaci\'on
de la relaci\'on de dispersi\'on, y la inclusi\'on de infinitos
estados con energ\'{\i}a por debajo del nivel de Fermi, el modelo
se vuelve tratable matem\'aticamente mediante bosonizaci\'on, pero
al costo de la introducci\'on de divergencias ultravioletas. Por
otro lado, el par\'ametro que se introduce para regular el modelo
[la constante $a$ en la ecuaci\'on (\ref{eq:camposBosonicos}), o
$\Lambda$ en (\ref{eq:bosExpSerie})], usualmente tiene un
significado f\'{\i}sico: est\'a asociada con la constante de red.
Por esta raz\'on es usual que se deje su dependencia
expl\'{\i}cita en funciones de correlaci\'on y otras cantidades de
inter\'es.

En la secci\'on anterior seguimos este \'ultimo camino y obtuvimos
una expresi\'on para el gap que depend\'{\i}a expl\'{\i}citamente
del cutoff $\Lambda$. En esta secci\'on seguiremos el camino usual
de la teor\'{\i}a de campos, es decir, introduciremos una
constante de acoplamiento renormalizada, y obtendremos expresiones
para el gap del modelo seno-Gordon que dependen de ella. Partimos
de la acci\'on

\begin{equation}
S\true=\int\difp\varphi(p)\frac{F(p)}{2}\varphi(-p) - \int d^2x\,
\frac{\alpha}{\beta^2}e^{\frac{1}{2}\beta^2I_1(\rho)}\cos(\beta\varphi)
\end{equation}
donde $\varphi(p)$ es un campo escalar y $\alpha$ es la constante
de acoplamiento renormalizada\footnote{Aqu\'{\i} el t\'ermino
renormalizado es un abuso de lenguaje, ya que en modelos en (1+1)
dimensiones con interacciones sin derivadas, el orden normal
elimina completamente todas las divergencias} mediante un orden
normal\cite{coleman75,naon85}; $\rho$ es el par\'ametro que
implementa el orden normal. Por simplicidad, en esta f\'ormula
hemos escrito el t\'ermino cin\'etico en espacio de Fourier, pero
hemos dejado el t\'ermino de interacci\'on en espacio de
coordenadas.

Como acci\'on de prueba, proponemos una acci\'on cuadr\'atica,

\begin{equation}
S\trial=\int\difp\left[\varphi(p)\frac{F(p)}{2}\varphi(-p)
+\frac{\Omega^2}{2}\varphi(p)\varphi(-p)\right],
\end{equation}
donde $\Omega$ es el par\'amtro de prueba. Con el objeto de
realizar el proceso de minimizaci\'on estandar, primero evaluamos
$\langle S\true-S\trial \rangle$. El resultado es

\begin{equation}\label{elegant}
-\bra S\true-S\trial \ket\trial =
\V\left[\frac{\alpha}{\beta^2}e^{-\unmedio\beta^2[I_0(\Omega)-I_0(\rho)]}
+\frac{\Omega^2}{2}I_0(\Omega)\right].
\end{equation}

Insertando ahora (\ref{elegant}) en la desigualdad de Feynman
(\ref{eq:Feynman}), y minimizando el lado derecho de dicha
ecuaci\'on con respecto a $\Omega$, finalmente obtenemos

\begin{equation}\label{gapscha}
\Omega^2-\alpha
e^{-\beta^2/2\left[I_0(\Omega)-I_0(\rho)\right]}=0.
\end{equation}
Esta ecuaci\'on del gap permite extraer una respuesta finita para
$\Omega$, dependiente del par\'ametro de masa $\rho$ (la
diferencia $I_0(\Omega)-I_0(\rho)$ es finita). N\'otese que el
valor de $\rho$ es completamente arbitrario, si uno lo elige igual
a la masa de prueba $\Omega$, la soluci\'on de la ecuaci\'on es

\begin{equation}\label{SGGaussian}
\Omega^2=\alpha .
\end{equation}
El mismo resultado se obtiene si en lugar de $\rho=\Omega$ se toma
$\rho=\sqrt{\alpha}$.

\section{Determinaci\'on de $\Omega$ mediante t\'ecnicas de
teor\'{\i}a de campos conformes}

Presentaremos a continuaci\'on una ruta alternativa para
determinar $\Omega$. Para este fin explotaremos una predicci\'on
cuantitativa de la invarianza conforme para sistemas en 2D en el
r\'egimen de escala, fuera del punto cr\'{\i}tico. Partiendo del
llamado `Teorema-c'\cite{zamolodchikov86},
Cardy\cite{cardy88a,cardy88b} mostr\'o que el valor de la
anomal\'{\i}a conforme $c$, que caracteriza al modelo en el punto
cr\'{\i}tico, y el segundo momento del correlador de densidad de
energ\'{\i}a en el r\'egimen de escala de la teor\'{\i}a no
cr\'{\i}tica est\'an relacionados por

\begin{equation} \label{Cardy}
\int d^2x\, |x|^2\, \bra\varepsilon(x)\varepsilon(0)\ket = \frac{c}{3\, \pi \,t^2\,(2
- \Delta_\varepsilon)^2},
\end{equation}
donde $\varepsilon$ es el operador de densidad de energ\'{\i}a,
$\Delta_\varepsilon$ es su dimensi\'on de escala y
$t\propto(T-T_c)$ es la constante de acoplamiento del t\'ermino de
interacci\'on que saca al sistema fuera del punto cr\'{\i}tico. La
validez de esta f\'ormula ha sido verificada expl\'{\i}citamente
para varios modelos \cite{cardy88a,cardy88b}. Para el modelo SG,
el operador de densidad de energ\'{\i}a est\'a dado por el
t\'ermino del coseno, su dimensi\'on conforme es
$\Delta_\varepsilon=\beta^2/4\pi$, $t$ es la constante de
acoplamiento $\alpha/\beta^2$ y la teor\'{\i}a conforme bos\'onica
libre asociada posee $c=1$.

\begin{figure}[ht]
\begin{center}
\includegraphics{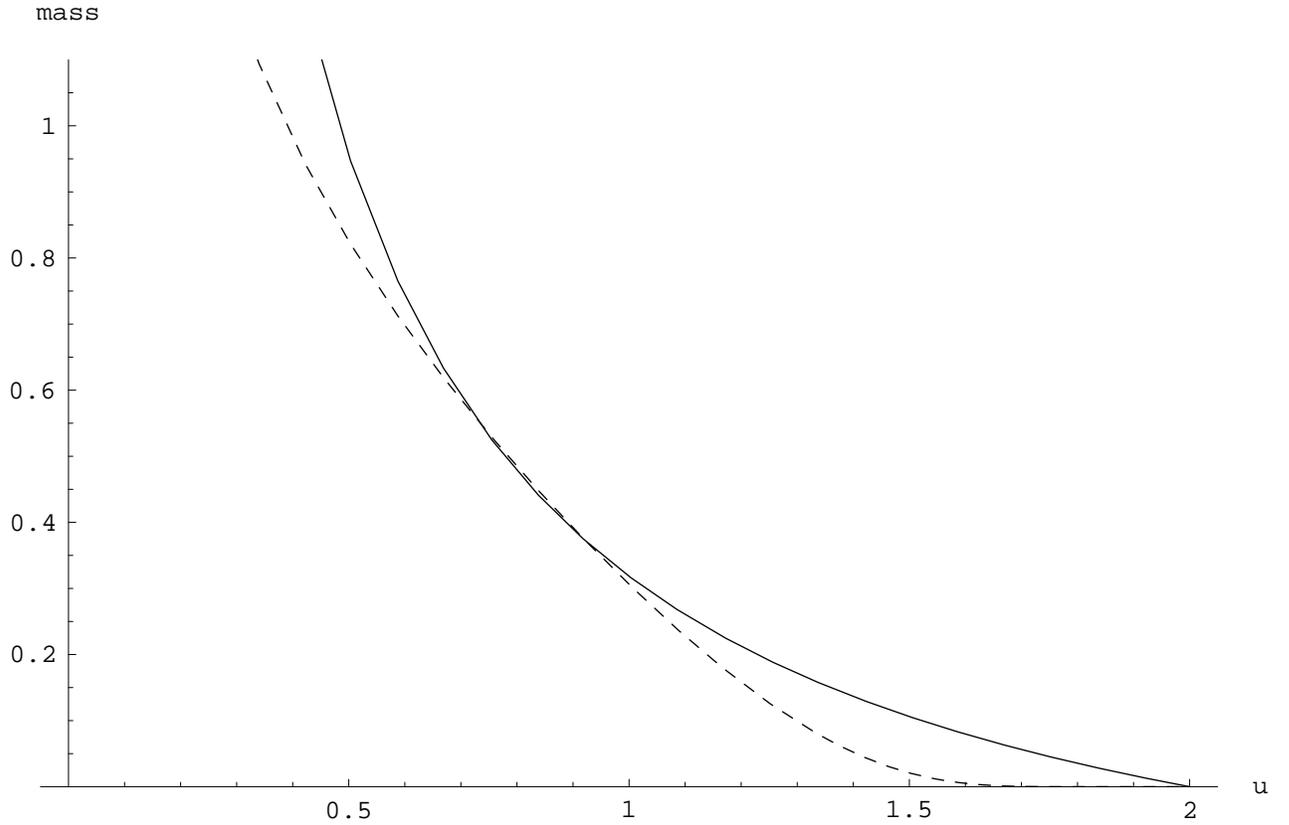}
\caption{Masas en unidades de $\sqrt{\alpha}$ como funci\'on de
$u$. La l\'{\i}nea punteada es $M_\text{sol}/\sqrt{\alpha}$,
mientras que la l\'{\i}nea llena representa
$\Omega/\sqrt{\alpha}$} \label{fig:mass}
\end{center}
\end{figure}

Nosotros afirmamos que $\Omega$ puede ser determinado de una forma
no variacional completamente diferente, forzando la validez de la
identidad conforme de arriba para la acci\'on de prueba. En otras
palabras, impondremos que se verifique la siguiente ecuaci\'on:

\begin{equation} \label{Main}
\frac{\alpha^2}{\beta^4}\int d^2x\, |x|^2\, \bra
\cos\beta\varphi(x)\,\cos\beta\varphi(0) \ket\trial = \frac{1}{3\, \pi \,\,(2 -
\frac{\beta^2}{4\pi})^2},
\end{equation}
que debe ser considerada como una ecuaci\'on para el par\'ametro
de masa $\Omega$. Por supuesto, si uno est\'a interesado en
comparar la respuesta dada por esta f\'ormula con el resultado
dado por la SCHA usual, al evaluar el lado derecho de
(\ref{Main}), se debe adoptar una prescripci\'on de
regularizaci\'on equivalente al orden normal implementado en la
SCHA. Un c\'alculo cuidadoso conduce a la siguiente ecuaci\'on del
gap:

\begin{equation}\label{gapcga}
(\frac{\Omega}{\rho})^{2(2-u)}=(\frac{\alpha}{\rho^2})^2\,\frac{3}{32}\,\frac{2-u}{u^2}
\end{equation}
donde hemos definido la variable $u=\beta^2/4\pi$ ($0\leq u <2$) y
$\rho$ es el par\'ametro de orden normal, como antes. Vemos que,
igual que en la ecuaci\'on SCHA estandar (\ref{gapscha}), hay
diferentes respuestas para diferentes elecciones de $\rho$, pero
en este caso, los resultados obtenidos para los valores
$\sqrt{\alpha}$ y $\Omega$ son diferentes. En cualquier caso se
obtiene una dependencia no trivial de $\Omega$ con $\beta^2$ en
contraste con la SCHA variacional. Esto es interesante si
recordamos el significado f\'{\i}sico de la masa del gap en el
contexto del modelo SG. Efectivamente, como es bien sabido,
Dashen, Hasslacher y Neveu (DHN) \cite{dashen75} calcularon
mediante t\'ecnicas semicl\'asicas el espectro de masa para el
modelo SG. Consiste en un solit\'on (asociado al fermi\'on del
modelo de Thirring) con masa

\begin{equation}\label{soliton}
M_{sol}=\frac{2-u}{\pi\,u}\,\sqrt{\alpha},
\end{equation}
y una secuencia de estados ligados con masas

\begin{equation}\label{doublets}
M_{N}=\frac{2(2-u)}{\pi\,u}\,\sin\left[N \frac{\pi\,u}{2(2-u)}\right]\, \sqrt{\alpha},
\end{equation}
con $N=1,2,...<(2-u)/u$. (De esta \'ultima condici\'on es f\'acil
ver que para tener $N$ estados ligados debemos tener $u<2/(N+1)$.
Como consecuencia no hay estado ligado para $u>1$). M\'as
recientemente, Zamolodchikov \cite{zamolodchikov95},
reinterpretando resultados obtenidos mediante el ansatz de Bethe,
dio expresiones exactas para este espectro. En particular para el
solit\'on su f\'ormula acuerda muy bien con (\ref{soliton}),
excepto para $u$ cercano a $2$, donde predice una divergencia. Por
simplicidad, aqu\'{\i} comparamos nuestros resultados con la
ecuaci\'on (\ref{soliton}). Lo primero que debe notarse es que las
masas en el espectro del modelo SG tambi\'en dependen de $u$,
igual que nuestra predicci\'on dada por la ecuaci\'on
(\ref{gapcga}). Por lo tanto, al respecto, nuestra propuesta
parece mejorar la predicci\'on gausiana est\'andar para el modelo
SG, al menos cualitativamente. Para efectuar una discusi\'on
cuantitativa m\'as espec\'{\i}fica  comparemos las ecuaciones
(\ref{gapcga}) y (\ref{soliton}) como funciones de $u$. Fijamos
$\rho=\sqrt{\alpha}$, lo que corresponde a la prescripci\'on
empleada por DHN al derivar (\ref{soliton}) y (\ref{doublets}). El
resultado se muestra en la figura \ref{fig:mass} donde se puede
observar una analog\'{\i}a cualitativa general entre ambas curvas.
En particular, para $0.7\leq u \leq 1$ ($u=1$ corresponde al punto
del fermi\'on libre del modelo de Thirring y al punto de Emery en
el modelo de dispersi\'on hacia atr\'as\cite{luther74b}) nuestra
predicci\'on est\'a de acuerdo con los valores de la masa del
solit\'on calculados por DHN. Queremos recalcar que para $u=1$
obtenemos $\Omega/\sqrt{\alpha}=\sqrt{3/32}\approx0.30$ mientras
que el valor dado por (\ref{soliton}) es $1/\pi\approx0.31$ (la
SCHA estandar da, por supuesto, $\Omega/\sqrt{\alpha}=1$).

\subsection{El modelo de Ising en 2D}

Habiendo verificado la admisibilidad de nuestra propuesta en un
modelo en el que se conocen resultados exactos, es ahora deseable
explorar un problema no trivial. Consideremos el modelo de Ising
en 2D fuera del punto cr\'{\i}tico ($T\neq T_c$ y $h\neq0$):

\begin{equation} \label{IsingAction}
S= S_{\text{M}} + \int d^2x \left[t \, \epsilon(x)\,+ h \,
\sigma(x)\right],
\end{equation}
donde $S_{\text{M}}$ es la acci\'on cr\'{\i}tica, $t
\propto(T-T_c)$, y $\epsilon(x)$ y $\sigma(x)$ son los operadores
de densidad de energ\'{\i}a y spin respectivamente. Usaremos la
representaci\'on fermi\'onica de la mencionada acci\'on. De este
modo, $S_{\text{M}}$ es la acci\'on de un fermi\'on de Majorana
libre y sin masa, y $\epsilon\propto \Psib \Psi$. Por otro lado,
la expresi\'on de $\sigma(x)$ en t\'erminos de los campos de
Majorana es m\'as complicada. Efectivamente, por medio de una
transformaci\'on de Jordan-Wigner puede escribirse como
exponencial de un bilineal fermi\'onico. En analog\'{\i}a con la
SCHA usual, proponemos la siguiente acci\'on cuadr\'atica de
prueba

\begin{equation}
S\trial= S_{\text{M}} + \Omega\, \int d^2x \, \epsilon(x),
\end{equation}
La ecuaci\'on conforme(\ref{Cardy}) para el presente caso toma la
forma

\begin{multline}\label{Cardy2}
\int d^2r\, r^2\,[t^2\,(2 - \Delta_\varepsilon)^2
\bra\varepsilon(r)\varepsilon(0)\ket\trial + h^2\,(2 -
\Delta_\sigma)^2 \bra\sigma(r)\sigma(0)\ket\trial \\ + 2\,t\,h\,(2
- \Delta_\varepsilon)\,(2 - \Delta_\sigma)
\bra\varepsilon(r)\sigma(0)\ket\trial]= \frac{1}{6\, \pi},
\end{multline}
donde fijamos $c=1/2$, que es la carga central de los fermiones
libres de Majorana, y $\Delta_\varepsilon =1$ y
$\Delta_\sigma=1/8$ son las dimensiones de escala de los
correspondientes operadores. Ahora debemos evaluar el valor medio
en la acci\'on de prueba. Esto nos dar\'a una ecuaci\'on para
$\Omega$ como funci\'on de $t$ y $h$. Las funciones de
correlaci\'on energ\'{\i}a-energ\'{\i}a y energ\'{\i}a-spin fueron
calculadas por Hecht \cite{hecht67} mientras que el correlador
spin-spin puede hallarse en el trabajo de Wu, McCoy, Tracy and
Barouch\cite{wu76}. Como es usual, se define la longitud de
correlaci\'on $\xi=1/4\Omega$ y se considera el l\'{\i}mite de
escala, dado por $\xi\rightarrow\infty$, $r\rightarrow\infty$, con
$r/\xi$ fijo. El siguiente paso es usar la expresi\'on de los
correladores para $(r/\xi)<<1$ y realizar las integrales
correspondientes. En este punto debemos tener en cuenta que las
funciones de correlaci\'on son proporcionales a ciertas funciones
de escala $F_{\pm}(r/{\xi})$ donde los signos $+$ y $-$
corresponden a los casos $\Omega>0$ y $\Omega<0$ respectivamente.
En otras palabras, el par\'ametro $\Omega$ puede verse como un
par\'ametro que define una nueva temperatura cr\'{\i}tica
\emph{efectiva}. Las funciones $F_{\pm}$ describen el r\'egimen de
escala por encima y por debajo de esta temperatura. Dado que
estamos aproximando una perturbaci\'on magn\'etica en el sistema,
es claro que debemos usar la funci\'on $F_{-}$. De este modo
obtenemos la siguiente ecuaci\'on que relaciona $\xi$, $h$ y $t$:

\begin{equation} \label{Main2}
t^2(4\xi)^2+C_1 h^2 (4\xi)^{15/4}+C2 t |h| (4\xi)^{23/8}=1
\end{equation}
donde hemos introducido las constantes num\'ericas $C_1=0.749661$
y $C_2=0.186966$. El valor absoluto del campo magn\'etico en el
segundo t\'ermino proviene del hecho de que
$\bra\epsilon\sigma\ket\propto\bra\sigma\ket$ y el producto
$\bra\sigma\ket h$ debe ser positivo dado que la magnetizaci\'on y
el campo magn\'etico tienen que tener la misma orientaci\'on. Para
$\xi$ fijo, esta ecuaci\'on da una dependencia simple de $h$ como
funci\'on de $t$. Efectivamente, para $h>0$ tenemos una semi
elipse levemente rotada en el plano $h - t$ superior, y para $h<0$
tenemos su reflexi\'on sobre el eje $t=0$.

Si reescribimos la ecuaci\'on \ref{Main2} en t\'erminos de la
longitud de correlaci\'on a campo nulo, $\xi_0=1/4t$, y la
combinaci\'on adimensional $\chi={\mid h \mid}^{-\frac{8}{15}}
/4\xi_0$ obtenemos

\begin{equation} \label{Main3}
\left(\frac{\xi}{\xi_0}\right)^2 + C_1
\,\chi^{\frac{-15}{4}}\,\left(\frac{\xi}{\xi_0}\right)^{\frac{15}{4}}\pm
C_2
\,\chi^{\frac{-15}{8}}\,\left(\frac{\xi}{\xi_0}\right)^{\frac{23}{8}}=1.
\end{equation}
Los signos $+$ y $-$ en el tercer t\'ermino del miembro izquierdo
corresponden al caso $t>0$ y $t<0$ respectivamente. La acci\'on
(\ref{IsingAction}) define una familia de teor\'{\i}as de campos
dependientes de un par\'ametro $\chi$\cite{mccoy95}.

\begin{figure}[ht]
\begin{center}
\includegraphics{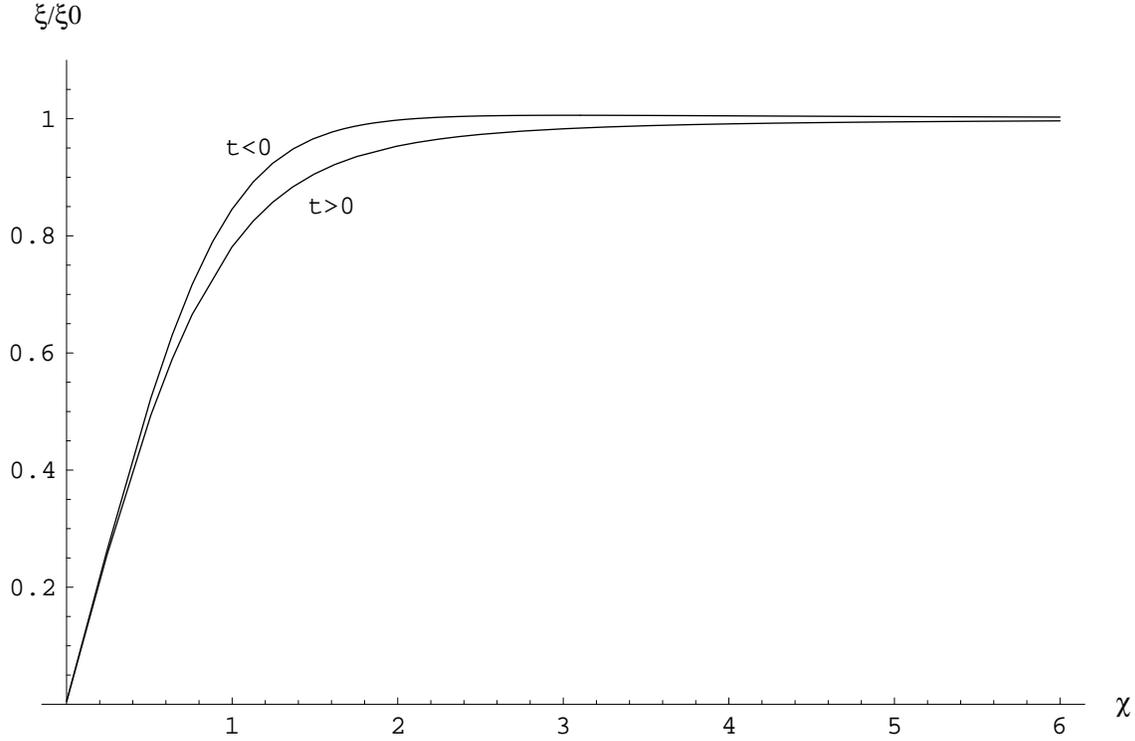}
\caption{Longitud de correlaci\'on en unidades de $\xi_0$ como
funci\'on de $\chi$, para ambos casos, $t>0$ y $t<0$.}
\label{fig:longCorrGrande}
\end{center}
\end{figure}

\begin{figure}[ht]
\begin{center}
\includegraphics{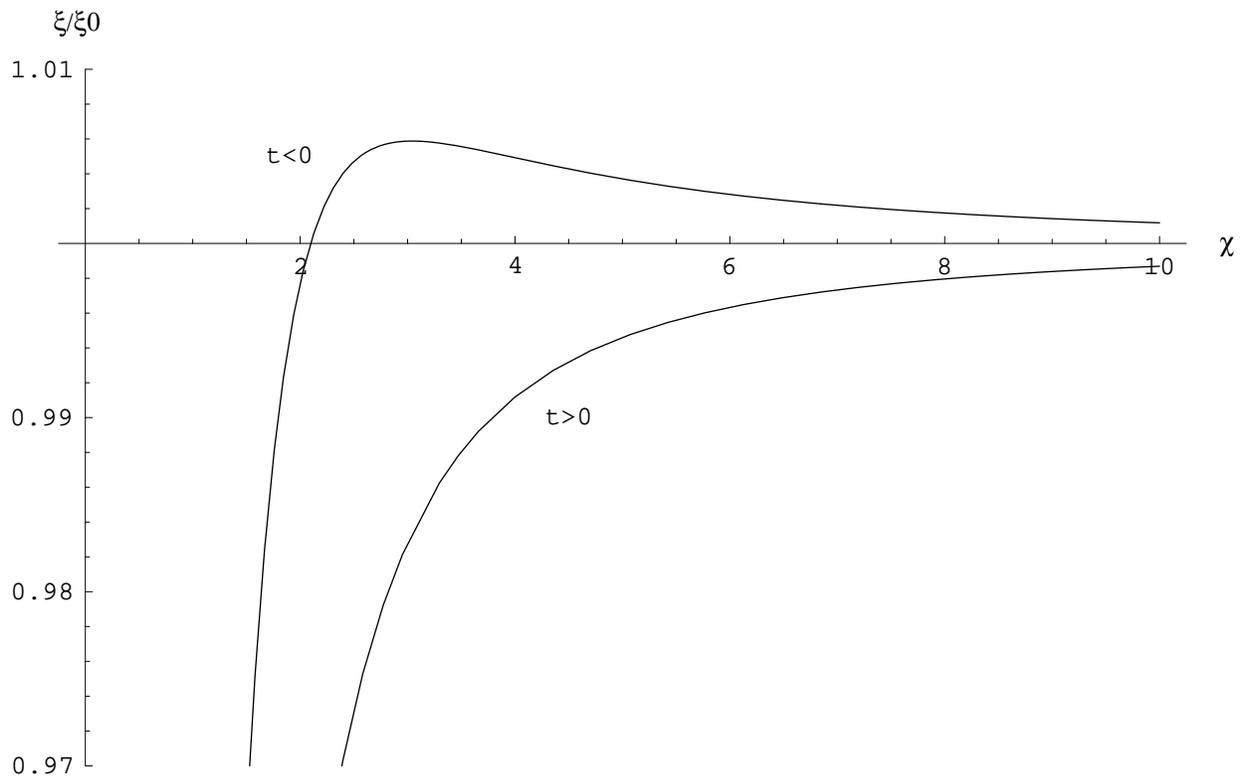}
\caption{Igual que la Fig. \ref{fig:longCorrGrande}. Al agrandar
la escala se observan los detalles del comportamiento de
$\xi(\chi)$ para $t>0$ y $t<0$.}\label{fig:longCorrChico}
\end{center}
\end{figure}

Con el objeto de verificar la consistencia de la ecuaci\'on de
arriba, consideramos los l\'{\i}mites $h\rightarrow0$ y
$t\rightarrow0$ en forma separada. El primer caso corresponde a
$\chi\rightarrow\infty$ y se obiene inmediatamente $\xi=\xi_0$,
como era de esperar. En el segundo caso tenemos $\chi\rightarrow0$
y obtenemos entonces $\xi\sim{\mid h \mid}^{-\frac{8}{15}}$, lo
cual est\'a en acuerdo con el resultado exacto obtenido en las
Ref.\citen{zamolodchikov90,fateev94}. Debemos mencionar que en
estas referencias la constante de proporcionalidad fue determinada
exactamente en el valor $4,4$, mientras que nuestro c\'alculo
aproximado conduce al valor $3,7$. Volviendo al caso general,
resolvimos la ecuaci\'on (\ref{Main2}) num\'ericamente para $\xi$
como funci\'on de $\chi$ para $t>0$ y $t<0$. Los resultados se
muestran en el gr\'afico de la figura \ref{fig:longCorrGrande}. En
el caso $t>0$ la longitud de correlaci\'on se incrementa en forma
mon\'otona desde cero y alcanza el valor correspondiente a $h=0$,
$\xi_0$ desde abajo cuando $\chi\rightarrow\Infinity$. En el caso
$t<0$, aunque el comportamiento de $\xi$ parece muy similar al
caso previo, una mirada cuidadosa muestra que presenta una
diferencia sutil, mostrada en la figura \ref{fig:longCorrChico}.
Para $\chi\approx 2$ la longitud de correlaci\'on pasa sobre el
valor $\xi_0$, alcanza un m\'aximo y entonces tiende a $\xi_0$
desde arriba cuando $\chi\rightarrow\Infinity$. Como este
comportamiento depende de los valores de las constantes $C_1$ y
$C_2$ no sabemos si es efectivamente una propiedad del modelo de
Ising o un artificio introducido por nuestra aproximaci\'on.

\section{Conclusiones}

En este cap\'{\i}tulo hemos reconsiderado el conocido m\'etodo de
aproximaci\'on arm\'onica autoconsistente, en el cual una acci\'on
comparativamente compleja es reemplazada por un sistema
cuadr\'atico m\'as simple dependiente de un par\'ametro de masa
$\Omega$ que se determina usualmente mediante un c\'alculo
variacional. Aplicamos este m\'etodo para la obtenci\'on del gap
del espectro de excitaciones del modelo de Thirring no local con
interacciones de inversi\'on de spin, formulado en el
cap\'{\i}tulo anterior.

Al trabajar con la SCHA advertimos que, para el caso de
teor\'{\i}as 1+1 dimensionales, el par\'ametro $\Omega$ se
pod\'{\i}a determinar de un modo alternativo, no variacional.
Nuestra propuesta se basa en una consecuencia del teorema-c de
Zamolodchikov\cite{zamolodchikov86} derivada por primera vez por
Cardy\cite{cardy88a}. Ilustramos la idea considerando el modelo
seno-Gordon. Mostramos que para este modelo nuestro m\'etodo da
una predicci\'on bastante buena para el comportamiento de la masa
del solit\'on como funci\'on de $\beta^2$ (ver las ecuaciones
(\ref{gapcga}) y (\ref{soliton}) y la Fig. \ref{fig:mass}).

Como aplicaci\'on no trivial consideramos el modelo de Ising en 2D
fuera del punto cr\'{\i}tico ($T\neq T_c$ and $h\neq0$). A partir
de una descripci\'on mediante teor\'{\i}a de campos en t\'erminos
de fermiones de Majorana, proponemos una acci\'on cuadr\'atica de
prueba dependiente de un par\'ametro $\Omega$ que define una
longitud de correlaci\'on aproximada $\xi$. Nuestro resultado
principal est\'a dado por la ecuaci\'on (\ref{Main2})(o su forma
alternativa (\ref{Main3})) que permite determinar el par\'ametro
$\Omega$ (o lo que es lo mismo, $\xi$) en t\'erminos de los
par\'ametros f\'{\i}sicos originales $t$ y $h$.

Ser\'{\i}a interesante probar nuestro enfoque en otros modelos
tales como la versi\'on continua del modelo de Ising
tricr\'{\i}tico, que puede describirse mediante el segundo modelo
de la serie minimal unitaria\cite{belavin84,friedan84} con carga
central $c=7/10$.


\chapter{Interacciones spin-\'orbita}\label{part:spinorbit}

\begin{center}
\begin{minipage}{5.6in}
\textsl{En este cap\' \i tulo calculamos funciones de
correlaci\'on en sistemas unidimensionales de electrones en
interacci\'on en los que los grados de libertad de carga y spin se
encuentran acoplados a trav\'es de la interacci\'on spin-\'orbita.
Estudiamos fluctuaciones de tipo ondas de densidad de carga y
spin, y de tipo superconductor singulete y triplete. Mostramos que
la interacci\'on spin-\'orbita modifica los exponentes del
decaimiento de las funciones de correlaci\'on y el diagrama de
fases del sistema. Adem\'as encontramos que susceptibilidades que
eran finitas a bajas temperaturas, se vuelven divergentes a causa
de la interacci\'on spin-\'orbita. Estos resultados constituyen
una contribuci\'on original de esta tesis\cite{iucci03}}
\end{minipage}
\end{center}

\section{Introducci\'on}
Al considerar el comportamiento de los electrones dentro de
materiales, debe tenerse en cuenta que \'estos se mueven en
presencia de campos el\'ectricos. Como consecuencia, experimentan
no s\'olo la fuerza electrost\'atica originada en estos campos,
sino una influencia relativista conocida como \emph{interacci\'on
spin-\'orbita} (SO) que rompe la simetr\'{\i}a de rotaci\'on de
spin SU(2). Su or\'{\i}gen se encuentra en el acoplamiento de
Pauli entre el momento magn\'etico de spin del electr\'on y un
campo magn\'etico que aparece en su sistema de referencia en
reposo debido al movimiento en un campo el\'ectrico. Una forma
general de describir la interacci\'on SO consiste en agregar el
siguiente t\'ermino al Hamiltoniano, que se obtiene a partir de la
expansi\'on cuadr\'atica en $v/c$ de la ecuaci\'on de Dirac:

\begin{equation}\label{eq:HSOgeneral}
H_\text{SO}=\frac{\hbar}{(2m_0c)^2}\nabla
V(\mathbf{r})(\mathbf{\hat{\boldsymbol{\sigma}}}\times\mathbf{\hat{p}}).
\end{equation}
Aqu\'{\i} $m_0$ es la masa en reposo del electr\'on,
$\mathbf{\hat{p}}$ es el operador momento,
$\mathbf{\hat{\boldsymbol{\sigma}}}=\{\sigma_x,\sigma_y,\sigma_z\}$
es el vector de las matrices de Pauli, y $V(\mathbf{r})$ es el
potencial de la part\'{\i}cula.

En materiales cristalinos tridimensionales, la energ\'{\i}a
$V(\mathbf{r})$ proviene exclusivamente del potencial cristalino
\emph{microsc\'opico}. Dresselhaus \cite{dresselhaus55} mostr\'o
que en estructuras cristalinas sin simetr\'{\i}a de inversi\'on
como la estructura de tipo zinc-blenda (por ejemplo el material
semiconductor GaAs posee esta estructura), la interacci\'on SO
conduce a un splitting de la banda de conducci\'on en dos
subbandas. La magnitud del splitting es proporcional al cubo del
n\'umero de onda $k$ del electr\'on.

Es posible obtener gases de electrones bidimensionales partiendo
de sistemas en 3D confinando el movimiento electr\'onico a dos
dimensiones mediante la aplicaci\'on de un campo el\'ectrico
perpendicular, generando un pozo cu\'antico. La reducci\'on de la
dimensi\'on efectiva disminuye la simetr\'{\i}a del cristal
subyacente, y agrega un t\'ermino adicional, lineal en $k$, al
splitting de las subbandas. M\'as a\'un, si el pozo cu\'antico es
suficientemente estrecho, el t\'ermino lineal en $k$ se vuelve
dominante. Por otro lado Rashba \cite{rashba60a} mostr\'o que este
potencial confinante \emph{macrosc\'opico} da lugar a un segundo
t\'ermino en el Hamiltoniano de interacci\'on SO que es
responsable de la aparici\'on de un t\'ermino lineal en $k$
adicional en el splitting de las subbandas. En diversos sistemas,
como heteroestructuras de InGaAs/InAlAs \cite{nitta97}, GaAs
\cite{hassenkam97} y GaAs/AlGaAs \'esta se convierte en la
contribuci\'on m\'as importante al acoplamiento SO. En este
\'ultimo caso, el splitting de las subbandas tambi\'en depende del
potencial confinante, hecho que ha sido comprobado
experimentalmente por ejemplo en las Ref.
\citen{hassenkam97,nitta97,meier02,miller03} mediante el estudio
de fen\'omenos de transporte.

Las investigaciones tendientes a la comprensi\'on y control de los
fen\'omenos que involucran el spin electr\'onico en materiales
semiconductores han cobrado un renovado inter\'es en los \'ultimos
a\~nos a partir de la idea de fabricar unidades
\emph{spintronicas}, en las que se utiliza el spin electr\'onico
en lugar de su carga para el manejo y almacenamiento de
informaci\'on. En ellas, la interacci\'on SO, y el efecto Rashba
en particular juegan un rol central
\cite{wolf01,aronov93,koga02,datta90}.

\begin{figure}[ht]
\begin{center}\label{fig:EspectroAsimLibre}
\includegraphics{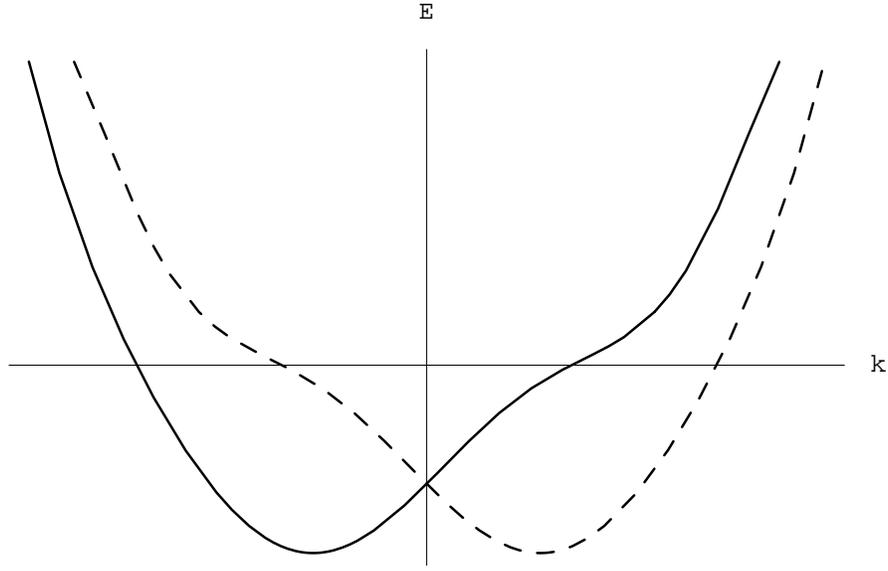}
\caption{Espectro de energ\'{\i}a de los electrones en un alambre
cu\'antico con interacciones spin-\'orbita.}
\end{center}
\end{figure}

Existen diversas t\'ecnicas para crear sistemas de electrones en
una dimensi\'on a partir de gases de electrones
bidimensionales\cite{yacoby96}. En esencia, todos utilizan un
confinamiento adicional mediante un potencial transversal. Este se
convierte as\'{\i} en una fuente adicional de potencial
macrosco\'opico, origen de la interacci\'on spin-\'orbita. Si el
confinamiento es suficientemente fuerte (angosto y profundo),
entonces este campo se vuelve importante frente al efecto Rashba,
y puede convertirse en dominante. Aunque hasta nuestro
conocimiento no existe evidencia experimental de la interacci\'on
SO que resulta de este potencial, estudios te\'oricos indican que
afecta cualitativamente el comportamiento de la energ\'{\i}a de
splitting como funci\'on del vector de onda $k$. En efecto, Moroz
y col. resolvieron la ecuaci\'on de Shr\"odinger que resulta de
situar un electr\'on estrictamente bidimensional (en el plano
$xy$) en un campo el\'ectrico transversal (en la direcci\'on del
eje $z$) y en un potencial cuadr\'atico en el eje
$x$\cite{moroz99}. Adem\'as consideraron los t\'erminos de
interacci\'on SO del tipo de la ecuaci\'on (\ref{eq:HSOgeneral})
con los potenciales anteriormente descriptos. Dado que el sistema
contin\'ua teniendo invarianza traslacional en el eje $y$, $k_y$
sigue siendo un buen n\'umero cu\'antico. Las energ\'{\i}as como
funci\'on de $k_y$ que se encuentran al resolver la ecuaci\'on de
Schr\"odinger se muestran en la figura
\ref{fig:EspectroAsimLibre}. Se observa un splitting entre las
subbandas de spin para arriba y spin para abajo, y adem\'as un
nueva caracter\'{\i}stica, propia del confinamiento a una
dimensi\'on, que es la deformaci\'on de cada una de las bandas
como funci\'on de $k_y$. La caracter\'{\i}stica m\'as importante
de esta deformaci\'on consiste en que cada banda pierde su eje de
simetr\'{\i}a vertical, y la velocidad de Fermi de los electrones
se vuelve diferente para cada direcci\'on de movimiento. Los
c\'alculos indican\cite{moroz99} que la diferencia de las
velocidades de Fermi se incrementa en forma mon\'otona con el
acoplamiento SO llegando a ser del orden de 10-20\%. Esta raz\'on
se encuentra dentro del rango en el es que es posible realizar
mediciones experimentales.

\section{Formulaci\'on del modelo y formalismo}

\subsection{Hamiltoniano fermi\'onico}

Como fue mencionado ya varias veces a lo largo de esta tesis, un
ingrediente central de los electrones en una dimensi\'on en
metales y materiales semiconductores es la alta correlaci\'on. De
modo que se vuelve necesario incorporar la interacci\'on
electr\'onica en modelos para interacciones SO. Para ello fue
propuesto el siguiente Hamiltoniano\cite{moroz00b,moroz00c}

\begin{equation}\label{eq:HamiltonianoFermionicoSO}
\H=\H_0+\H_\text{int},
\end{equation}
donde el Hamiltoniano libre es

\begin{equation}
\H_0=-iv_1\int
dx\,\left(\psid_{R,\uparrow}\partial_x\psi_{R,\uparrow}
-\psid_{L,\downarrow}\partial_x\psi_{L,\downarrow}\right)
-iv_2\int
dx\,\left(\psid_{R,\downarrow}\partial_x\psi_{R,\downarrow}
-\psid_{L,\uparrow}\partial_x\psi_{L,\uparrow}\right).
\end{equation}
$\psid_{r,\alpha}$ crea un fermi\'on de spin $\alpha$ que se mueve
a la derecha ($r=+1$) o a la izquierda ($r=-1$). Este Hamiltoniano
es similar al correspondiente al modelo de Tomonaga-Luttinger con
spin, con la diferencia de que en este caso, los fermiones que se
mueven a izquierda y derecha en ambas bandas, poseen diferentes
velocidades de Fermi ($v_1\neq v_2$), reflejando la asimetr\'{\i}a
del espectro libre. El t\'ermino de interacci\'on describe
interacciones de dispersi\'on hacia adelante, y posee la forma
usual:

\begin{multline}
\H\Tint=\sum_{\alpha, \beta}\int dx\,
\left(g_{2\parallel}\delta_{\alpha\beta}+g_{2\perp}\delta_{\alpha,-\beta}\right)
\psid_{L,\alpha}\psi_{R,\alpha}\psid_{R,\beta}\psi_{L,\beta}\\
+\frac{1}{2}\sum_{r, \alpha, \beta}\int dx\,
\left(g_{4\parallel}\delta_{\alpha\beta}+g_{4\perp}\delta_{\alpha,-\beta}\right)
\psid_{r,\alpha}\psi_{r,\alpha}\psid_{r,\beta}\psi_{r,\beta}.
\end{multline}
Los t\'erminos de dispersi\'on hacia atr\'as y umklapp son
irrelevantes si nos restringimos a interacciones repulsivas en el
primer caso, y estamos lejos del llenado medio en el segundo.

En este cap\'{\i}tulo calculamos las funciones de correlaci\'on de
los operadores que representan ondas de densidad de carga (CDW) y
spin (SDW), y superconductividad de tipo singulete (SS) y triplete
(TS) en el modelo anteriormente presentado. Las funciones de
correlaci\'on para estos operadores son bien conocidas en ausencia
de acoplamiento SO\cite{voit95,solyom79,emery79}, incluyendo
factores de correci\'on logar\'{\i}tmica que se originan en
t\'erminos irrelevantes\cite{giamarchi89,voit92} y dependencia con
el tiempo y la temperatura\cite{emery79}. En el presente
cap\'{\i}tulo se extienden estos c\'alculos al caso en que las
interacciones SO se encuentran presentes, y se estudia como se
modifican los exponentes de sus decaimientos algebraicos. Como
resultado encontramos interesantes modificaciones del diagrama de
fases del sistema. Para ciertas regiones del espacio de
par\'ametros, la interacci\'on SO cambia la fuctuaci\'on
dominante, y hace que susceptibilidades que eran finitas, ahora se
vuelvan tambi\'en divergentes a bajas temperaturas.

\subsection{Bosonizaci\'on}

El Hamiltoniano (\ref{eq:HamiltonianoFermionicoSO}) puede
estudiarse mediante la t\'ecnica de bosonizaci\'on, como en la
Ref. \citen{moroz00b,moroz00c}. Aunque es indistinto emplear
bosonizaci\'on funcional u operacional, elegiremos esta \'ultima
para llevar un paralelismo lo m\'as estrecho posible con dicha
referencia, aunque finalmente los valores medios ser\'an
calculados utilizando t\'ecnicas funcionales. Por conveniencia
definimos una velocidad promedio $v_0=(v_1+v_2)/2$ y la diferencia
$\delta v=v_2-v_1$. A partir de la figura
\ref{fig:EspectroAsimLibre} se observa que los momentos de Fermi
tambi\'en son diferentes para ambas ramas, por esa raz\'on,
hacemos lo mismo para los momentos de Fermi, definiendo el
promedio $k_0=(k_1+k_2)/2$ y la diferencia $\delta k=k_2-k_1$.
Para efectuar la bosonizaci\'on introducimos los campos de fase
usuales $\phi_\rho$ y $\phi_\sigma$ para los grados de libertad de
carga y spin, y los correspondientes campos duales $\Pi_\rho$ y
$\Pi_\sigma$ siguiendo el procedimiento explicado en el
cap\'{\i}tulo \ref{part:bosonizaciónOp}. En t\'erminos de los
campos bos\'onicos, el Hamiltoniano puede representarse como

\begin{multline}\label{eq:HamiltonianoBosónico}
\H=\frac{v_\rho}{2}\int dx
\left[\frac{1}{K_\rho}(\partial_x\phi_\rho)^2+K_\rho\Pi_\rho^2\right]
+\frac{v_\sigma}{2}\int dx
\left[\frac{1}{K_\sigma}(\partial_x\phi_\sigma)^2+K_\sigma\Pi_\sigma^2\right]\\
+ \delta v\int dx
\left[(\partial_x\phi_\rho)\Pi_\sigma+(\partial_x\phi_\sigma)\Pi_\rho\right].
\end{multline}
$v_{\rho,\sigma}$ son las velocidades de propagaci\'on de los
modeos colectivos del modelo desacoplado ($\delta v=0$), y
$K_{\rho,\sigma}$ son las constantes de dureza. El acoplamiento SO
aparece como un efecto que rompe la separaci\'on spin-carga, lo
cual se manifiesta en la presencia del tercer t\'ermino en la
\'ultima ecuaci\'on. Sin embargo, este Hamiltoniano puede ser
diagonalizado en t\'erminos de dos nuevos campos de fase, que
portan una mezcla de carga y spin. Dado que el Hamiltoniano posee
t\'erminos cruzados en los campos y los momentos, su
diagonalizaci\'on no es trivial, por ejemplo no puede
diagonalizarse por una transformaci\'on de similitud, porque
resulta no can\'onica. Dejamos para el ap\'endice \ref{part:diago}
los detalles. Baste mencionar que las velocidades de propagaci\'on
de estos nuevos modos colectivos son

\begin{equation}
v_\pm^2=\frac{v_\sigma^2+v_\rho^2}{2}+\delta v^2
\pm\sqrt{\left(\frac{v_\rho^2-v_\sigma^2}{2}\right)^2+\delta
v^2\left[v_\sigma^2+v_\rho^2+v_\rho
v_\sigma\left(\frac{K_\rho}{K_\sigma}+\frac{K_\sigma}{K_\rho}\right)\right]}.
\end{equation}

A medida que $\delta v\rightarrow0$,
$v_+\rightarrow\max(v_\rho,v_\sigma)$ y
$v_-\rightarrow\min(v_\rho,v_\sigma)$. A medida que $\delta v$ se
incrementa, $v_-$ disminuye hasta anularse en los puntos

\begin{align}
\delta v_\rho^2=&v_\rho v_\sigma\frac{K_\sigma}{K_\rho},\\
\delta v_\sigma^2=&v_\rho v_\sigma\frac{K_\rho}{K_\sigma}.
\end{align}

En estos puntos, el congelamiento del modo bos\'onico m\'as lento
est\'a acompa\~nado por una divergencia en las funciones respuesta
de carga y spin. La compresibilidad de carga est\'atica $\kappa$
diverge para $\delta v = \delta v_\rho$ y para $\delta v = \delta
v_\sigma$ ocurre una divergencia en la susceptibilidad de spin
est\'atica $\chi$. Sus comportamientos son

\begin{align}\label{eq:instabilities1}
\kappa=&\kappa_0\left[1-\frac{\delta v}{\delta
v_\rho}\right]^{-1},\espacio\kappa_0=\frac{2K_\rho}{\pi v_\rho},\\
\chi=&\chi_0\left[1-\frac{\delta v}{\delta
v_\sigma}\right]^{-1},\espacio\chi_0=\frac{2K_\sigma}{\pi
v_\sigma}\label{eq:instabilities2},
\end{align}
donde $\kappa_0$ y $\chi_0$ son los valores de $\kappa$ y $\chi$
en ausencia de acoplamiento SO. M\'as all\'a de estos puntos las
susceptibilidades se vuelven negativas. Este comportamiento de las
funciones respuesta est\'aticas junto con el hecho de que las
velocidades de los modos colectivos se anulen indican que el
sistema se ha vuelto inestable\cite{voit95,drut03} y que realiza
una transici\'on de fase de primer orden\cite{voit92}. Para
$K_\rho>K_\sigma$, $\delta v_\rho$ resulta menor que $\delta
v_\sigma$, y a medida que $\delta v$ crece desde $0$, la
divergencia f\'{\i}sica tiene lugar en la compresibilidad de
carga. Esta inestabilidad se conoce como separaci\'on de fases y
se ha mostrado que ocurre en el modelo de Hubbard
extendido\cite{penc94} y en el modelo $t-J$\cite{ogata91}. En el
caso en que $K_\rho<K_\sigma$, la inestabilidad tiene lugar en el
subsistema de spin, y est\'a relacionada a la llamada transici\'on
metamagn\'etica, observada por ejemplo en el compuesto cuasi
unidimensional
$\text{Ba}_3\text{Cu}_2\text{O}_4\text{Cl}_2$\cite{eckert98}.
Tambi\'en ocurre en el diagrama de fases del modelo XXZ con
segundos vecinos\cite{gerhardt98}. En presencia de un potencial
qu\'{\i}mico (campo magn\'etico), la regi\'on en la cual $\kappa$
($\chi$) se vuelve negativa est\'a asociada a la coexistencia de
dos fases con diferente concentraci\'on de agujeros
(magnetizaci\'on). La divergencia en $\kappa$ fue hallada
tambi\'en en otros modelos con dispersi\'on asim\'etrica y sin
interacciones SO \cite{fernandez02b}.

\section{Funciones de correlaci\'on}

Enfoquemos ahora nuestra atenci\'on en las funciones de
correlaci\'on. Los operadores de inter\'es son

\begin{align}
\O_{\text{CDW}}&=\sum_{r,\alpha}\psid_{r\alpha}\psi_{-r\alpha}e^{-2irk\f
x},\\
\O_{\text{SDW},x}&=\sum_{r,\alpha}\psid_{r\alpha}\psi_{-r,-\alpha}e^{-2irk\f
x},\\
\O_{\text{SDW},y}&=\sum_{r,\alpha}(-i\alpha)\psid_{r\alpha}\psi_{-r,-\alpha}e^{-2irk\f
x},\\
\O_{\text{SDW},z}&=\sum_{r,\alpha}\alpha\psid_{r\alpha}\psi_{-r\alpha}e^{-2irk\f
x},\\
\O_{\text{SS}}&=\frac{1}{\sqrt{2}}\sum_\alpha\alpha\psi_{L\alpha}\psi_{R,-\alpha},\\
\O_{\text{TS},0}&=\frac{1}{\sqrt{2}}\sum_\alpha\psi_{L\alpha}\psi_{R,-\alpha},\\
\O_{\text{TS},\alpha}&=\psi_{L\alpha}\psi_{R\alpha}.
\end{align}

Para obtener su forma bos\'onica basta utilizar las expresiones de
equivalencia entre campos fermi\'onicos y bos\'onicos,
(\ref{eq:bosPsi}) y (\ref{eq:camposCargaSpin}). El resultado es

\begin{align}\label{eq:operadoresBosónicos}
\O_{\text{CDW}}&=\frac{2}{\pi
a}\cos(2k_0x+\sqrt{2\pi}\phi_\rho)\cos\sqrt{2\pi}\phi_\sigma,\\
\O_{\text{SDW},x}&=\frac{2}{\pi a}
\cos(2k_0x+\sqrt{2\pi}\phi_\rho)\cos(\delta k
x+\sqrt{2\pi}\theta_\sigma),\\
\O_{\text{SDW},y}&=\frac{2}{\pi a}
\cos(2k_0x+\sqrt{2\pi}\phi_\rho)\sin(\delta k
x+\sqrt{2\pi}\theta_\sigma),\\
\O_{\text{SDW},z}&=\frac{2}{\pi a}
\sin(2k_0x+\sqrt{2\pi}\phi_\rho)\sin\sqrt{2\pi}\phi_\sigma,\\
\O_{\text{SS}}&=\frac{-i}{\sqrt{2}\pi a}
e^{-i\sqrt{2\pi}\theta_\rho}\sin\sqrt{2\pi}\phi_\sigma,\\
\O_{\text{TS},0}&=\frac{1}{\sqrt{2}\pi a}
e^{-i\sqrt{2\pi}\theta_\rho}\cos\sqrt{2\pi}\phi_\sigma,\\
\O_{\text{TS},\pm1}&=\frac{1}{2\pi a}e^{\pm i\delta k
x}e^{-i\sqrt{2\pi}(\theta_\rho\pm\theta_\sigma)}.
\end{align}
Recordamos que, seg\'un se defini\'o en el cap\'{\i}tulo
\ref{part:bosonizaciónOp}, $\theta_\nu$ est\'a vinculado con
$\Pi_\nu$ por la relaci\'on $\Pi_\nu=\partial_x\theta_\nu$.

\subsection{Funciones de correlaci\'on a temperatura finita}

Las funciones de correlaci\'on se calcularon en el marco de la
integral funcional dentro del formalismo de tiempo imaginario de
Matsubara\cite{negele88}. En este formalismo, se definen como

\begin{multline}\label{eq:funcionesR}
R_i(x,\tau;\beta)=\bra\O_i(x,\tau)\O^\dagger_i(0,0)\ket\\
=\frac{1}{\Z_0}
\int\D\Pi_\rho\D\Pi_\sigma\D\phi_\rho\D\phi_\sigma\O_i(x,\tau)\O^\dagger_i(0,0)
\exp\{-S[\Pi_\nu,\phi_\nu]\},
\end{multline}
donde $\Z_0$ es la funci\'on de partici\'on

\begin{equation}
\Z_0=\int\D\Pi_\rho\D\Pi_\sigma\D\phi_\rho\D\phi_\sigma
\exp\{-S[\Pi_\nu,\phi_\nu]\},
\end{equation}
$S$ es la acci\'on eucl\'idea

\begin{equation} S[\Pi_\nu,\phi_\nu]=\int_0^\beta
d\tau\,\H(\tau)-i\int_0^\beta d\tau\int
dx\,\Pi_\nu(x,\tau)\partial_\tau\phi_\nu(x,\tau),
\end{equation}
y $\tau$ es el tiempo imaginario. Las propiedades de tiempo real
se obtienen por continuaci\'on anal\'{\i}tica $\tau\rightarrow
it$. Como los operadores (\ref{eq:operadoresBosónicos}) est\'an
expresados en t\'erminos de $\phi_\nu$ y $\theta_\nu$, es
conveniente trabajar en t\'erminos de $\theta_\nu$ en lugar de
$\Pi_\nu$.  La acci\'on expresada en las nuevas variables es

\begin{multline}\label{eq:accionNueva}
S[\theta_\nu, \phi_\nu]=\frac{v_\rho}{2}\int_0^\beta d\tau\,\int
dx \left[\frac{1}{K_\rho}\left(\partial_x\phi_\rho\right)^2+
K_\rho\left(\partial_x\theta_\rho\right)^2\right]\\+
\frac{v_\sigma}{2}\int_0^\beta d\tau\,\int dx
\left[\frac{1}{K_\sigma}\left(\partial_x\phi_\sigma\right)^2+
K_\sigma\left(\partial_x\theta_\sigma\right)^2\right]\\+
\int_0^\beta d\tau\,\int
dx\left[\partial_x\theta_\rho\left(i\partial_\tau\phi_\rho+\delta
v\partial_x\phi_\sigma\right) +
\partial_x\theta_\sigma\left(i\partial_\tau\phi_\sigma+\delta
v\partial_x\phi_\rho\right)\right].
\end{multline}

Por la forma que poseen los operadores
(\ref{eq:operadoresBosónicos}), sus funciones de correlaci\'on
pueden escribirse de forma general como una combinaci\'on de
t\'erminos de la forma

\begin{equation}\label{eq:valorMedio}
\bra\exp\left\{i\sum_k\beta_k\left[\varphi_k(x)-\varphi_k(0)\right]\right\}\ket
\end{equation}
donde definimos el campo $\varphi$ de la siguiente manera:

\begin{equation}
(\varphi_j)=\begin{pmatrix}
    \phi_\rho \\
    \theta_\rho \\
    \phi_\sigma \\
    \theta_\sigma \\
    \end{pmatrix},
\end{equation}
con $\beta_k$ constantes apropiadas, y $j=1,2,3,4$. La expresi\'on
(\ref{eq:valorMedio}) es igual a (f\'ormula obtenida en el
ap\'endice \ref{part:identity})

\begin{equation}
\exp \left\{
\sum_{i,j}\beta_i\beta_j\left[\Delta_{ij}^{-1}(x)-\Delta_{ij}^{-1}(0)\right]
\right\},
\end{equation}
donde
\begin{equation}\label{eq:meanValues}
\bra\varphi_i(x)\varphi_j(y)\ket=\Delta^{-1}_{ij}(x-y).
\end{equation}

De modo tal que las funciones de correlaci\'on que nos interesan
(exponenciales de los campos) quedan expresadas en t\'erminos de
funciones de correlaci\'on de los campos $\phi_\nu$ y
$\theta_\nu$. \'Estas se pueden calcular siguiendo el
procedimiento est\'andar de construir una funcional generatriz con
la acci\'on (\ref{eq:accionNueva}), y derivar funcionalmente
respecto de las fuentes externas. Mediante este procedimiento se
encuentra

\begin{multline}\label{eq:integrales}
\R_{\varphi_i,\varphi_j}(x,\tau;\beta)\equiv\bra\left[\varphi_i(x)-\varphi_i(0)\right]
\left[\varphi_j(x)-\varphi_j(0)\right]\ket\\=
\frac{1}{\pi\beta}\sum_{n=-\Infinity}^{\Infinity}
\int_{-\Infinity}^\Infinity
dk\left(1-e^{-ikx-i\omega_n\tau}\right)G_{\varphi_i,\varphi_j}(\omega_n,k)e^{-\epsilon|k|}.
\end{multline}
donde $\omega_n$ son las frecuencias de Matsubara $\omega_n=2\pi
n/\beta$. El factor $e^{-\epsilon|k|}$ act\'ua como regulador en
el U.V. Las funciones $G_{\varphi_i,\varphi_j}(\omega_n,k)$,
sim\'etricas en el intercambio de $\varphi_i$ con $\varphi_j$,
resultan

\begin{align}
G_{\phi_\rho,\phi_\rho}(\omega_n,k)=&\frac{K_\rho
v_\rho(\omega_n^2+k^2 v_s^2)-\delta v K_\sigma v_\sigma k^2}
{(\omega_n^2+v_+^2 k^2)(\omega_n^2+v_-^2 k^2)}\\
G_{\phi_\sigma,\phi_\sigma}(\omega_n,k)=&\frac{K_\sigma
v_\sigma(\omega_n^2+k^2 v_s^2)-\delta v K_\rho v_\rho k^2}
{(\omega_n^2+v_+^2 k^2)(\omega_n^2+v_-^2 k^2)}\\
G_{\theta_\rho,\theta_\rho}(\omega_n,k)=&\frac{K_\sigma
v_\rho(\omega_n^2+k^2 v_s^2)-\delta v K_\rho v_\sigma k^2}
{K_\sigma K_\rho(\omega_n^2+v_+^2 k^2)(\omega_n^2+v_-^2 k^2)}\\
G_{\theta_\sigma,\theta_\sigma}(\omega_n,k)=&\frac{K_\rho
v_\sigma(\omega_n^2+k^2 v_s^2)-\delta v K_\sigma v_\rho k^2}
{K_\rho K_\sigma(\omega_n^2+v_+^2 k^2)(\omega_n^2+v_-^2 k^2)}\\
G_{\phi_\rho,\theta_\rho}(\omega_n,k)=&
i\frac{\omega_n\left[\omega_n^2+k^2\left(v_\sigma^2+\delta
v^2\right)\right]}{k(\omega_n^2+v_+^2 k^2)(\omega_n^2+v_-^2
k^2)}\\
G_{\phi_\sigma,\theta_\sigma}(\omega_n,k)=&
i\frac{\omega_n\left[\omega_n^2+k^2\left(v_\rho^2+\delta
v^2\right)\right]}{k(\omega_n^2+v_+^2 k^2)(\omega_n^2+v_-^2
k^2)}\\
G_{\phi_\rho,\phi_\sigma}(\omega_n,k)=&\delta v\frac{i\omega_n k
(K_\rho v_\rho + K_\sigma v_\sigma)}{(\omega_n^2+v_+^2
k^2)(\omega_n^2+v_-^2k^2)}\\
G_{\theta_\rho,\theta_\sigma}(\omega_n,k)=&\delta v\frac{i\omega_n
k (K_\rho v_\sigma + K_\sigma v_\rho)}{K_\rho
K_\sigma(\omega_n^2+v_+^2k^2)(\omega_n^2+v_-^2k^2)}\\
G_{\phi_\rho,\theta_\sigma}(\omega_n,k)=&\delta v\frac{k^2(K_\rho
v_\rho v_\sigma-K_\sigma \delta v^2)/K_\sigma -
\omega_n^2}{(\omega_n^2+v_+^2
k^2)(\omega_n^2+v_-^2k^2)}\\
G_{\phi_\sigma,\theta_\rho}(\omega_n,k)=&\delta
v\frac{k^2(K_\sigma v_\sigma v_\rho-K_\rho \delta v^2)/K_\rho -
\omega_n^2}{(\omega_n^2+v_+^2
k^2)(\omega_n^2+v_-^2k^2)}.\label{eq:funcionesDeCorrelación}
\end{align}
Debe notarse que las funciones que involucran mezclas de campos de
carga con campos de spin son proporcionales a $\delta v$, de modo
que se anulan cuando no hay acoplamiento spin-\'orbita,
restaurando la separaci\'on spin-carga. Dejaremos los detalles del
c\'alculo de las integrales para el ap\'endice
\ref{part:integrales}, y aqu\'{\i} presentaremos los resultados
finales para las funciones (\ref{eq:funcionesR}):

\begin{multline}
R_{\text{CDW}}(x,\tau;\beta)=R_{\text{SDW},z}(x,\tau;\beta)=\\\frac{\cos{2k_0x}}{2(\pi{a})^2}
\left(z_+ \zb_+\right)^
{-(K_\rho\nu_+^\rho+K_\sigma\nu_+^\sigma)/2}
\left(z_-\zb_-\right)^
{-(K_\rho\nu_-^\rho+K_\sigma\nu_-^\sigma)/2}
\left[\left(\frac{\zb_+z_-}{z_+\zb_-}\right)^{H\sign(x\tau)}+h.c.\right]
\end{multline}

\begin{multline}\label{eq:SDWxy}
R_{\text{SDW},xy}(x,\tau;\beta)=\frac{\cos{2k_1x}}{2(\pi{a})^2}
\left(z_+\zb_+\right)^{-(K_\rho\nu_+^\rho+\mu_+^\sigma/K_\sigma)/2-\theta_+^\sigma}
\left(z_-\zb_-\right)^{-(K_\rho\nu_-^\rho+\mu_-^\sigma/K_\sigma)/2-\theta_-^\sigma}\\+
\frac{\cos{2k_2x}}{2(\pi{a})^2}
\left(z_+\zb_+\right)^{-(K_\rho\nu_+^\rho+\mu_+^\sigma/K_\sigma)/2+\theta_+^\sigma}
\left(z_-\zb_-\right)^{-(K_\rho\nu_-^\rho+\mu_-^\sigma/K_\sigma)/2+\theta_-^\sigma}
\end{multline}

\begin{multline}
R_{\text{SS}}(x,\tau;\beta)=R_{\text{TS},0}(x,\tau;\beta)\\
=\frac{1}{2(2\pi{a})^2}
\left(z_+\zb_+\right)^{-(\mu_+^\rho/K_\rho+K_\sigma\nu_+^\sigma)/2+\theta_+^\rho}
\left(z_-\zb_-\right)^{-(\mu_-^\rho/K_\rho+K_\sigma\nu_-^\sigma)/2+\theta_-^\rho}+
(\theta_\pm^\rho\rightarrow -\theta_\pm^\rho)
\end{multline}

\begin{equation}
R_{\text{TS},\pm1}(x,\tau;\beta)=\frac{e^{\pm i\delta
kx}}{(2\pi{a})^2}
\left(z_+\zb_+\right)^{-(\mu_+^\rho/K_\rho+\mu_+^\sigma/K_\sigma)/2}
\left(z_-\zb_-\right)^{-(\mu_-^\rho/K_\rho+\mu_-^\sigma/K_\sigma)/2}
\left(\frac{\zb_+z_-}{z_+\zb_-}\right)^{\pm G\sign(x\tau)}
\end{equation}
donde

\begin{align}
z_\pm=&\frac{\sin\frac{\pi}{v_\pm\beta}(\epsilon+v_\pm\tau+ix)}
{\sin{\frac{\pi\epsilon}{v_\pm\beta}}}\\
\zb_\pm=&\frac{\sin\frac{\pi}{v_\pm\beta}(\epsilon+v_\pm\tau-ix)}
{\sin{\frac{\pi\epsilon}{v_\pm\beta}}}
\end{align}
y los exponentes dependen de las constantes $K$ multiplicadas por
factores que incluyen dependencias en las velocidades. Est\'an
dados por

\begin{align}
\nu_\pm^\lambda=&\pm\frac{v_\lambda}{v_\pm}\frac{v^2_\pm-v^2_{-\lambda}\left(1-\delta
v^2/\delta v_{-\lambda}^2\right)}{v_+^2-v_-^2}\\
\mu_\pm^\lambda=&\pm\frac{v_\lambda}{v_\pm}\frac{v^2_\pm-v^2_{-\lambda}\left(1-\delta
v^2/\delta v_\lambda^2\right)}{v_+^2-v_-^2}\\
\theta_\pm^\lambda=&\pm\frac{\delta
v}{v_\pm}\frac{v^2_\pm-\left(\delta v_\lambda^2- \delta
v^2\right)}{v_+^2-v_-^2}\label{eq:exptheta}
\end{align}
con $\lambda=\rho,\sigma$, y

\begin{align}
H=&\delta v\frac{K_\rho v_\rho+K_\sigma v_\sigma}{v_+^2-v_-^2}\\
G=&\delta v\frac{v_\rho/K_\rho + v_\sigma/K_\sigma
}{v_+^2-v_-^2}\label{eq:expG}.
\end{align}
$\nu_\pm^\lambda$ y $\mu_\pm^\lambda$ son positivos, y
$\theta_\pm^\lambda$, $G$ y $H$ tienen el mismo signo que $\delta
v$. En el modelo sin acoplamiento SO, la simetr\'{\i}a SU(2) puede
restaurarse fijando el valor de $K_\sigma=1$, valor que emerge
naturalmente si el modelo bajo estudio es el l\'{\i}mite
cont\'{\i}nuo de un modelo en la red con solamente interacciones
de tipo densidad de carga. En nuestro caso esta simetr\'{\i}a
est\'a expl\'{\i}citamente rota desde el principio, y no es
posible restaurarla. Esta ruptura se manifiesta en las diferencias
en los decaimientos entre las funciones de correlaci\'on de
operadores SDW en la direcci\'on $z$ y las direcciones $x$ e $y$.

Igual que en el caso en que no hay acoplamiento SO, las funciones
de correlaci\'on para operadores SDW en la direcci\'on $z$ y
operadores CDW son iguales. Lo mismo ocurre con las funciones de
correlaci\'on para operadores TS y SS. Esta degeneraci\'on se
rompe al incluir correcciones logar\'{\i}tmicas que surgen si se
tienen en cuenta t\'erminos irrelevantes de umklapp y
backscattering\cite{giamarchi89}.

Un punto interesante de observar es la aparici\'on de dos
t\'erminos en la funci\'on SDW,$xy$ [Ec. (\ref{eq:SDWxy})] donde
las modulaciones poseen diferentes frecuencias y decaen con
diferentes exponentes. Como $\theta_\pm^\lambda$ tiene el mismo
signo que $\delta v$ [ver la Ec. (\ref{eq:exptheta}) y el
comentario debajo de la Ec. (\ref{eq:expG})] para $v_2>v_1$
($v_2<v_1$) el t\'ermino dominante  es el de frecuencia $k_2$
($k_1$). En otras palabras, la mayor frecuencia domina. Adem\'as
$R_{\text{TS},\pm1}$ se vuelve oscilante.

\subsection{Funciones de correlaci\'on instant\'aneas a temperatura
cero} Hasta aqu\'{\i} hemos obtenido f\'ormulas muy generales para
las funciones de correlaci\'on dependientes del espacio, tiempo
imaginario y la temperatura. Podemos lograr una mayor
comprensi\'on de la f\'{\i}sica del problema observando el
decaimiento algebraico de las funciones de correlaci\'on
instant\'aneas ($\tau=0$) y a temperatura cero y estudiar como sus
exponentes se modifican respecto del caso con acoplamiento SO
nulo. El comportamiento general de estas funciones es

\begin{equation}
R_i(x)\sim |x|^{-2+\alpha_i}.
\end{equation}
Los exponentes $\alpha_i'\text{s}$ determinan la divergencia de la
correspondiente susceptibilidad en el espacio de Fourier cuando
$T\rightarrow 0$, $\chi_i(T)\sim T^{-\alpha_i}$\cite{voit95}. De
este modo, estas inestabilidades resultan de una naturaleza
completamente diferente a las descriptas en las Ecs.
(\ref{eq:instabilities1}) y (\ref{eq:instabilities2}). Las
expresiones obtenidas para los $\alpha_i$ son

\begin{figure}[ht]
\begin{center}
\includegraphics{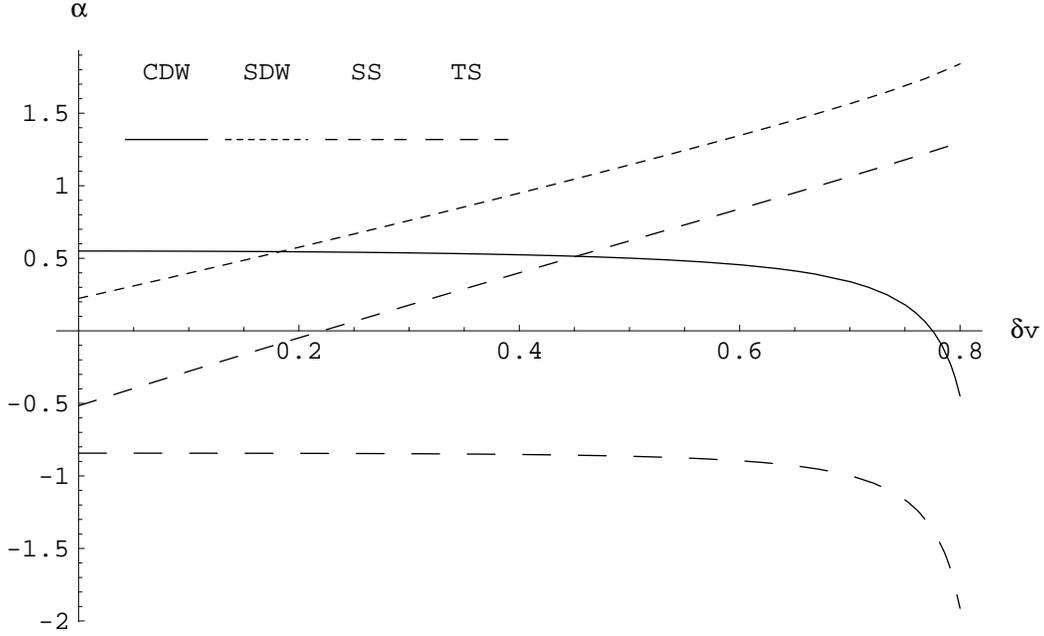}
\caption{\label{fig:exponents} Comportamiento de los exponentes
$\alpha_i'\text{s}$ como funci\'on de $\delta v$ (en unidades de
$v_0$). Para $v_\rho=1.2 v_0$, $v_\sigma=0.8 v_0$, $K_\rho=0.6$ y
$K_\sigma=0.85$. Para $\delta v\gtrsim 0.16$ las fluctuaciones
$\text{SDW},xy$ se vuelven dominantes, y para $\delta v\gtrsim
0.25$ $\alpha_{\text{SS}}$ se vuelve positivo, y
$\chi_{\text{SS}}$ divergente para $T\rightarrow 0$.}
\end{center}
\end{figure}

\begin{align}
\alpha_{\text{CDW}}=\alpha_{\text{SDW},z}=&2-K_\rho\nu^\rho-K_\sigma\nu^\sigma\\
\alpha_{\text{SDW},x}=\alpha_{\text{SDW},y}=&2(1+|\theta^\sigma|)-K_\rho\nu^\rho-\mu^\sigma/K_\sigma\\
\alpha_{\text{SS}}=\alpha_{\text{TS},0}=&2(1+|\theta^\rho|)-\mu^\rho/K_\rho-K_\sigma\nu^\sigma\\
\alpha_{\text{TS},\pm 1}=&2-\mu^\rho/K_\rho-\mu^\sigma/K_\sigma.
\end{align}
Estos son los nuevos exponentes, que retienen la misma estructura
que en el caso de SO nulo, pero modificados por los factores

\begin{gather}
\mu^\lambda=\mu_+^\lambda+\mu_-^\lambda\\
\nu^\lambda=\nu_+^\lambda+\nu_-^\lambda\\
\theta^\lambda=\theta_+^\lambda+\theta_-^\lambda.
\end{gather}
Cuando $\delta v\rightarrow 0$, se verifica que
$\theta^\lambda\rightarrow 0$, y
$\mu^\lambda,\nu^\lambda\rightarrow 1$, de modo que reproducimos
los resultados correctos para el caso SO nulo.

\begin{figure}
\begin{center}
\includegraphics{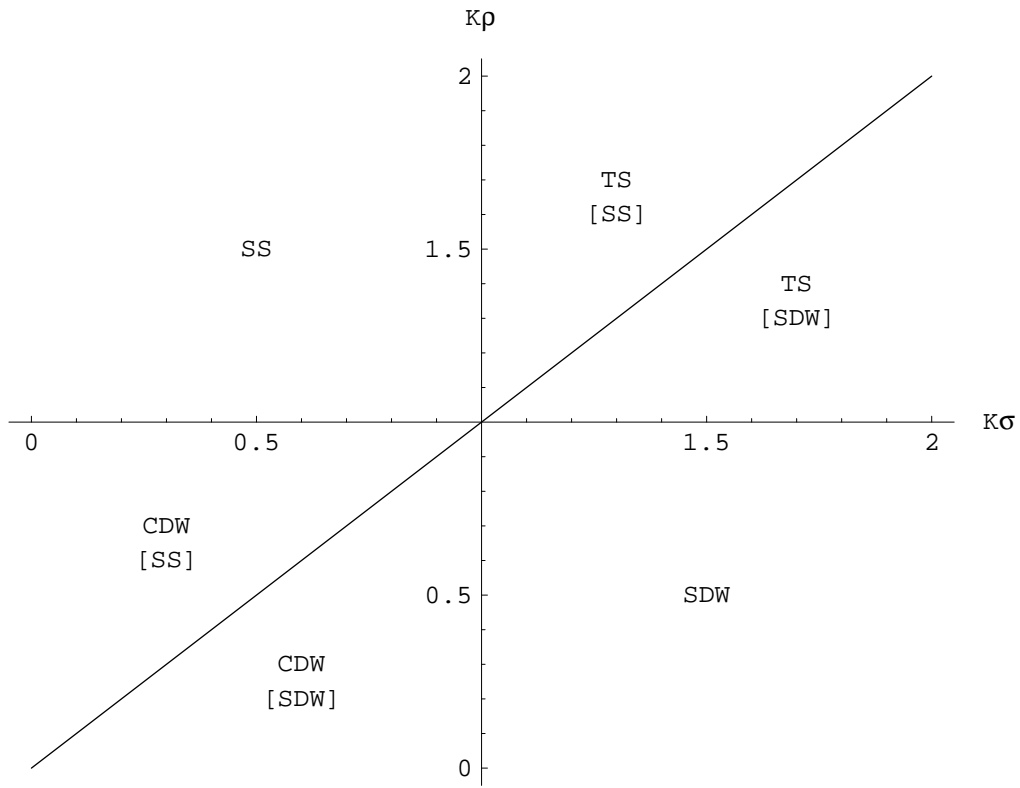}
\caption{\label{fig:phaseDiag} Diagrama de fases en el espacio
$K_\rho-K_\sigma$. Las fases entre corchetes son subdominantes,
que se vuelven dominantes para acoplamiento SO suficientemente
fuerte.}
\end{center}
\end{figure}

\begin{figure}
\begin{center}
\includegraphics{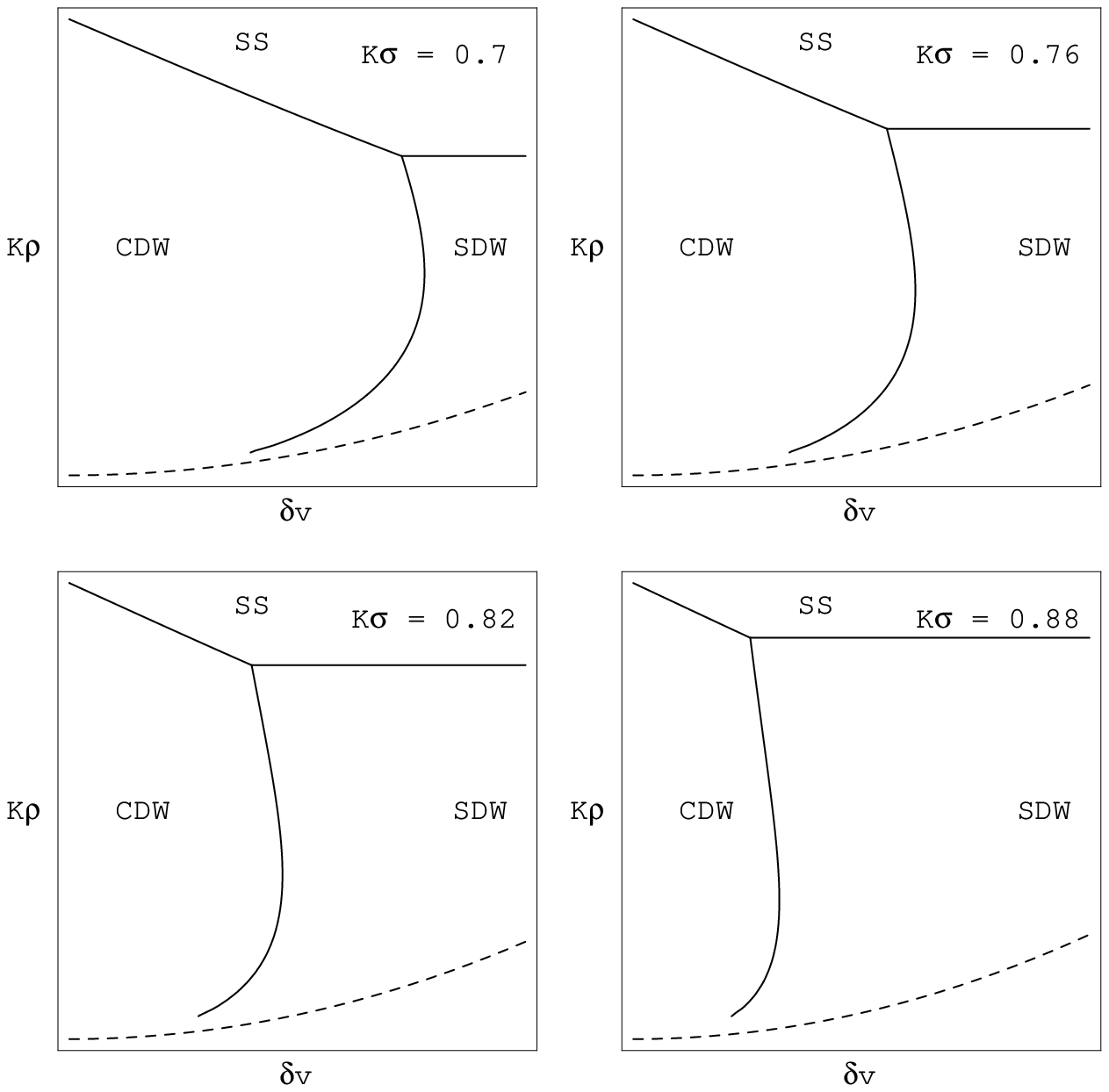}
\caption{\label{fig:crossSections} Diagrama de fases en el espacio
$K_\rho-\delta v$ para $v_\rho=1.2 v_0$, $v_\sigma=0.8 v_0$ y
diferentes valores de $K_\sigma$. $\delta v>\delta v_\sigma$
debajo de la l\'{\i}nea punteada y ocurre el metamagnetismo.}
\end{center}
\end{figure}

Para acoplamiento SO finito, $\delta v$ es un par\'ametro que
juega un rol en determinar cu\'al es la funci\'on de correlaci\'on
que decae m\'as lentamente y cu\'ales son las susceptibilidades
divergentes. En la Fig. \ref{fig:exponents} observamos, a modo de
ejemplo, el comportamiento de los exponentes como funci\'on de
$\delta v$ para $v_\rho=1.2v_0$, $v_\sigma=0.8v_0$, $K_\rho=0.6$ y
$K_\sigma=0.85$. Para $\delta v$ peque\~no, las fluctuaciones CDW
son dominantes, pero para $\delta v\gtrsim0.16v_0$ las
correlaciones SDW,$xy$ decaen m\'as lentamente. Para $\delta v$
peque\~no, las fluctuaciones CDW y SDW son las \'unicas
fluctuaciones divergentes para $T\rightarrow 0$, pero para $\delta
v \gtrsim0.25v_0$, $\alpha_\text{SS}$ se vuelve positivo y
$\chi_\text{SS}$ divergente para $T\rightarrow 0$. C\'alculos de
la estructura de bandas electr\'onicas modificadas por el
acoplamiento SO muestran que estos valores de $\delta v$
deber\'{\i}an corresponder a sistemas de electrones cuasi
unidimensionales t\'{\i}picos.

Un an\'alisis cuidadoso de los exponentes nos permite construir un
diagrama de fases en el espacio $K_\rho-K_\sigma$. (Fig.
\ref{fig:phaseDiag}). En cada regi\'on indicamos las fluctuaciones
dominantes para $\delta v$ peque\~no, y entre corchetes las
dominantes para $\delta v$ mayores. Otras fluctuaciones
subdominantes no est\'an indicadas. En la Fig.
\ref{fig:crossSections} se observan cortes del diagrama de fases
en el espacio $K_\rho-\delta v$ para $K_\rho<1$ y diferentes
valores de $K_\sigma$. Para $\delta v$ peque\~no las fluctuaciones
CDW dominan, y para $\delta v$ m\'as grande , el sistema puede
encontrarse en la fase SDW o SS dependiendo de los valores de
$K_\rho$ y $K_\sigma$. En la regi\'on debajo de la l\'{\i}nea
punteada, $\delta v<\delta v _\sigma$, la susceptibilidad
est\'atica de spin se vuelve negativa, y tiene lugar el
metamagnetismo.

\section{Conclusiones}

En este cap\'{\i}tulo hemos calculado funciones de correlaci\'on
para operadores de fluctuaciones de ondas de densidad de carga y
spin, y superconductividad singulete y triplete en un modelo de
electrones altamente correlacionados en una dimensi\'on con
acoplamiento spin-\'orbita. El c\'alculo se realiz\'o en funci\'on
de la temperatura, y al final se estudiaron las funciones
instant\'aneas y a temperatura cero. El acoplamiento spin-\'orbita
destruye la separaci\'on spin-carga como se mostr\'o en las Ref.
\citen{moroz00b,moroz00c} y modifica los exponentes de los
decaimientos de las correlaciones. Como consecuencia se modifica
el diagrama de fases del sistema. Un acoplamiento spin-\'orbita
suficientemente fuerte, es responsable de un cambio en las
fluctuaciones dominantes y de promover que nuevas
susceptibilidades se vuelvan divergentes para $T\rightarrow0$.


\chapter{Conclusiones}\label{part:conclusionesFinales}

En esta tesis hemos estudiado diversos aspectos te\'oricos de los
sistemas de electrones altamente correlacionados en una
dimensi\'on espacial. En particular hicimos hincapi\'e en el
tratamiento del modelo de Tomonaga-Luttinger con spin y algunas de
sus extensiones, mediante el procedimiento de la bosonizaci\'on.
Al aplicar este m\'etodo a sistemas de materia condensada, el
primer punto en el que hay que detenerse es en el c\'alculo del
determinante fermi\'onico asociado a la teor\'{\i}a. Como el
determinante est\'a mal definido se debe implementar un mecanismo
de regularizaci\'on. El m\'etodo elegido viene usualmente dictado
por las simetr\'ias que posea el modelo en cuesti\'on. En
teor\'{\i}as de materia condensada la ausencia de invarianza de
Lorentz, otorga una libertad a\'un mayor para elegir el regulador.
En el cap\'{\i}tulo \ref{part:jacobiano} mostramos que al utilizar
el regulador usual que preserva la invarianza de Gauge y de
Lorentz en teor\'{\i}as de campos relativistas se arriba a
resultados que no concuerdan con los obtenidos mediante el camino
usual de bosonizaci\'on operacional en materia condensada, y
encontramos la forma exacta que debe tener este regulador para
reproducirlos. Como corolario obtenemos el jacobiano de las
transformaciones quirales, base del desacople fermi\'onico
mediante integrales funcionales\cite{iucci04}. Desafortunadamente,
un principio f\'{\i}sico que sirva de gu\'{\i}a para elegir a
priori el regulador todav\'{\i}a falta, aunque lo mismo ocurre en
el enfoque operacional.

En el cap\'{\i}tulo \ref{part:spinflip} empleamos el m\'etodo de
bosonizaci\'on funcional, mejorado de acuerdo a lo dicho
anteriormente, para estudiar el modelo de Thirring no local con
interacciones que invierten el spin electr\'onico (los
tratamientos previos de las interacciones de inversi\'on de spin
condujeron a un modelo no abeliano en el cual el an\'alisis del
contenido f\'isico se vuelve engorroso). Restringi\'endonos al
caso de interacciones de inversi\'on de spin locales, obtuvimos
una acci\'on bos\'onica efectiva cuyos grados de libertad de carga
coinciden con los encontrados en el cap\'{\i}tulo
\ref{part:jacobiano} al tratar el problema de dispersi\'on hacia
adelante sin spin. Respecto del sector de spin, que es el de mayor
inter\'es, hallamos que se corresponde con un modelo seno-Gordon
no local, cuya integrabilidad, en contraste con el caso local, no
ha sido demostrada hasta el momento. A pesar de este hecho,
pudimos mostrar expl\'{\i}citamente que ambos sectores se
desacoplan dando lugar a la separaci\'on spin-carga\cite{iucci01}.

En el cap\'{\i}tulo \ref{part:aaac} revemos la aproximaci\'on
arm\'onica autoconsistente, y su aplicaci\'on a sistemas de
materia condensada. En particular hacemos una extensi\'on que
permite atacar problemas no locales, y obtenemos una expresi\'on
para el gap del sector de spin del modelo introducido en el
cap\'{\i}tulo \ref{part:spinflip} como funci\'on de los
potenciales de dispersi\'on hacia adelante. Por otro lado, como
una leve digresi\'on con respecto a la l\'{\i}nea principal de
esta tesis, proponemos un camino diferente al usual para la
determinaci\'on del par\'ametro asociado a la aproximaci\'on,
basado en una consecuencia del teorema c de Zamolodchikov en
teor\'{\i}as de campos conformes. Verificamos la validez de la
nueva t\'ecnica en el modelo seno-Gordon, en el que obtenemos una
mejora en los valores aproximados para la masa del soliton como
funci\'on de la constante de acoplamiento $\beta$ del modelo, con
respecto a los predichos mediante el procedimiento usual. Por
\'ultimo aplicamos la nueva t\'ecnica al estudio de un problema no
trivial, para el cual existen escasos resultados anal\'{\i}ticos:
el modelo de Ising bidimensional a $T\neq T_c$ y $h\neq0$.
Hallamos una expresi\'on aproximada que relaciona la longitud de
correlaci\'on $\xi$, $T-T_c$ y $h$ y que resolvemos
num\'ericamente para obtener $\xi$ como funci\'on de $T-T_c$ y
$h$.\cite{iucci02}.

Finalmente en el cap\'{\i}tulo \ref{part:spinorbit} calculamos
funciones de correlaci\'on en sistemas de electrones altamente
correlacionados en una dimensi\'on en los que los grados de
libertad de carga y spin se encuentran acoplados a trav\'es de la
interacci\'on spin-\'orbita. Esta interacci\'on rompe la
separaci\'on spin-carga \cite{moroz00b,moroz00c} y modifica los
exponentes de los decaimientos de las correlaciones. Estudiamos
fluctuaciones de tipo ondas de densidad de carga y spin, y de tipo
superconductor singulete y triplete como funci\'on del
espacio-tiempo eucl\'{\i}deo y de la temperatura. Adem\'as
investigamos las funciones de correlaci\'on instant\'aneas a
temperatura cero, que nos permiten extraer los exponentes
cr\'{\i}ticos. Mostramos que la interacci\'on spin-\'orbita
modifica los exponentes del decaimiento de las funciones de
correlaci\'on y el diagrama de fases del sistema. Adem\'as
encontramos que susceptibilidades que eran finitas a bajas
temperaturas, pueden ahora volverse divergentes a causa de la
interacci\'on spin-\'orbita\cite{iucci03}.


\appendix

\chapter{Propagador bos\'onico a temperatura
finita}\label{part:integrales}

En este ap\'endice describiremos el c\'alculo del
propagador\cite{mattis63,emery79,vondelft98}

\begin{equation}
\Delta^{-1}(x,\tau;\beta)=\frac{1}{\beta}\sum_{n=-\Infinity}^\Infinity
\int_{-\Infinity}^\Infinity\frac{dk}{2\pi}\frac{e^{-ikx-i\omega_n\tau}}{\omega_n^2+v^2
k^2}
\end{equation}
donde $\omega_n=2n\pi/\beta$ son las frecuencias de Matsubara y
$\tau\in[0,\beta]$ ($\Delta^{-1}(x,\tau;\beta)$ es peri\'odica en
$\tau$ con per\'{\i}odo $\beta$). Esta funci\'on es divergente en
el infrarojo, pero nosotros en realidad estamos interesados en la
combinaci\'on

\begin{equation}
\Delta^{-1}(x,\tau;\beta)-\Delta^{-1}(0,0;\beta)=\frac{1}{\beta}\sum_{n=-\Infinity}^\Infinity
\int_{-\Infinity}^\Infinity\frac{dk}{2\pi}e^{-\epsilon
|k|}\frac{e^{-ikx-i\omega_n\tau}-1}{\omega_n^2+v^2 k^2}
\end{equation}
que es regular en esa regi\'on. El t\'ermino sustra\'{\i}do, sin
embargo, es divergente en el ultravioleta, raz\'on por la cual
hemos agregado un factor $e^{-\epsilon|k|}$ que regulariza dicha
divergencia.

A continuaci\'on escribimos $e^{-i \omega_n \tau}$ en t\'erminos
del seno y el coseno y nos quedamos \'unicamente con el t\'ermino
del coseno, ya que el del seno es nulo por simetr\'{\i}a. La
sumatoria resultante puede ser f\'acilmente evaluada mediante el
m\'etodo de los residuos, o consultada la Ref.
\citen{gradshteyn94}:

\begin{equation}\label{eq:sum}
\sum_{n=1}^\Infinity\frac{\cos nx}{n^2+a^2}=\frac{\pi\cosh
a(\pi-x)}{2a\sinh a\pi}-\frac{1}{2a^2}.
\end{equation}

Utilizando este resultado, obtenemos entonces
\begin{equation}
\Delta^{-1}(x,\tau;\beta)-\Delta^{-1}(0,0;\beta)=
\int_{-\Infinity}^\Infinity\frac{dk}{4\pi v|k|}
\frac{e^{-ikx}\cosh\frac{|k|v}{2}(\beta-2\tau)-\cosh\frac{|k|v\beta}{2}}
{\sinh\frac{|k|v\beta}{2}}e^{-\epsilon |k|}.
\end{equation}

Para calcular la integral antes debemos partir la integral en 0 y
en el t\'ermino integrado en la regi\'on negativa de $k$ hacer el
cambio $k\rightarrow-k$. Adem\'as reescribimos las funciones
hiperb\'olicas en t\'erminos de exponenciales, obteniendo

\begin{align}
\Delta^{-1}(x,\tau;\beta)-\Delta^{-1}(0,0;\beta)=&
\int_0^\Infinity\frac{dk}{4\pi v k}e^{-\epsilon k}
\frac{(e^{-ikx}+e^{ikx})\left(e^{-kv\tau}+e^{-kv(\beta-\tau)}\right)
-2\left(1+e^{-kv\beta}\right)} {1-e^{-kv\beta}}\\
=&-\int_0^\Infinity\frac{dk}{4\pi v k}e^{-\epsilon k}
\frac{(1-e^{-s k})(1-e^{(s-r)k})}{1-e^{-rk}}+(s\rightarrow s^*)\\
=&-\int_0^\Infinity\frac{dk}{4\pi v k}e^{-\epsilon k/r}
\frac{(1-e^{-s k/r})(1-e^{-(1-s/r)k})}{1-e^{-k}}+(s\rightarrow
s^*)
\end{align}
donde $s=v\tau+ix$ y $r=v\beta$. Esta \'ultima integral tiene
forma estandar, (3.413(.1)) en la Ref. \citen{gradshteyn94}. Con
este resultado,

\begin{align}
\Delta^{-1}(x,\tau;\beta)-\Delta^{-1}(0,0;\beta)&=\frac{1}{4\pi
v}\ln\frac {\Gamma(\epsilon/r+s/r)\Gamma(1+\epsilon/r-s/r)}
{\Gamma(\epsilon/r)\Gamma(1+\epsilon/r)}+(s\rightarrow
s^*)
\end{align}
donde $\Gamma$ es la funci\'on gamma. Para simplificar este
resultado reemplazamos en la ecuaci\'on anterior $1+\epsilon/r$
por $1-\epsilon/r$, cambio chico si $\epsilon$ es chico, adem\'as
usamos que $\Gamma(z)\Gamma(1-z)=\pi/\sin\pi z$ para dar
finalmente.

\begin{align}
\Delta^{-1}(x,\tau;\beta)-\Delta^{-1}(0,0;\beta)=&\frac{1}{4\pi
v}\ln\frac{\sin\pi\epsilon/r}{\sin\pi(\epsilon/r+s/r)}+(s\rightarrow
s^*)\\[6 pt]
=&\frac{1}{4\pi
v}\ln\frac{\sin\frac{\pi\epsilon}{v\beta}}{\sin\frac{\pi}{v\beta}
(\epsilon+v\tau+ix)}+(\text{c.c.}).
\end{align}

Las restantes integrales
(\ref{eq:integrales})-(\ref{eq:funcionesDeCorrelación}) se
eval\'uan utilizando el m\'etodo de fracciones simples, seguido
del procedimiento descripto en este ap\'endice. A continuaci\'on
damos los resultados:

\begin{multline}
I_1=\frac{1}{\pi\beta}\int_{-\Infinity}^\Infinity
dk\sum_{n=-\Infinity}^\Infinity\left(1-e^{-ikx-i\omega_n\tau}\right)e^{-\epsilon|k|}
\frac{(\omega_n^2+a^2k^2)}{(\omega_n^2+b^2k^2)(\omega_n^2+c^2k^2)}\\
=\frac{1}{2\pi b}\left(\frac{b^2-a^2}{b^2-c^2}\right)
\ln\frac{\sin\frac{\pi}{b\beta}(\epsilon+b\tau+ix)}{\sin\frac{\pi\epsilon}{b\beta}}
+\frac{1}{2\pi c}\left(\frac{c^2-a^2}{c^2-b^2}\right)
\ln\frac{\sin\frac{\pi}{c\beta}(\epsilon+c\tau+ix)}{\sin\frac{\pi\epsilon}{c\beta}}+c.c.
\end{multline}

\begin{multline}
I_2=\frac{1}{\pi\beta}\int_{-\Infinity}^\Infinity
dk\sum_{n=-\Infinity}^\Infinity\left(1-e^{-ikx-i\omega_n\tau}\right)e^{-\epsilon|k|}
\frac{i\omega_n k}{(\omega_n^2+b^2k^2)(\omega_n^2+c^2k^2)}\\
=\frac{\sign x\tau}{2\pi} \left[\frac{1}{b^2-c^2}
\ln\sin\frac{\pi}{b\beta}(\epsilon+b\tau-ix) + \frac{1}{c^2-b^2}
\ln\sin\frac{\pi}{c\beta}(\epsilon+c\tau-ix)\right]+c.c.
\end{multline}

\begin{multline}
I_3=\frac{1}{\pi\beta}\int_{-\Infinity}^\Infinity
dk\sum_{n=-\Infinity}^\Infinity\left(1-e^{-ikx-i\omega_n\tau}\right)e^{-\epsilon|k|}
\frac{i\omega_n}{k(\omega_n^2+b^2k^2)(\omega_n^2+c^2k^2)}\\
=\frac{\sign x\tau}{2\pi} \left[
\left(\frac{b^2-a^2}{b^2-c^2}\right)
\ln\sin\frac{\pi}{b\beta}(\epsilon+b\tau-ix) +
\left(\frac{c^2-a^2}{c^2-b^2}\right)
\ln\sin\frac{\pi}{c\beta}(\epsilon+c\tau-ix)\right]+c.c.
\end{multline}

\chapter{Diagonalizaci\'on del Hamiltoniano
bos\'onico}\label{part:diago}

Supongamos por simplicidad, una versi\'on de dos grados de
libertad de nuestro Hamiltoniano bos\'onico
\ref{eq:HamiltonianoBosónico} de la forma

\begin{equation}
H=\frac{p_1^2}{2 m_1}+\frac{m_1 \omega_1^2}{2}q_1^2+
\frac{p_2^2}{2 m_2}+\frac{m_2 \omega_2^2}{2}q_2^2 + \delta(q_1 p_2
+ q_2 p_1).
\end{equation}
Para diagonalizar este Hamiltoniano no podemos emplear una
transformaci\'on de similitud porque resulta no can\'onica debido
a los t\'erminos que mezclan coordenadas e impulsos. En su lugar
emplearemos m\'etodos simpl\'ecticos como se describen en la
segunda edici\'on del libro de Goldstein\cite{goldstein96}.

En forma general, tomamos un Hamiltoniano de $n$ grados de
libertad y construimos un vector $\xi$ con $2n$ elementos
compuesto por las $n$ coordenadas $q_i$ y los $n$ momentos $p_i$.
Las ecuaciones de Hamilton en notaci\'on simpl\'ectica se escriben

\begin{equation}
\dot{\xi}=\mathsf{J}\frac{\partial H}{\partial\xi}
\end{equation}
donde \textsf{J}, que es la matriz cuadrada de $2n\times 2n$
compuesta por las matrices de $n\times n$ nula e identidad,
seg\'un el esquema

\begin{equation}
\begin{pmatrix}
    \mathsf{0} & \mathsf{1} \\
    -\mathsf{1} & \mathsf{0} \\
\end{pmatrix}.
\end{equation}

Mencionamos algunas propiedades de estas matrices. Su cuadrado es
la identidad cambiada de signo

\begin{equation}\label{eq:propunit}
\mathsf{J}^2=-\mathsf{1}.
\end{equation}
Es tambi\'en ortogonal:

\begin{equation}
\mathsf{\tilde{J}J}=\mathsf{1}
\end{equation}
con lo cual

\begin{equation}
\mathsf{\tilde{J}}=-\mathsf{J}=\mathsf{J}^{-1}.
\end{equation}
De su ortogonalidad se deduce que el cuadrado de su determinante
es 1, si bien se puede probar la afirmaci\'on m\'as fuerte

\begin{equation}
\det\mathsf{J}=+1.
\end{equation}

Una transformaci\'on can\'onica es una tranformaci\'on de las
coordenadas y los impulsos de la forma

\begin{equation}
Q_i=Q_i(q, p),\qquad P_i=P_i(p,q)
\end{equation}
donde $P_i$ y $Q_i$ son las nuevas coordenadas e impulsos, que
preserva sus conmutadores cu\'anticos (o corchetes de Poisson en
el caso cl\'asico). En notaci\'on simpl\'ectica, si definimos al
vector $\eta$ de las nuevas coordenadas e impulsos, la
transformaci\'on resulta

\begin{equation}
\xi=\xi(\eta),
\end{equation}
y la condici\'on para que la transformaci\'on resulte can\'onica
(llamada condici\'on simpl\'ectica) se escribe

\begin{equation}\label{eq:condSimplectica}
\mathsf{MJ\tilde{M}=J}
\end{equation}
donde $\mathsf{M}$ es la matriz jacobiana de la transformaci\'on y
$\mathsf{\tilde{M}}$ su matriz traspuesta.

Sea $H$ un Hamiltoniano cuadr\'atico general de $n$ grados de
libertad

\begin{equation}
H=\frac{1}{2}\,\xi\,\mathsf{S}\,\xi
\end{equation}
donde $\mathsf{S}$ es una matriz cuadrada sim\'etrica constante.
Para hallar la transformaci\'on can\'onica lineal que lo
diagonaliza, apelaremos a las ecuaciones cl\'asicas de movimiento.
Las ecuaciones de Hamilton para este sistema resultan

\begin{equation}\label{eq:ecsHamilton}
\dot{\xi}=\mathsf{JS}\xi.
\end{equation}
Para resolverlas se puede proponer el ansatz

\begin{equation}
\xi=a e^{i\omega t}
\end{equation}
donde $a$ es un vector constante. La ecuaci\'on para $a$
resultante de reemplazar este ansatz en (\ref{eq:ecsHamilton}) es

\begin{equation}\label{eq:eigen}
-i\mathsf{JS}a=\omega a
\end{equation}
es una ecuaci\'on de autovectores, donde los autovalores $\omega$
resultan las frecuencias caracter\'{\i}sticas del sistema. Sea
$\mathsf{U}$ una matriz de un conjunto posible de vectores propios
de $-i\mathsf{JS}$ ordenados en columnas tales que s\'olo sus
direcciones est\'an fijas, no sus m\'odulos, y sea
$\textsf{D}=\diag(\omega_1,...,\omega_n,-\omega_1,...,-\omega_n)$
la matriz de los autovalores. La ecuaci\'on (\ref{eq:eigen}) se
escribe en t\'erminos de $\mathsf{U}$ y $\mathsf{D}$ como

\begin{equation}\label{eq:eigenMatrix}
-i\mathsf{JSU}=\mathsf{UD}.
\end{equation}
Ahora bien, la matriz $\mathsf{U}$ no es en general la matriz de
ninguna transformaci\'on can\'onica, ya que no cumple la
condici\'on simpl\'ectica (\ref{eq:condSimplectica}). Mostraremos
que es posible fijar condiciones sobre los m\'odulos $c_k$ de los
autovectores de modo tal que la nueva matriz de autovectores
cumpla con dicha condici\'on. Para ello introducimos la matriz de
los m\'odulos $\mathsf{C}=\diag{c_k}$, y definimos $\mathsf{M=CU}$
como una nueva matriz de autovectores, donde cada autovector
est\'a multiplicado por una constante $c_k$ a determinar. Si le
exijimos a $\mathsf{M}$ que cumpla la condici\'on simpl\'ectica,
entonces

\begin{equation}
\mathsf{\tilde{U}\tilde{C}JCU=J},
\end{equation}
y multiplicando por $\mathsf{\tilde{U}}^{-1}$ a izquierda y por
$\mathsf{U}^{-1}$ a derecha obtenemos

\begin{equation}
\mathsf{\tilde{C}JC=\tilde{U}^{-1}JU^{-1}}
\end{equation}
que resultan ecuaciones cuadr\'aticas para los $c_k$.

Una vez obtenida la matriz de la transformaci\'on can\'onica,
podemos transformar el Hamiltoniano reemplazando
$\xi=\mathsf{M}\eta$, lo que resulta

\begin{equation}
H=\frac{1}{2}\xi\,\mathsf{S}\,\xi=\frac{1}{2}\eta\,\mathsf{\tilde{M}SM}\eta.
\end{equation}
Por la propiedad (\ref{eq:propunit}), $H$ se puede escribir

\begin{align}
H=&\frac{-i}{2}\,\eta\,\mathsf{\tilde{M}J}(-i\mathsf{JS})\mathsf{M}\,\eta\\
=&\frac{-i}{2}\,\eta\,\mathsf{\tilde{M}JMD}\,\eta\\
=&\frac{-i}{2}\,\eta\,\mathsf{JD}\,\eta,
\end{align}
donde hemos usado la ecuaci\'on de autovectores
(\ref{eq:eigenMatrix}), que tambi\'en satisface $\mathsf{M}$, y la
condici\'on simpl\'ectica para $\mathsf{M}$. Finalmente si
empleamos la definici\'on de $\mathsf{D}$ y $\mathsf{J}$, en
t\'erminos de las nuevas coordenadas e impulsos $\tilde{q}_i$ y
$\tilde{p}_i$ (que componen $\eta$), el Hamiltoniano se escribe

\begin{equation}
H=i\omega_i \tilde{q}_i \tilde{p}_i
\end{equation}
Por \'ultimo, mediante la transformaci\'on can\'onica

\begin{align}
\tilde{q}_i=&\,Q_i-i\frac{P_i}{\omega_i},\\[5 pt]
\tilde{p}_i=&\frac{P_i}{2}-\frac{i\omega_iQ_i}{2},
\end{align}
$H$ resulta diagonal:

\begin{equation}
H=P_i^2+\frac{\omega_i^2}{2}Q_i^2.
\end{equation}

\chapter{Identidad \'util para el c\'alculo de valores medios
bos\'onicos}\label{part:identity}

En este ap\'endice probaremos la identidad

\begin{equation}\label{eq:identity}
\bra\exp\left\{i\sum_k\beta_k\left[\varphi_k(x)-\varphi_k(0)\right]\right\}\ket=
\exp \left\{
\sum_{i,j}\beta_i\beta_j\left[\Delta_{ij}^{-1}(x)-\Delta_{ij}^{-1}(0)\right]
\right\}
\end{equation}
donde $\varphi$ es un campo bos\'onico y $\Delta_{ij}^{-1}(x)$ es
el valor medio bos\'onico

\begin{equation}
\bra\varphi_i(x)\varphi_j(y)\ket=\Delta^{-1}_{ij}(x-y).
\end{equation}
La identidad (\ref{eq:identity}) es v\'alida para valores medios
calculados con acci\'ones bos\'onicas cuadr\'aticas en el campo
$\varphi$. Sea $S$ una acci\'on de este tipo

\begin{equation}
S=\frac{1}{2}\int d^2x\,\varphi_i(x)\Delta_{ij}\,\varphi_j(x)
\end{equation}
donde $\Delta_{ij}$ es un operador sim\'etrico en los \'{\i}ndices
$i$ y $j$. La funci\'on $\Delta_{ij}^{-1}(x)$ resulta ser entonces
la funci\'on de Green de este operador. El lado izquierdo de
(\ref{eq:identity}) se puede calcular mediante el siguiente
procedimiento:

\begin{align}
&\bra\exp\left\{i\sum_k\beta_k\left[\varphi_k(x)-\varphi_k(0)\right]\right\}\ket=\\
&=\frac{1}{\Z_0}\int\D\varphi\exp\left\{-\frac{1}{2}\int
d^2x'\,\varphi_i(x')\Delta_{ij}\,\varphi_j(x')+i\sum_k\beta_k\left[\varphi_k(x)-\varphi_k(0)\right]\right\}\\
&=\frac{1}{\Z_0}\int\D\varphi\exp\left\{-\frac{1}{2}\int
d^2x'\,\varphi_i(x')\Delta_{ij}\,\varphi_j(x')+ \int d^2x'\,\varphi_k(x')i\beta_k\left[\delta(x'-x)-\delta(x')\right]\right\}\\
&=\frac{1}{\Z_0}\int\D\varphi\exp\left\{-\frac{1}{2}\int
d^2x'\,\varphi_i(x')\Delta_{ij}\,\varphi_j(x')-\int
d^2x'\,\varphi_k(x')j_k(x',x)\right\}
\end{align}
donde $\Z_0$ es la funci\'on de partici\'on

\begin{equation}
\Z_0=\int\D\varphi\exp\left\{-\frac{1}{2}\int
d^2x'\,\varphi_i(x')\Delta_{ij}\,\varphi(x')\right\}
\end{equation}
y definimos

\begin{equation}\label{eq:defj}
j_k(x',x)=-i\beta_k\left[\delta(x'-x)-\delta(x')\right].
\end{equation}
que es independiente de los campos. A continuaci\'on hacemos una
traslaci\'on en el campo $\varphi$

\begin{equation}
\varphi_i(x')\rightarrow\varphi_i(x')+\int d^2y\,
\Delta^{-1}_{ik}(x'-y)j_k(y,x).
\end{equation}
Este es el \'unico cambio en la integral funcional porque la
medida de integraci\'on funcional es invariante frente a
traslaciones. Finalmente, el valor medio queda

\begin{align}
&\bra\exp\left\{i\sum_k\beta_k\left[\varphi_k(x)-\varphi_k(0)\right]\right\}\ket\\
&=\exp\left\{\frac{1}{2}\int
d^2x'\,d^2y\,j_i(x',x)\Delta^{-1}_{ij}(x'-y)j_j(y,x)\right\}\\
&=\exp \left\{
\sum_{i,j}\beta_i\beta_j\left[\Delta_{ij}^{-1}(x)-\Delta_{ij}^{-1}(0)\right]
\right\}
\end{align}
donde en el \'ultimo paso hemos reemplazado $j_k(x',x)$ por su
definici\'on (\ref{eq:defj}), e integramos en $x'$ e $y$.


\begin{thebibliography}{100}

\bibitem{sommerfeld28}
A.~Sommerfeld,
\newblock Z. Physik {\bf 47}, 1 (1928).

\bibitem{pauli26}
W.~Pauli,
\newblock Z. Physik {\bf 41}, 81 (1926).

\bibitem{bloch29}
F.~Bloch,
\newblock Z. Physik {\bf 57}, 545 (1929).

\bibitem{wigner34}
E.~P. Wigner,
\newblock Phys. Rev. {\bf 46}, 1002 (1934).

\bibitem{landau56}
L.~D. Landau,
\newblock Zh. Eksp. Teor. Fiz. {\bf 30}, 1058 (1956).

\bibitem{landau57}
L.~D. Landau,
\newblock Zh. Eksp. Teor. Fiz. {\bf 32}, 59 (1957).

\bibitem{nozieres64}
P. Nozi\`eres,
\newblock {\em Interacting Fermi Systems},
\newblock W. A. Benjamin Inc., 1964.

\bibitem{bockrath99}
M.~Bockrath, D.~H. Cobden, J.~Lu, A.~G. Rinzler, R.~E. Smalley,
T.~Ballents, y
  P.~L. McEuen,
\newblock Nature {\bf 397}, 598 (1999).

\bibitem{wen90a}
X.~G. Wen,
\newblock Phys. Rev. B {\bf 41}, 12838 (1990).

\bibitem{wen90b}
X.~G. Wen,
\newblock Phys. Rev. Lett {\bf 64}, 2206 (1990).

\bibitem{kang00}
W.~Kang, H.~L. Stomer, K.~W. Baldwin, L. N. Pfeiffer, y K.~W.
West,
\newblock Nature {\bf 403}, 59 (2000).

\bibitem{jerome82}
D.~J\'erome y H.~J. Schulz,
\newblock Adv. Phys. {\bf 31}, 299 (1982).

\bibitem{tarucha95}
S.~Tarucha, T.~Honda, y T.~Saku,
\newblock Solid State Comm. {\bf 94}, 413 (1995).

\bibitem{tomonaga50}
S.~Tomonaga,
\newblock Prog. Theor. Phys. {\bf 5}, 544 (1950).

\bibitem{luttinger63}
J.~M. Luttinger,
\newblock J. Math. Phys. {\bf 4}, 1154 (1963).

\bibitem{mattis63}
D.~C. Mattis y E.~H. Lieb,
\newblock J. Math, Phys {\bf 6}, 304 (1963).

\bibitem{voit95}
J.~Voit,
\newblock Rep. Prog. Phys. {\bf 58}, 977 (1995).

\bibitem{haldane81}
F.~D.~M. Haldane,
\newblock J. Phys. C {\bf 14}, 2585 (1981).

\bibitem{anderson90a}
P.~W. Anderson,
\newblock Phys. Rev. Lett. {\bf 64}, 1839 (1990).

\bibitem{anderson90b}
P.~W. Anderson,
\newblock Phys. Rev. Lett. {\bf 65}, 2306 (1990).

\bibitem{postma00}
H.~W.~C. Postma, M.~de~Jonge, Z.~Yao, y C.~Dekker,
\newblock Phys. Rev. B {\bf 62}, R10653 (2000).

\bibitem{gao03}
B.~Gao, A.~Komnik, R.~Egger, D.~C. Glattli, y A.~Bachtold,
\newblock cond-mat/0311645 .

\bibitem{hilke03}
M.~Hilke, D.~C. Tsui, L. N. Pfeiffer, y K.~W. West,
\newblock J. Phys. Soc. Jpn. {\bf 72}, 92 (2003).

\bibitem{tserkovnyak03}
Y.~Tserkovnyak, B.~I. Halperin, O.~M. Auslaender, y A.~Yacoby,
\newblock cond-mat/0312159 .

\bibitem{mattis74}
D.~C. Mattis,
\newblock J. Math, Phys {\bf 15}, 609 (1974).

\bibitem{luther74a}
A.~Luther y I.~Peschel,
\newblock Phys. Rev. B {\bf 9}, 2911 (1974).

\bibitem{klaiber68}
B.~Klaiber,
\newblock en {\em Lectures in Theoretical Physics}, editado por A.~Barut,
  Gordon and Reach, 1968.

\bibitem{coleman75}
S.~Coleman,
\newblock Phys. Rev. D {\bf 11}, 2088 (1975).

\bibitem{mandelstam75}
S.~Mandelstam,
\newblock Phys. Rev. D {\bf 11}, 3026 (1975).

\bibitem{fujikawa79}
K.~Fujikawa,
\newblock Phys. Rev. Lett. {\bf 42}, 1195 (1979).

\bibitem{furuya82}
K.~Furuya, R.~E.~G. Sarav\'{\i}, y F.~A. Schaposnik,
\newblock Nucl. Phys. B {\bf 208}, 159 (1982).

\bibitem{naon85}
C.~M. Na\'on,
\newblock Phys. Rev. D {\bf 31}, 2035 (1985).

\bibitem{fogebdy76}
H.~C. Fogebdy,
\newblock J. Phys. C {\bf 9}, 3757 (1976).

\bibitem{lee88}
D.~K.~K. Lee y Y.~Chen,
\newblock J. Phys. A {\bf 21}, 4155 (1988).

\bibitem{naon95}
C.~M. Na\'on, M.~C. von Reichenbach, y M.~L. Trobo,
\newblock Nucl. Phys. B {\bf 435[FS]}, 567 (1995).

\bibitem{iucci00}
A.~Iucci y C. Na\'on,
\newblock Phys. Rev. B {\bf 61}, 15530 (2000).

\bibitem{moroz00b}
A.~V. Moroz, K.~V. Samokhin, y C.~H.~W. Barnes,
\newblock Phys. Rev. Lett {\bf 84}, 4164 (2000).

\bibitem{moroz00c}
A.~V. Moroz, K.~V. Samokhin, y C.~H.~W. Barnes,
\newblock Phys. Rev. B {\bf 62}, 16900 (2000).

\bibitem{vondelft98}
J.~von Delft y H.~Schoeller,
\newblock Ann. Phys. (Leipzig) {\bf 7}, 225 (1998).

\bibitem{varma02}
C.~M. Varma, Z. Nussinov, y W.~van Saarloos,
\newblock Phys. Rep. {\bf 361}, 267 (2002).

\bibitem{iucci04}
A.~Iucci y C.~M. Na\'on,
\newblock hep-th/0311128 .

\bibitem{iucci01}
A.~Iucci, K.~Li, y C.~M. Na\'on,
\newblock Nucl. Phys. B {\bf 601[FS]}, 607 (2001).

\bibitem{iucci02}
A.~Iucci y C.~M. Na\'on,
\newblock J. Phys. A {\bf 35}, 8005 (2002).

\bibitem{iucci03}
A.~Iucci,
\newblock Phys. Rev. B {\bf 68}, 075107 (2003).

\bibitem{fernandez02a}
V.~I. Fern\'andez, A.~Iucci, y C.~M. Na\'on,
\newblock Nucl. Phys. B {\bf 636[FS]}, 514 (2002).

\bibitem{fernandez02b}
V.~I. Fern\'andez, A.~Iucci, y C.~M. Na\'on,
\newblock Eur. Phys. J. B {\bf 30}, 53 (2002).

\bibitem{negele88}
J.~W. Negele y H.~Orland,
\newblock {\em Quantum Many-Particle Systems},
\newblock Addison-Wesley, 1988.

\bibitem{emery79}
V.~J. Emery,
\newblock en {\em Higly Conducting One-Dimensional Solids}, editado por J.~T.
  Devreese et~al., page 247, Plenum, New York, 1979.

\bibitem{bethe31}
H.~Bethe,
\newblock Z. Phys. {\bf 71}, 205 (1931).

\bibitem{solyom79}
J.~Solyom,
\newblock Adv. Phys. {\bf 28}, 209 (1979).

\bibitem{ramond96}
P.~Ramond,
\newblock {\em Field Theory, a modern premier},
\newblock Cambridge University Press, 1996.

\bibitem{fujikawa80a}
K.~Fujikawa,
\newblock Phys. Rev. D {\bf 21}, 2848 (1980).

\bibitem{fujikawa80b}
K.~Fujikawa,
\newblock Phys. Rev. D erratum {\bf 22}, 1499 (1980).

\bibitem{roskies81}
R.~Roskies y F.~Schaposnik,
\newblock Phys. Rev. D {\bf 23}, 558 (1981).

\bibitem{gamboa81}
R.~E.~G. Sarav\'{\i}, F.~A. Schaposnik, y J.~E. Solomin,
\newblock Nucl. Phys. B {\bf 185}, 239 (1981).

\bibitem{stone94}
M.~Stone,
\newblock {\em Bosonization},
\newblock World Scientific Pub. Co. Pte. Ltd., 1994.

\bibitem{rubin86}
M.~Rubin,
\newblock J. Phys. A {\bf 19}, 2105 (1986).

\bibitem{cabra89}
D.~Cabra y F.~Schaposnik,
\newblock J. Math. Phys. {\bf 30}, 816 (1989).

\bibitem{jackiw85a}
R.~Jackiw y R.~Rajaraman,
\newblock Phys. Rev. Lett. {\bf 54}, 1219 (1985).

\bibitem{jackiw85b}
R.~Jackiw y R.~Rejaraman,
\newblock Phys. Rev. Lett. {\bf 55}, 224 (1985).

\bibitem{fujikawa03}
K.~Fujikawa y H.~Suzuki,
\newblock hep-th/0305008 .

\bibitem{schulz00}
H.~Schulz, G.~Cuniberti, y P.~Pieri,
\newblock en {\em Field Theories for Low-Dimensional Condensed Matter Systems},
  editado por G.~Morandi et~al., Springer, 2000.

\bibitem{rao01}
S.~Rao y D.~Sen,
\newblock en {\em Field Theories in Condensed Matter Systems}, editado por
  S.~Rao, page 239, Hindustan Book Agency, New Delhi, 2001.

\bibitem{gamboa84}
R.~E.~G. Sarav\'{\i}, M.~A. Muschietti, F.~A. Schaposnik, y J.~E.
Solomin,
\newblock Ann. Phys. (NY) {\bf 157}, 360 (1984).

\bibitem{giamarchi91}
T.~Giamarchi,
\newblock Phys. Rev. B {\bf 44}, 2905 (1991).

\bibitem{naon97}
C.~M. Na\'on, M.~C. von Reichenbach, y M.~L. Trobo,
\newblock Nucl.Phys. B {\bf 485[FS]}, 665 (1997).

\bibitem{manias98}
M.~V. Man\'{\i}as, C.~M. Na\'on, y M.~L. Trobo,
\newblock Nucl. Phys. B {\bf 525[FS]}, 721 (1998).

\bibitem{zinn-justin97}
J.~Zinn-Justin,
\newblock {\em Quantum Field Theory and Critical Phenomena},
\newblock Oxford University Press, 1997.

\bibitem{grinstein79}
G.~Grinstein, P.~Minnhagen, y A.~Rosengren,
\newblock J. Phys. C {\bf 12}, 1271 (1979).

\bibitem{li98}
K.~Li y C.~M. Na\'on,
\newblock J. Phys. A {\bf 31}, 7929 (1998).

\bibitem{luther74b}
A.~Luther y V.~J. Emery,
\newblock Phys. Rev. Lett. {\bf 33}, 589 (1974).

\bibitem{saito78}
Y.~Saito,
\newblock Z. Phys. B {\bf 32}, 75 (1978).

\bibitem{fisher85}
M.~P.~A. Fisher y W.~Zwerger,
\newblock Phys. Rev. B {\bf 32}, 6190 (1985).

\bibitem{gogolin93}
A.~O. Gogolin,
\newblock Phys. Rev. Lett. {\bf 71}, 2995 (1993).

\bibitem{prokof'ev94}
N.~V. Prokof'ev,
\newblock Phys. Rev. B {\bf 49}, 2243 (1994).

\bibitem{egger95}
R.~Egger y H.~Grabert,
\newblock Phys. Rev. Lett. {\bf 75}, 3505 (1995).

\bibitem{xu96}
B.~W. Xu y Y.~M. Zhang,
\newblock J. Phys. A {\bf 29}, 7349 (1996).

\bibitem{stevenson84}
M.~Stevenson,
\newblock Phys. Rev. D {\bf 30}, 1712 (1984).

\bibitem{stevenson85}
M.~Stevenson,
\newblock Phys. Rev. D {\bf 32}, 1389 (1985).

\bibitem{ingermanson86}
R.~Ingermanson,
\newblock Nucl. Phys. B {\bf 266}, 620 (1986).

\bibitem{stevenson81}
M.~Stevenson,
\newblock Phys. Rev. D {\bf 23}, 2916 (1981).

\bibitem{kauffman84}
S.~K. Kauffman y S.~M. Perez,
\newblock J. Phys. A {\bf 17}, 2027 (1984).

\bibitem{dashen75}
R.~Dashen, B.~Hasslacher, y A. Neveu,
\newblock Phys. Rev. D {\bf 11}, 3424 (1975).

\bibitem{delfino98}
G.Delfino y G.~Mussardo,
\newblock Nucl. Phys. B {\bf 516}, 675 (1998).

\bibitem{voit92}
J.~Voit,
\newblock Phys. Rev. B {\bf 45}, 4027 (1992).

\bibitem{schulz96}
H.~J. Schulz,
\newblock Phys. Rev. B {\bf 53}, R2959 (1996).

\bibitem{boyanovsky90}
D.~Boyanovsky y C. Na\'on,
\newblock Riv. Nuovo Cimento {\bf 13}, 1 (1990).

\bibitem{difrancesco99}
P.~D. Francesco, P.~Mathieu, y D.~S\'en\'echal,
\newblock {\em Conformal Field Theory},
\newblock Springer, 1999.

\bibitem{zamolodchikov95}
A.~B. Zamolodchikov,
\newblock Int. J. Mod. Phys. A {\bf 10}, 1125 (1995).

\bibitem{mccoy78}
B.~M. McCoy y T.~T. Wu,
\newblock Phys. Rev. D {\bf 18}, 1259 (1978).

\bibitem{mccoy95}
B.~M. McCoy,
\newblock en {\em Statistical Mechanics and Field Theory}, editado por V.~V.
  Bazhanov y C.~J. Burden, pages 26--128, World Scientific, Singapore, 1995.

\bibitem{delfino96}
G.~Delfino, G.~Mussardo, y P.~Simonetti,
\newblock Nucl. Phys. B {\bf 473}, 469 (1996).

\bibitem{feynman72}
R.~P. Feynman,
\newblock {\em Statistical Mechanics - A set of Lectures},
\newblock W. A. Benjamin Inc., 1972.

\bibitem{zamolodchikov86}
A.~B. Zamolodchikov,
\newblock JETP Lett. {\bf 43}, 730 (1986).

\bibitem{cardy88a}
J.~L. Cardy,
\newblock Phys. Rev. Lett. {\bf 60}, 2709 (1988).

\bibitem{cardy88b}
J.~L. Cardy,
\newblock J. Phys. A {\bf 21}, L797 (1988).

\bibitem{hecht67}
R.~Hecht,
\newblock Phys. Rev. {\bf 158}, 557 (1967).

\bibitem{wu76}
T.~T. Wu, B.~M. McCoy, C.~A. Tracy, y E.~Barouch,
\newblock Phys. Rev. D {\bf 13}, 316 (1976).

\bibitem{zamolodchikov90}
A.~B. Zamolodchikov,
\newblock Nucl. Phys. B {\bf 342}, 695 (1990).

\bibitem{fateev94}
V.~A. Fateev,
\newblock Phys. Lett. B {\bf 324}, 45 (1994).

\bibitem{belavin84}
A.~A. Belavin, A.~M. Polyakov, y A.~B. Zamolodchikov,
\newblock Nucl. Phys. B {\bf 241}, 333 (1984).

\bibitem{friedan84}
D.~Friedan, Z.~Qiu, y S.~Shenker,
\newblock Phys. Rev. Lett. {\bf 52}, 1575 (1984).

\bibitem{dresselhaus55}
G.~Dresselhaus,
\newblock Phys. Rev. {\bf 100}, 580 (1955).

\bibitem{rashba60a}
E.~I. Rashba,
\newblock Fiz. Tverd. Tela (Leningrad) {\bf 2}, 1224 (1960).

\bibitem{nitta97}
J. Nitta, T.~Akazaki, H.~Takayanagi, y T.~Enoki,
\newblock Phys. Rev. Lett. {\bf 78}, 1335 (1997).

\bibitem{hassenkam97}
T.~Hassenkam, S.~Pedersen, K.~Baklanov, K.~Kristensen, C.~B.
Sorensen, P.~E.
  Lindelof, F.~G. Pikus, y G.~E. Pikus,
\newblock Phys. Rev. B {\bf 55}, 9298 (1997).

\bibitem{meier02}
G.~Meier, T.~Matsuyama, y U.~Merkt,
\newblock Phys. Rev. B {\bf 65}, 125327 (2002).

\bibitem{miller03}
J.~B. Miller, D.~M. Zymbuhl, C.~M. Marcus, Y.~B. Lyanda-Geller,
  D.~Goldhaber-Gordon, K.~Campman, y A.~C. Gossard,
\newblock Phys. Rev. Lett. {\bf 90}, 076807 (2003).

\bibitem{wolf01}
S.~A. Wolf,
\newblock Science {\bf 294}, 1488 (2001).

\bibitem{aronov93}
A.~G. Aronov y Y.~B. Lyanda-Geller,
\newblock Phys. Rev. Lett. {\bf 70}, 343 (1993).

\bibitem{koga02}
T.~Koga, J. Nitta, H.~Takayanagi, y S.~Datta,
\newblock Phys. Rev. Lett. {\bf 88}, 126601 (2002).

\bibitem{datta90}
S.~Datta y B.~Das,
\newblock Appl. Phys. Lett.,
\newblock  {\bf 56}, 665 (1990).

\bibitem{yacoby96}
A.~Yacoby, H.~L. Stormer, N.~S. Wingreen, K.~W. Baldwin, y K.~W.
West,
\newblock Phys. Rev. Lett {\bf 77}, 4612 (1996).

\bibitem{moroz99}
A.~V. Moroz y C.~H.~W. Barnes,
\newblock Phys. Rev. B {\bf 60}, 14272 (1999).

\bibitem{giamarchi89}
T.~Giamarchi y H.~J. Schulz,
\newblock Phys. Rev. B {\bf 39}, 4620 (1989).

\bibitem{drut03}
J.~Drut y D.~Cabra,
\newblock J. Phys.:Condens. Matter {\bf 15}, 1445 (2003).

\bibitem{penc94}
K.~Penc y F.~Mila,
\newblock Phys. Rev. B {\bf 49}, 9670 (1994).

\bibitem{ogata91}
M.~Ogata, M.~U. Luchini, S.~Sorella, y F.~F. Assaad,
\newblock Phys. Rev. Lett. {\bf 66}, 2388 (1991).

\bibitem{eckert98}
D.~Eckert, K.~Ruck, M.~Wolf, G.~Krabbes, y K.~H. M{\"u}tter,
\newblock J. Appl. Phys. {\bf 83}, 7240 (1998).

\bibitem{gerhardt98}
C.~Gerhardt, K.~H. M{\"u}tter, y H.~Kr{\"o}ger,
\newblock Phys. Rev. B {\bf 57}, 11504 (1998).

\bibitem{gradshteyn94}
I.~S. Gradshteyn y I.~M. Ryzhik,
\newblock {\em Table of Integrals, Series, and Products},
\newblock Academic Press, 1994.

\bibitem{goldstein96}
H.~Goldstein,
\newblock {\em Mec\'anica Cl\'asica},
\newblock Revert\'e, 1996.

\end{thebibliography}

\end{document}